# Development of Scientific Skills via IYPT

## How does YPT participation lead to hard-skill development?

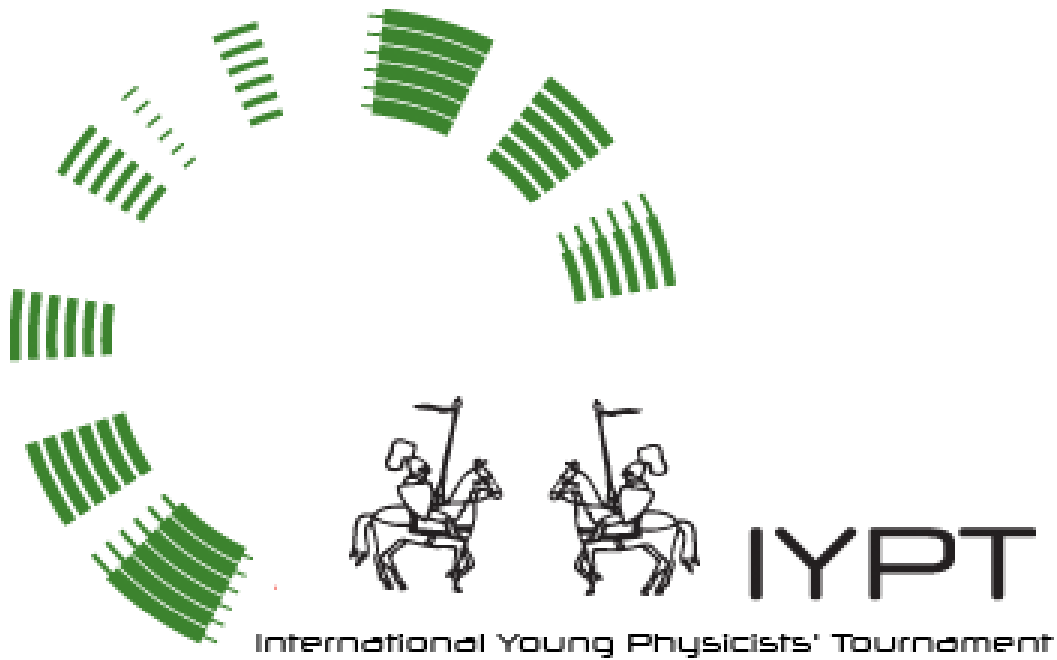

**Title**: Development of Scientific Skills via IYPT

**Subtitle**: How does YPT participation lead to hard-skill development?

**Authors**: Boyka Aneva[1], Sergej Faletič[2], Mihály Hömöstrei[3], Péter Jenei[3], Isza Éva[3], František Kundracik[4], Assen Kyuldjiev[1], Thomas Lindner[5], Hynek Němec[6], Harald Puhr[6], and Martin Plesch[7]*

**Reviewed:** Alan Allinson[8]

[1] Institute of Nuclear Research and Nuclear Energy, Bulgarian Academy of Sciences, Sofia, Bulgaria
[2] University of Ljubljana Faculty of Mathematics and Physics, Ljubljana, Slovenia
[3] ELTE Institute of Physics, Budapest, Hungary
[4] Faculty of Mathematics, Physics and Informatics, Comenius University, Bratislava, Slovakia
[5] Faculty of Business, University of Innsbruck, Innsbruck, Austria
[6] Institute of Physics of the Czech Academy of Sciences, Praha, Czech Republic
[7] Institute of Physics, Slovak Academy of Sciences, Bratislava, Slovakia
* martin.plesch@savba.sk
[8] Brisbane Girls Grammar School, Brisbane, Australia

The European Commission's support for the production of this publication does not constitute an endorsement of the contents, which reflect the views only of the authors, and the Commission cannot be held responsible for any use which may be made of the information contained therein.

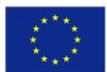

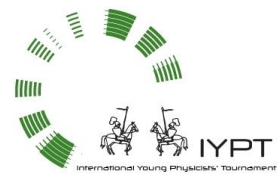

# Content

The European Commission's support for the production of this publication does not constitute an endorsement of the contents, which reflect the views only of the authors, and the Commission cannot be held responsible for any use which may be made of the information contained therein.

# The relationship between inquiry-based learning in YPT and the development of hard skills

*IO3 Dibali: 2019-1-SK01-KA201-060798*

## REPORT

## What is this project about?

Students from about 35 countries around the world regularly participate in the IYPT (International Young Physicists' Tournament) research-based physics competition. Compared to other traditional physics competitions, IYPT differs not only in that students work on open-ended problems instead of solving closed-ended problems, but also in that their results need to be not only presented but even discussed in English – which is mostly not the mother tongue of the students –, all as part of a team. In the following, we refer to the competitions based on the IYPT methodology as YPT (type) competitions which are mainly the national qualifiers and IYPT. In our project, we investigated the effect of YPT preparation and participation on students' hard (scientific) skills compared to RPC (regular physics classes) and Non-YPT type competitions.

This intellectual output is concerned with the question how inquiry-based learning contributes to the development of hard skills in high school students. To this end, two research activities were conducted. First, we investigated how students perceive the role of YPT participation in their development of hard skills. Second, we investigated how students' teachers assess the contribution of YPT participation to students' hard skills development. Taken together, the two steps, investigating the relationship between inquiry-based learning and hard-skill development, allow building inference about how inquiry-based learning helps students to build hard skills, and how these hard skills influence student performance in research tasks. The data for the two stages includes 308 student responses for stage one, 33 teacher responses for stage two. Condensing the detailed findings from our analysis, we suggest eight guidelines for developing hard skills of students below. In the supplementary materials that complement this report, we present our findings in  detail. These supplementary materials consist of four sections. The first section shows survey results on students' assessment of hard-skill development through regular physics classes, YPT-related activities, and other, non-YPT extracurricular activities. The second section present results from a survey of teachers' assessment of hard-skill development through these three types of activities. The third section shows a detailed comparison of students' and teachers' answers. In the fourth section  the two master theses of the analysis and its conclusion can be read Throughout the report, we refer to the respective sections in the supplementary materials.

## The Investigated Scientific skills

In the case of scientific skills, we are dealing with a basically ample set. We used this list of competencies to select those that can play an important role in the analysis. We compared the effect of YPT-type competitions with the effect of RPC and non-YPT competitions on the competences we have chosen. The hard skills we examined are the following:

- "High school mathematics"

The European Commission's support for the production of this publication does not constitute an endorsement of the contents, which reflect the views only of the authors, and the Commission cannot be held responsible for any use which may be made of the information contained therein.

- "High school physics"
- "Solving close-ended problems in physics"
- "Designing experiments"
- "Conducting experiment (based on clear manual) = Cookbook experiments"
- "Interpreting experimental data, data analysis"
- "Developing own theoretical model"
- "Numerical simulations"
- "Interdependent research in scientific literature"
- "Critical assessment of others' results"

# Guidelines for Developing Hard Skills through Inquiry-Based Learning in YPT

## I. YPT participation reinforces hard-skill development

In the survey, students were asked to evaluate their own hard skills, they had to indicate their opinion on a 5-point Likert-scale.. In addition to this evaluation, students also indicated the usefulness of RPC , YPT-related activities, and other Non-YPT activities that may develop these hard skills. All students had to indicate their opinion on the influence of skill-development of RCP, YPT-related activities and Non-YPT extracurricular activities by answering a set of questions for each type of activities. . Yet within these groups of questions, not all students responded to all questions about all hard skills. Therefore, between hard skills the number of responses varies between 140 and 280.

On average, students evaluated their hard skills very positively (median = 4). The lowest mean evaluations were on "Critical assessment of others' results" (3,31). The highest self-evaluations were on "High school mathematics" (4,18) and "Independent research in scientific literature" (4,18). Self-evaluations for all categories of hard skills were positively correlated (r ~ 0,3-0,6).

In the survey, students responded that regular physics classes, YPT-related activities, as well as other Non-YPT activities were mostly useful for the development of their hard skills (median ≥ 3). However, according to the results, YPT-related activities and other Non-YPT activities were perceived more useful in increasing hard skills than regular physics classes (see tables of tests below). For all three types of activities, usefulness was positively correlated across categories of hard skills. These correlations were highest for YPT-related activities (r ~ 0,5-0,8), indicating that YPT-related activities have the most holistic impact on hard skills.

<table>
<tr><td colspan="5" align="center"><strong>Comparison of Usefulness of RPC vs. YPT</strong></td></tr>
<tr><th>Hard Skills in RPC</th><th>Hard Skills in YPT</th><th>t</th><th>df</th><th>P</th></tr>
<tr><td>High school mathematics</td><td>High school mathematics</td><td>0,288</td><td>136</td><td>0,774</td></tr>
<tr><td>High school physics</td><td>High school physics</td><td>0,524</td><td>184</td><td>0,601</td></tr>
<tr><td><strong>Solving close-ended problems</strong></td><td>Solving close-ended problems</td><td><strong>4,409</strong></td><td><strong>178</strong></td><td><strong>0,000</strong></td></tr>
<tr><td>Designing experiments</td><td><strong>Designing experiments</strong></td><td><strong>-3,157</strong></td><td><strong>131</strong></td><td><strong>0,002</strong></td></tr>
<tr><td>Conducting experiment</td><td>Conducting experiment</td><td>-0,095</td><td>176</td><td>0,924</td></tr>
<tr><td>Interpreting experimental data, data analysis</td><td><strong>Interpreting experimental data, data analysis</strong></td><td><strong>-3,593</strong></td><td><strong>180</strong></td><td><strong>0,000</strong></td></tr>
<tr><td>Developing own theoretical model</td><td><strong>Developing own theoretical model</strong></td><td><strong>-8,185</strong></td><td><strong>173</strong></td><td><strong>0,000</strong></td></tr>
</table>

The European Commission's support for the production of this publication does not constitute an endorsement of the contents, which reflect the views only of the authors, and the Commission cannot be held responsible for any use which may be made of the information contained therein.

| | | | | |
|---|---|---|---|---|
| Numerical simulations | **Numerical simulations** | **-7,447** | **170** | **0,000** |
| Independent research in scientific literature | **Independent research in scientific literature** | **-1,760** | **169** | **0,080** |
| Critical assessment of others' results | **Critical assessment of others' results** | **-4,323** | **173** | **0,000** |

*Note: Student's t-Test, coefficients with p ≤ .05 highlighted bold, positive t-value means better in RPC, negative in YPT.*

To verify the descriptive statistics (see 1. Supplement 1.2.2), we use t-tests to test differences between the perceived usefulness of RPC (regular physics classes), YPT-related activities, and other Non-YPT activities. The results give a highly differentiated picture. While regular physics classes seem to be more useful for "Solving close-ended problems" (p = 0,000) than YPT-related activities, we find that YPT-related activities and other activities are more useful than regular physics classes for "Designing experiments", "Interpreting experimental data, data analysis", "Developing own theoretical model", "Numerical simulations", "Independent research in scientific literature", and "Critical assessment of others' results".

For teachers, this result helps a lot to allocate the available time frame and resources optimally. Students do not find YPT-type activities primarily useful in expanding their basic professional knowledge. Therefore, among students, that are not quite interested in phyisics and the aim of the teacher is to mediate basic physics, we recommend that teachers use simpler and more familiar traditional physics teachingmethods. However, if the goal is (also) to develop hard skills beyond basic knowledge, then YPT-type activities seem to be much more effective than traditional methods.

## II. Addition of inquiry-based learning and other extracurricular activities

As part of the survey (see 1. Supplement 1.2.3), students also evaluated the usefulness of other Non-YPT-like extracurricular activities (e.g., Physics Olympiad, IJSO, EUSO, or Project Science Competition). Overall, students considered these extracurricular activities as useful to develop their hard skills as YPT-activities. We find that, based on students' self-evaluation, Non-YPT extracurricular activities had a significantly stronger impact on hard skills than RPC: "Designing experiments" (p=0,000), "Interpreting experimental data, data analysis" (p=0,032), "Developing own theoretical model" (p=0,000), "Numerical simulations" (p=0,000), "Independent research in scientific literature" (p=0,000), "Critical assessment of others' results" (p=0,000).

**Usefulness of RPC vs. Non-YPT activities**

| **Hard Skills in RPC** | **Hard Skills in Non-YPT** | **t** | **df** | **P** |
|---|---|---|---|---|
| High school mathematics | High school mathematics | -1,160 | 185 | 0,248 |
| High school physics | High school physics | 0,419 | 262 | 0,676 |
| Solve close-ended problems | Solve close-ended problems | 1,425 | 240 | 0,156 |
| Designing experiments | **Designing experiments** | **-4,715** | **240** | **0,000** |
| Conducting experiment | Conducting experiment | 1,108 | 232 | 0,269 |
| Interpreting experimental data, data analysis | **Interpreting experimental data, data analysis** | **-2,156** | **238** | **0,032** |
| Developing own theoretical model | **Developing own theoretical model** | **-5,971** | **228** | **0,000** |
| Numerical simulations | **Numerical simulations** | **-6,490** | **216** | **0,000** |
| Independent research in scientific literature | **Independent research in scientific literature** | **-8,060** | **238** | **0,000** |
| Critical assessment of others' results | **Critical assessment of others' results** | **-4,315** | **233** | **0,000** |

*Note: Student's t-Test, coefficients with p ≤ .05 highlighted bold, positive t-value means better in RPC, negative in Non-YPT.*

In comparison to YPT-related activities, we observe greater perceived usefulness for Non-YPT extracurricular activities for some types of hard skills, which might not be surprising.

### Usefulness of YPT activities vs. Non-YPT activities

| Hard Skills in YPT | Hard Skills in Non-YPT | t | df | p |
|---|---|---|---|---|
| High school mathematics | High school mathematics | -1,000 | 128 | 0,319 |
| High school physics | High school physics | -0,495 | 178 | 0,621 |
| Solve close-ended problems | **Solve close-ended problems** | **-2,588** | **169** | **0,010** |
| Designing experiments | Designing experiments | -0,076 | 127 | 0,939 |
| Conducting experiment | Conducting experiment | 0,648 | 168 | 0,518 |
| **Interpreting experimental data, data analysis** | Interpreting experimental data, data analysis | **2,970** | **175** | **0,003** |
| **Developing own theoretical model** | Developing own theoretical model | **4,345** | **162** | **0,000** |
| **Numerical simulations** | Numerical simulations | **3,765** | **166** | **0,000** |
| Independent research in scientific literature | **Independent research in scientific literature** | **-4,069** | **171** | **0,000** |
| **Critical assessment of others' results** | Critical assessment of others' results | **2,079** | **168** | **0,039** |

*Note: Student's t-Test, coefficients with p ≤ .05 highlighted bold, positive t-value means better in YPT, negative in Non-YPT.*

We also observed that YPT-related activities are perceived more useful than other activities to develop skills, e.g. "Interpreting experimental data, data analysis" (p = 0,003), "Developing own theoretical model" (p = 0,000), "Numerical simulations" (p = 0,000), and "Critical assessment of others' results" (p = 0,039). On the other hand, other Non-YPT activities are considered more useful than YPT-related activities in developing abilities to "Solve close-ended problems" (p = 0,010) and to conduct "Independent research in scientific literature" (p = 0,000).

### Usefulness of RPC vs. YPT activities

| Hard Skills in RPC | Hard Skills in YPT | t | df | p |
|---|---|---|---|---|
| High school mathematics | High school mathematics | 0,288 | 136 | 0,774 |
| High school physics | High school physics | 0,524 | 184 | 0,601 |
| **Solve close-ended problems** | Solve close-ended problems | **4,409** | **178** | **0,000** |
| Designing experiments | **Designing experiments** | **-3,157** | **131** | **0,002** |
| Conducting experiment | Conducting experiment | -0,095 | 176 | 0,924 |
| Interpreting experimental data, data analysis | **Interpreting experimental data, data analysis** | **-3,593** | **180** | **0,000** |
| Developing own theoretical model | **Developing own theoretical model** | **-8,185** | **173** | **0,000** |
| Numerical simulations | **Numerical simulations** | **-7,447** | **170** | **0,000** |
| Independent research in scientific literature | Independent research in scientific literature | -1,760 | 169 | 0,080 |
| Critical assessment of others' results | **Critical assessment of others' results** | **-4,323** | **173** | **0,000** |

*Note: Student's t-Test, coefficients with p ≤ .05 highlighted bold, positive t-value means better in RPC, negative in YPT*

While RPC seems to be way more useful for "Solving close-ended problems" (p = 0,000) than YPT-related activities, we find that YPT-related activities and other Non-YPT activities are more useful than RPC for "Designing experiments", "Interpreting experimental data, data analysis"(by p < .1), "Developing own theoretical model", "Numerical simulations", "Independent research in scientific literature", and "Critical assessment of others' results".

For teachers, these findings imply that YPT-related activities and other extracurricular activities may complement each other, and seems to be rational to use and support both of them to maximize the hard

The European Commission's support for the production of this publication does not constitute an endorsement of the contents, which reflect the views only of the authors, and the Commission cannot be held responsible for any use which may be made of the information contained therein.

skills development: apart from "High school mathematics", "High school physics", "Conducting experiment by clear manuals" are YPT and Non-YPT activities significantly better then RPC, where Non-YPT activities can have positive effect on "Solving close-ended problems" and "Independent research in scientific literature", and YPT activities for all the rest. As a result, we suggest that teachers reinforce inquiry-based learning activities in regular physics classes and encourage students to participate in YPT-related activities, because YPT has such a positive effect on hard skills as YPT-like activities.

## III. YPT activities built on existing hard skills

We test the hypothesis that the perceived usefulness of regular physics classes, YPT-related activities, and other activities depends on the students' level of knowledge—the number of years to their final exam. Below and in the 1. Supplement (see in 1. Supplement 1.3), we show regression results for the perceived usefulness with the responses of students in their final year as baseline.

For RPC, we find lower perceived usefulness for "High school mathematics" (p = 0,046) for students who still had two years until their final exam. Students that were three or more years away from their final exam indicated lower usefulness to develop skills for "High school physics" (p = 0,064) and to "Solve close-ended problems" (p = 0,052). At the same time, students who had only one or two years left until the final exam considered RPC more useful for "Developing own theoretical model", "Numerical simulations", "Independent research in scientific literature", and "Critical assessment of others' results".

With few exceptions, students that were three or more years away from their final exam considered YPT-related activities as less useful to develop their hard skills than students who were closer to their final exams. With the exception of "High school mathematics", "Conducting experiment", and "Critical assessment of others' results", we did not find any differences in the perceived usefulness of participation in other activities based on time schoolyears to final exam.

**Differences in usefulness of YPT activities based on years to final exam**

| Hard Skills - YPT | 1 | 2 | 3+ | R² |
|---|---|---|---|---|
| High school mathematics | -0,188 | **-0,851** | **-0,877** | 0,152 |
| Std. Error | 0,184 | **0,188** | **0,206** | |
| p-value | 0,310 | **0,000** | **0,000** | |
| High school physics | -0,108 | **-0,386** | **-0,690** | 0,069 |
| Std. Error | 0,187 | **0,190** | **0,210** | |
| p-value | 0,566 | **0,044** | **0,001** | |
| Solve close-ended problems | 0,094 | -0,223 | -0,230 | 0,021 |
| Std. Error | 0,194 | 0,200 | 0,216 | |
| p-value | 0,628 | 0,266 | 0,288 | |
| Designing experiments | -0,100 | -0,394 | -0,311 | 0,025 |
| Std. Error | 0,221 | 0,233 | 0,268 | |
| p-value | 0,653 | 0,094 | 0,249 | |
| Conducting experiment | -0,157 | **-0,645** | **-0,775** | 0,094 |
| Std. Error | 0,193 | **0,198** | **0,220** | |
| p-value | 0,417 | **0,001** | **0,001** | |
| Interpreting experimental data, data analysis | -0,222 | **-0,580** | **-0,862** | 0,095 |
| Std. Error | 0,188 | **0,193** | **0,216** | |

The European Commission's support for the production of this publication does not constitute an endorsement of the contents, which reflect the views only of the authors, and the Commission cannot be held responsible for any use which may be made of the information contained therein.

| | | | | |
|---|---|---|---|---|
| p-value | 0,240 | **0,003** | **0,000** | |
| Developing own theoretical model | -0,159 | **-0,536** | **-0,659** | 0,071 |
| Std. Error | 0,188 | **0,193** | **0,216** | |
| p-value | 0,400 | **0,006** | **0,003** | |
| Numerical simulations | -0,133 | **-0,673** | **-0,790** | 0,087 |
| Std. Error | 0,216 | **0,222** | **0,243** | |
| p-value | 0,537 | **0,003** | **0,001** | |
| Independent research in scientific literature | -0,160 | -0,290 | -0,395 | 0,016 |
| Std. Error | 0,228 | 0,229 | 0,254 | |
| p-value | 0,484 | 0,207 | 0,121 | |
| Critical assessment of others' results | -0,163 | **-0,432** | **-0,769** | 0,072 |
| Std. Error | 0,199 | **0,200** | **0,225** | |
| p-value | 0,413 | **0,032** | **0,001** | |

*Note: Linear regression, baseline: year of final exam, coefficients with p ≤ .05 highlighted bold.*

For teachers, these findings imply that participation in YPT-related activities may constitute a "capstone" element in student education – or as a "bridge" element to the university level education. It seems as if teachers need to ensure sufficient levels of skills for students to make the most from participation in YPT and the most for the optimal developing. Teachers should therefore carefully and consciously build on existing hard skills of students in order to maximize hard skill development in the last year(s) before students take the matura exam. For students who still have some time to their final exams, these findings indicate that there is ant additional need for guidance by teachers, and the importance of selecting the appropriate level of the investigated problems and of the desired results. In this case, teachers should ensure that students get sufficient preparation and support for YPT-related activities in order to avoid feeling overwhelmed by the events' requirements. This step will help to allow junior students to maximize their benefits of YPT-related activities.

## IV. YPT activities can be useful independently from the number of RPC

We test the hypothesis that the perceived usefulness of RPC, YPT-related activities, and other Non-YPT activities depends on the students' weekly physics classes. Below (1. Supplement 1.4), we show regression results for the perceived usefulness with the responses of students with weekly physics classes as baseline.

Contrary to our expectations, we observe that students perceive their RPC as more useful to develop skills for "Numerical simulations" when they attend only few (≤ 3 hours) weekly physics classes. At the same time, except for "High school mathematics ", we find no differences in the perceived usefulness of YPT-related activities contingent on the number of weekly physics classes. However, we observe lower perceived usefulness of participation in other activities for students who take only a few (1- hour-long) physics classes a week..

| **Hard Skills – YPT** | **1** | **2** | **3** | **4** | **5+** | **R²** |
|---|---|---|---|---|---|---|
| High school mathematics | 0,917 | 0,545 | 0,652 | **1,105** | 0,883 | 0,055 |
| Std. Error | 0,630 | 0,499 | 0,511 | **0,518** | 0,549 | |
| p-value | 0,147 | 0,276 | 0,203 | **0,034** | 0,109 | |
| High school physics | 0,143 | -0,230 | 0,023 | 0,167 | 0,400 | 0,046 |

Differences in usefulness of YPT activities based on regular physics classes per week

The European Commission's support for the production of this publication does not constitute an endorsement of the contents, which reflect the views only of the authors, and the Commission cannot be held responsible for any use which may be made of the information contained therein.

| | | | | | | |
|---|---|---|---|---|---|---|
| Std. Error | 0,589 | 0,481 | 0,491 | 0,500 | 0,529 | |
| p-value | 0,809 | 0,633 | 0,962 | 0,739 | 0,450 | |
| Solve close-ended problems | 0,433 | 0,072 | 0,325 | 0,406 | 0,529 | 0,030 |
| Std. Error | 0,590 | 0,448 | 0,462 | 0,470 | 0,508 | |
| p-value | 0,464 | 0,873 | 0,483 | 0,388 | 0,299 | |
| Designing experiments | 0,500 | -0,323 | 0,161 | 0,290 | 0,067 | 0,077 |
| Std. Error | 0,691 | 0,577 | 0,593 | 0,600 | 0,618 | |
| p-value | 0,471 | 0,577 | 0,786 | 0,630 | 0,914 | |
| Conducting experiment | 0,393 | -0,098 | 0,250 | 0,485 | 0,450 | 0,059 |
| Std. Error | 0,629 | 0,513 | 0,525 | 0,530 | 0,565 | |
| p-value | 0,533 | 0,848 | 0,635 | 0,361 | 0,427 | |
| Interpreting experimental data, data analysis | 0,250 | 0,008 | 0,440 | 0,656 | 0,500 | 0,070 |
| Std. Error | 0,611 | 0,498 | 0,510 | 0,517 | 0,545 | |
| p-value | 0,683 | 0,987 | 0,389 | 0,206 | 0,360 | |
| Developing own theoretical model | 0,200 | -0,092 | 0,122 | 0,323 | 0,000 | 0,026 |
| Std. Error | 0,613 | 0,446 | 0,459 | 0,467 | 0,501 | |
| p-value | 0,745 | 0,837 | 0,791 | 0,491 | 1,000 | |
| Numerical simulations | 0,417 | -0,272 | 0,250 | 0,350 | 0,250 | 0,064 |
| Std. Error | 0,708 | 0,561 | 0,575 | 0,584 | 0,617 | |
| p-value | 0,557 | 0,628 | 0,664 | 0,550 | 0,686 | |
| Independent research in scientific literature | 1,083 | 0,506 | 0,957 | 0,650 | 1,036 | 0,046 |
| Std. Error | 0,717 | 0,568 | 0,582 | 0,591 | 0,630 | |
| p-value | 0,133 | 0,375 | 0,102 | 0,273 | 0,102 | |
| Critical assessment of others' results | 0,417 | 0,394 | 0,869 | 0,853 | 0,821 | 0,060 |
| Std. Error | 0,635 | 0,503 | 0,514 | 0,524 | 0,557 | |
| p-value | 0,512 | 0,435 | 0,093 | 0,105 | 0,142 | |

Note: Linear regression, baseline: no weekly physics classes, coefficients with $p \leq .05$ highlighted bold.

For teachers, based on the result, we can say that along with the other activities, it is also true for YPT activities that basically students studying in any group, regardless of the number of lessons, can try out the YPT activities. Based on these, we encourage teachers to work with YPT methods for all groups of students in addition to the appropriate methods and goals.

## V. Earlier (and also former) participation in YPT leads to positive bias towards YPT

We test the hypothesis that the perceived usefulness of RPC, YPT-related activities, and other Non-YPT activities depends on the students' most recent or earlier participation in YPT-related activities. Below (1. Supplement 1.5), we show regression results for the perceived usefulness with the responses of students who never participated in YPT-related activities as baseline. It is *very important* to mention that the most of the students, who are participating "This year", are answering the survey in October or November, so mostly in the first phase of their first participation. That can mean, that they do not still have as much experience, as students, who have already participated earlier and are still involved in YPT activity – most likely because of their former success in YPT activities. Depending on the year of the survey, the year of reference — "This year"— is either 2021 or 2020.

The European Commission's support for the production of this publication does not constitute an endorsement of the contents, which reflect the views only of the authors, and the Commission cannot be held responsible for any use which may be made of the information contained therein.

**Differences in usefulness of regular classes based on most recent participation in YPT activities**

| Hard Skills - RPC | Earlier | This year | R² |
|---|---|---|---|
| High school mathematics | **-0,900** | 0,481 | 0,135 |
| Std. Error | **0,264** | 0,531 | |
| p-value | **0,001** | 0,368 | |
| High school physics | **-0,747** | 0,279 | 0,070 |
| Std. Error | **0,217** | 0,451 | |
| p-value | **0,001** | 0,537 | |
| Solve close-ended problems | **-0,979** | 0,493 | 0,119 |
| Std. Error | **0,218** | 0,442 | |
| p-value | **0,000** | 0,266 | |
| Designing experiments | **-1,310** | -0,250 | 0,130 |
| Std. Error | **0,266** | 0,576 | |
| p-value | **0,000** | 0,665 | |
| Conducting experiment | **-1,219** | -0,528 | 0,125 |
| Std. Error | **0,255** | 0,553 | |
| p-value | **0,000** | 0,341 | |
| Interpreting experimental data, data analysis | **-1,621** | -0,771 | 0,206 |
| Std. Error | **0,254** | 0,621 | |
| p-value | **0,000** | 0,216 | |
| Developing own theoretical model | **-1,191** | -0,341 | 0,123 |
| Std. Error | **0,256** | 0,624 | |
| p-value | **0,000** | 0,586 | |
| Numerical simulations | **-1,262** | -1,962 | 0,134 |
| Std. Error | **0,279** | 1,164 | |
| p-value | **0,000** | 0,094 | |
| Independent research in scientific literature | **-1,405** | -0,355 | 0,144 |
| Std. Error | **0,271** | 0,806 | |
| p-value | **0,000** | 0,661 | |
| Critical assessment of others' results | **-1,349** | 0,051 | 0,116 |
| Std. Error | **0,297** | 0,630 | |
| p-value | **0,000** | 0,936 | |

*Note: Linear regression, baseline: no participation, coefficients with $p \leq .05$ highlighted bold.*

For all types of hard skills, we observe that students that participated in YPT-related activities consider RPC - and also other Non-YPT activities -   less useful to develop these hard skills. What interesting is that we observe these effects only for students that participated in YPT-related activities some time ago, but not for students that recently participated in these activities. This may suggest that synergies between the YPT-related activities and RPC as well as other Non-YPT activities are limited. We observe no differences in the perceived usefulness of YPT-related activities based on the most recent participation.

For teachers, theses results show that participating in YPT can lead to a long-term strong positive bias towards the inquiry-based physics activities, and also that these activities have very positive effects on the further university studies. Therefore, we mostly recommend to encourage those students for the participation in YPT activities, who are interested in physics, but the traditional Non-YPT activities and competitions have limited their motivations to make efforts for getting better in physics.

## VI. Cross-national differences matter

As our data includes responses from students and teachers from several countries, we are interested in how cross-national differences affect our findings (see 1.7 and 2.3). We observe that responses by students as well as teachers differ by country. Home country factors (e.g., education system, curricula, teaching style) seem to affect how students and teachers consider the usefulness of the different activities in the development of students' hard skills.

To test the impact of country differences on our results, we use ANOVA to test for differences in self-evaluation and perceived usefulness of RPC, YPT-related activities, and other Non-YPT activities contingent on the student's home country. This kind of test indicates only, if there is a significant effect of the country – for deeper analysis see 1. Supplement 1.7.2. We observe that students' self-evaluations for 5 types of hard skills differ by country. We find across-country differences in the perceived usefulness of RPC for seven out of ten hard skills. In the case of YPT-related activities, however, we observe that the perceived usefulness for all types of hard skills depends on students' home countries. We observe only weak ($p \leq .10$) country dependent differences for two out of ten types of hard skills for the perceived usefulness of participation in Non-YPT activities (see 1. Supplement 1.7.1).

### Differences in self-evaluation based on country

| Hard Skills – self-evaluation | df | F | p |
|---|---|---|---|
| **High school mathematics** | **12,899** | **2,231** | **0,026** |
| High school physics | 2,230 | 0,322 | 0,957 |
| Solve close-ended problems | 5,077 | 1,031 | 0,410 |
| **Designing experiments** | **12,170** | **4,770** | **0,001** |
| **Conducting experiment** | **13,233** | **4,432** | **0,002** |
| **Interpreting experimental data, data analysis** | **13,632** | **3,829** | **0,005** |
| Developing own theoretical model | 1,349 | 0,453 | 0,770 |
| **Numerical simulations** | **23,109** | **5,440** | **0,000** |
| Independent research in scientific literature | 6,652 | 1,267 | 0,284 |
| Critical assessment of others' results | 2,050 | 0,592 | 0,669 |

*Note: ANOVA (Value ~ Country), coefficients with $p \leq .05$ highlighted bold.*

### Differences in usefulness of RPC based on country

| Hard Skills – RPC | df | F | p |
|---|---|---|---|
| **High school mathematics** | **8,264** | **2,742** | **0,030** |
| High school physics | 1,226 | 0,365 | 0,833 |
| Solve close-ended problems | 2,683 | 0,785 | 0,536 |
| **Designing experiments** | **13,066** | **2,579** | **0,038** |
| **Conducting experiment** | **38,296** | **9,334** | **0,000** |
| **Interpreting experimental data, data analysis** | **21,258** | **4,719** | **0,001** |
| Developing own theoretical model | 5,550 | 1,154 | 0,332 |
| **Numerical simulations** | **48,752** | **9,621** | **0,000** |
| **Independent research in scientific literature** | **27,770** | **5,158** | **0,001** |
| **Critical assessment of others' results** | **56,740** | **11,722** | **0,000** |

*Note: ANOVA (Value ~ Country), coefficients with $p \leq .05$ highlighted bold.*

The European Commission's support for the production of this publication does not constitute an endorsement of the contents, which reflect the views only of the authors, and the Commission cannot be held responsible for any use which may be made of the information contained therein.

### Differences in usefulness of YPT activities based on country

| Hard Skills – YPT | df | F | p |
|---|---|---|---|
| **High school mathematics** | **52,205** | **7,542** | **0,000** |
| **High school physics** | **53,342** | **7,649** | **0,000** |
| **Solve close-ended problems** | **17,245** | **2,685** | **0,011** |
| **Designing experiments** | **18,976** | **5,350** | **0,000** |
| **Conducting experiment** | **24,766** | **6,651** | **0,000** |
| **Interpreting experimental data, data analysis** | **24,654** | **6,972** | **0,000** |
| **Developing own theoretical model** | **11,134** | **3,104** | **0,017** |
| **Numerical simulations** | **35,661** | **8,236** | **0,000** |
| **Independent research in scientific literature** | **14,385** | **2,985** | **0,020** |
| **Critical assessment of others' results** | **13,743** | **3,636** | **0,007** |

*Note: ANOVA (Value ~ Country), coefficients with $p \leq .05$ highlighted bold.*

### Differences in usefulness of Non-YPT based on country

| Hard Skills – Non-YPT | df | F | p |
|---|---|---|---|
| **High school mathematics** | **8,691** | **2,221** | **0,068** |
| High school physics | 4,938 | 1,378 | 0,242 |
| Solve close-ended problems | 3,788 | 0,976 | 0,422 |
| Designing experiments | 1,866 | 0,450 | 0,772 |
| **Conducting experiment** | **8,101** | **2,039** | **0,090** |
| Interpreting experimental data, data analysis | 6,305 | 1,770 | 0,135 |
| Developing own theoretical model | 4,499 | 1,068 | 0,373 |
| Numerical simulations | 0,763 | 0,144 | 0,965 |
| Independent research in scientific literature | 2,472 | 0,655 | 0,624 |
| Critical assessment of others' results | 6,371 | 1,488 | 0,207 |

*Note: ANOVA (Value ~ Country), coefficients with $p \leq ,10$ highlighted bold.*

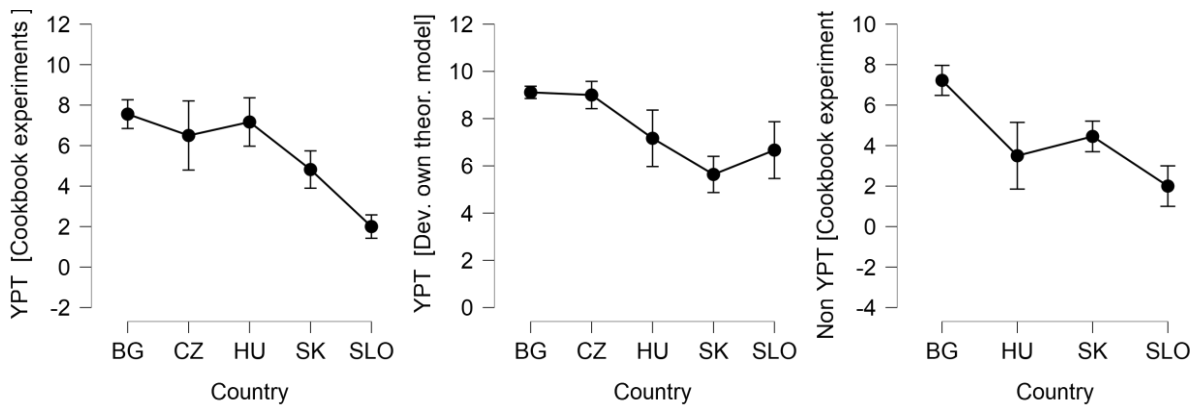

However, the analysis shows differences in the country effects reported by students and teachers. While we find country-differences in the student-reported usefulness of YPT-related activities for all types of hard skills, for responses by teachers, we observe differences only in the case of the "Conducting experiments by clear manual/cookbook experiments" and "Development of own theoretical model".

Teachers report also only one difference for the usefulness of Non-YPT activities, we find basically also no differences in the student survey.

For teachers, these findings imply that preparatory classes for YPT may require more adaption than teachers initially assume. Students from different countries reported varying perceived usefulness for YPT to develop their hard skills meanwhile teachers' responses do not show substantial differences. Therefore, it seems to be important that the adaption of foreign good practices has to be well considered and well built-in in the national curricula.

## VII. Teachers take positive perspective on YPT participation

In our second survey, we analyze teachers' evaluation of the usefulness of YPT-related activities to develop students' hard – and also soft - skills. Teachers generally considered YPT-related activities highly beneficial for students' hard-skill development (6 out of 10, see 2. Supplement 2.2.2). This result does not seem very strong when compared to the usefulness ascribed to RPC (5 out of 10 are beneficial, see 2. Supplement 2.2.1), but the comparison is quite clear. Results from paired t-tests/Wilcoxon tests confirm these differences. Across all types of hard skills, we observe greater perceived usefulness for YPT-related activities than for regular physics classes.

In the following, we can see the results of the Hard Skills in both RPC and YPT and the comparison.

### Comparison: Teachers - RPC vs. YPT

| RPC | YPT | Test | Statistic | df | p |
|---|---|---|---|---|---|
| **RPC [High school mathematics]** | - YPT [High school mathematics] | **Wilcoxon** | **93.500** | | **0.037** |
| **RPC [Solving close-ended problems]** | - YPT [Solving close-ended problems] | **Student** | **5.010** | **32** | **< .001** |
| RPC [Cookbook experiments ] | - YPT [Cookbook experiments ] | Student | 0.291 | 32 | 0.773 |
| RPC [Dev. own theor. model] | - **YPT [Dev. own theor. model]** | **Student** | **-9.332** | **32** | **< .001** |
| RPC [Indep. research in scientific litr.] | - **YPT [Indep. research in scientific lit.]** | **Student** | **-9.891** | **32** | **< .001** |
| RPC [High school physics] | - YPT [High school physics] | Wilcoxon | 116.000 | | 1.000 |
| RPC [Designing experiments] | - **YPT [Designing experiments]** | **Student** | **-8.269** | **32** | **< .001** |
| RPC [Interp. exp. data, data analysis] | - **YPT [Interp. exp. data, data analysis]** | **Student** | **-7.187** | **32** | **< .001** |
| RPC [Numerical simulations] | - **YPT [Numerical simulations]** | **Student** | **-8.505** | **32** | **< .001** |
| RPC [Crit. asses. of other's res.] | - **YPT [Crit. asses. of other's res.]** | **Student** | **-9.336** | **32** | **< .001** |

*Note: with $p \leq .05$ highlighted bold*

In the field of "High school physics" and "Cookbook experiments" there is no significant difference between YPT and RPC, in the case of "High school mathematics", "Solving close-ended problems in physics" there is a negative significant difference in the case of YPT compared to RPC. However, there are significant positive differences in "Designing experiment", "Interpreting experimental data, data analysis", "Developing own theoretical model", "Numerical simulations, Independent research in scientific literature", and "Critical assessment of others' results".

As RPC is a form of education developed for all high school students, we get much more useful and more information, especially for hard skills, by comparing YPT and Non-YPT type competitions. As the competitions are already open to interested and / or talented students, the result of comparing them can be useful for teachers, as we want to turn as many interested students with different backgrounds to physics and research activities in general. The results presented below show well what additional opportunities YPT-type competitions have for interested and talented students compared to traditional ones.

The European Commission's support for the production of this publication does not constitute an endorsement of the contents, which reflect the views only of the authors, and the Commission cannot be held responsible for any use which may be made of the information contained therein.

**Comparison: Hard Skills (YPT vs. Non-YPT)**

| YPT | Non-YPT | Test | Statistic | df | p |
|---|---|---|---|---|---|
| YPT [High school mathematics] | - Non YPT [High school mathematics] | Wilcoxon | 81.500 | | 0.828 |
| YPT [Solving close-ended problems] | - **Non YPT [Solving close-ended problems]** | Student | -3.841 | 28 | < .001 |
| YPT [Conducting experiment ] | - Non YPT [Conducting experiment ] | Student | 1.629 | 28 | 0.115 |
| **YPT [Dev. own theor. model]** | - Non YPT [Dev. own theor. model] | Student | 5.554 | 28 | < .001 |
| **YPT [Indep. research in sci. litr.]** | - Non YPT [Indep. research in sci. litr.] | Student | 4.400 | 27 | < .001 |
| | | Wilcoxon | 259.500 | | < .001 |
| YPT [High school physics] | - Non YPT [High school physics] | Wilcoxon | 35.500 | | 0.855 |
| **YPT [Designing experiments]** | - Non YPT [Designing experiments] | Student | 8.267 | 28 | < .001 |
| **YPT [Interp. exp. data, data analysis]** | - Non YPT [Interp. exp. data, data analysis] | Student | 5.953 | 27 | < .001 |
| | | Wilcoxon | 325.000 | | < .001 |
| **YPT [Numerical simulations]** | - Non YPT [Numerical simulations] | Student | 6.841 | 28 | < .001 |
| **YPT [Crit. asses. of other's res.]** | - Non YPT [Crit. asses. of other's res.] | Student | 9.374 | 28 | < .001 |

*Note: with $p \leq .05$ highlighted bold*

There is no difference between "High school mathematics", "High school physics" development, and "Conducting experiments (based on clear manual)". Non-YPT is significantly better in "Solving close-ended problems in physics", and in all other hard skills, the developmental impact of YPT is considered to be quite serious by the teachers interviewed.

While this finding attests to the usefulness of YPT-related activities to develop students' hard skills, an important caveat appears. Only teachers who have some experience with YPT activities participated in the teacher survey for IO3. Therefore, we have to consider the possibility of a self-selection bias by teachers. This may explain the differences in the perceived usefulness of YPT-related activities by students (see 3. Supplement) and teachers.

For teachers, these findings show that colleagues, who are active in YPT-related activities are tending to have positive bias towards YPT activities. It is important not to forget, how strong students' motivation relies on the teachers' enthusiasm. And even compared to other, Non-YPT activities, where the answering colleagues also most likely have have positive bias, the scores of YPT show in the case of many hard skills positive difference for YPT. This very strong scoring shows that working on YPT-like problems can be very motivating also for teachers, which can help to convince other colleagues to make a try on this kind of activities.

## VIII. Students and Teachers see YPT not the same but in the same way

In an additional analysis (see more details in 3. Supplement), we investigate how students compare the usefulness of their regular physics classes and YPT-related events in the development of their hard skills and how teachers compare the two activities. To make the results between students and teachers comparable, first we had to clean the data from students. Only 77 students answered all the questions, which are needed for the investigation of differences. Students' answers do not show normal distributions: Mann-Whitney test is needed. Because of the originally different scales of students (1-5) and teachers (1-10) we had to rescale the scores of the students to (1-10) scale for comparison. The comparison is not country specific, because of the quite small number of the teachers in the given countries.

We can observe that teachers perceive YPT-related activities significantly more useful across 6 types of out of 10 hard skills. Students reported YPT-related activities more useful for 4 out of 10 hard skills (see details in 3. Supplement 3.1), and 3 out of 10 more useful in RPC, even though they were using a smaller range of scores.

### Comparison: Teachers - RPC vs. YPT

| RPC | YPT | Test | Statistic | df | p |
|---|---|---|---|---|---|
| **RPC [High school mathematics]** | - YPT [High school mathematics] | **Wilcoxon** | **93.500** | | **0.037** |
| **RPC [Solving close-ended problems]** | - YPT [Solving close-ended problems] | **Student** | **5.010** | **32** | **< .001** |
| RPC [Cookbook experiments ] | - YPT [Cookbook experiments ] | Student | 0.291 | 32 | 0.773 |
| RPC [Dev. own theor. model] | - **YPT [Dev. own theor. model]** | **Student** | **-9.332** | **32** | **< .001** |
| RPC [Indep. research in scientific litr.] | - **YPT [Indep. research in scientific lit.]** | **Student** | **-9.891** | **32** | **< .001** |
| RPC [High school physics] | - YPT [High school physics] | Wilcoxon | 116.000 | | 1.000 |
| RPC [Designing experiments] | - **YPT [Designing experiments]** | **Student** | **-8.269** | **32** | **< .001** |
| RPC [Interp. exp. data, data analysis] | - **YPT [Interp. exp. data, data analysis]** | **Student** | **-7.187** | **32** | **< .001** |
| RPC [Numerical simulations] | - **YPT [Numerical simulations]** | **Student** | **-8.505** | **32** | **< .001** |
| RPC [Crit. asses. of other's res.] | - **YPT [Crit. asses. of other's res.]** | **Student** | **-9.336** | **32** | **< .001** |

*Note: with $p \leq .05$ highlighted bold*

### Comparison (Wilcoxon): 77 Students - RPC vs. YPT

| RPC | YPT | W | p |
|---|---|---|---|
| **High sch. math. - RPC** | **- High sch. math. -YPT** | **355.000** | **0.002** |
| **High sch. phy.- RPC** | **- High sch. phy.-YPT** | **619.000** | **0.003** |
| **Solv. clos-end. prob. - RPC** | **- Solv. clos-end. prob. -YPT** | **570.500** | **< .001** |
| Des. exp.- RPC | **- Des. exp.-YPT** | **270.000** | **0.012** |
| Cookbook exp. - RPC | - Cookbook exp. -YPT | 376.000 | 0.163 |
| Int. exp. data - RPC | - Int. exp. data -YPT | 406.500 | 0.410 |
| Dev. own. th. mod. - RPC | **- Dev. own. th. mod. -YPT** | **296.000** | **0.029** |
| Num. sim. - RPC | **- Num. sim. -YPT** | **175.500** | **0.019** |
| Research in sci. lit. - RPC | **- Research in sci. lit. -YPT** | **169.500** | **0.002** |
| Crit. ass. - RPC | - Crit. ass. -YPT | 321.500 | 0.222 |

*Note. Wilcoxon signed-rank test. Highlighted bold if $p \leq .05$*

### Comparison: Hard Skills in RPC and YPT of Students and Teachers

| | W | p |
|---|---|---|
| **High sch. math. - RPC (Students)** | **734.500** | **0.005** |
| High sch. math. -YPT | 437.000 | 0.159 |
| High sch. phy.- RPC | 1169.000 | 0.660 |
| High sch. phy.-YPT | 998.000 | 0.108 |
| Solv. clos-end. prob. - RPC | 1415.500 | 0.200 |
| **Solv. clos-end. prob. -YPT (Students)** | **1894.500** | **< .001** |
| **Des. exp.- RPC (Students)** | **1833.500** | **< .001** |
| Des. exp.-YPT | 973.000 | 0.073 |
| **Cookbook exp. - RPC (Students)** | **1925.500** | **< .001** |
| **Cookbook exp. -YPT (Students)** | **1697.000** | **0.001** |
| **Int. exp. data - RPC (Students)** | **1581.500** | **0.017** |
| **Int. exp. data -YPT (Teachers)** | **801.500** | **0.004** |
| **Dev. own. theo. mod. - RPC (Students)** | **2078.500** | **< .001** |
| Dev. own. theo. mod. -YPT | 1155.500 | 0.679 |
| **Num. sim. - RPC (Students)** | **2042.000** | **< .001** |
| Num. sim. -YPT | 1061.000 | 0.241 |
| **Research in sci. lit. - RPC (Students)** | **2006.000** | **< .001** |
| Research in sci. lit. -YPT | 1128.500 | 0.477 |
| **Crit. ass. - RPC (Students)** | **2007.500** | **< .001** |
| **Crit. ass. -YPT (Teachers)** | **816.000** | **0.004** |

*Note. Mann-Whitney U test. highlighted bold if $p \leq .05$ In parentheses the direction of positively biased group..*

The European Commission's support for the production of this publication does not constitute an endorsement of the contents, which reflect the views only of the authors, and the Commission cannot be held responsible for any use which may be made of the information contained therein.

In the case of all hard skills in YPT there was only one significantly positive difference for students: "Solving close-ended problems", where students find YPT more improving than teachers considered.k. It was also clear to see, that students tend to give significantly higher scores in RPC for hard skills compared to the teachers. It can rely on many different effects, such as:

1. Students were not only focusing on the physics lessons but all the classes in school – e.g. the differences for English skills can explained by this interpretation.
2. Students and teachers understand slightly different thing under the investigated expressions – e.g. "Numerical analysis" can mean very different things depending on the reached educational level, which obviously different for teachers and students.
3. The students who are answering the survey are already somewhat different from the average – e.g. more interested and practiced in physics –, therefore they answers can be different from the teachers', who are according their answers on the whole – average – classes.

If we observe the comparisons of teachers and students in RPC and YPT it is easy to see, that even if their scorings are different, their results are tending mostly in the same directions. The more detailed comparison can be seen in 3. Supplement. But to have a generally better view of the comparison of the evaluation of usefulness in RPC and YPT we have compared the differences between scores in RPC and YPT both for students and teachers. The next table and Figure 2. show the result of this comparison.

**Differences between YPT and RPC (positive value means better in YPT)**

|  | Group | N | Mean | SD | SE |
|---|---|---|---|---|---|
| Diff. Math. | Student | 34 | -1.412 | 2.388 | 0.410 |
|  | Teacher | 32 | 0.938 | 2.711 | 0.479 |
| Diff. Phys. | Student | 77 | -0.649 | 1.790 | 0.204 |
|  | Teacher | 32 | -0.156 | 2.112 | 0.373 |
| Diff. Solv. Cl. Pr. | Student | 77 | -0.987 | 1.990 | 0.227 |
|  | Teacher | 32 | -3.094 | 3.383 | 0.598 |
| Diff. Des. Exp. | Student | 77 | 0.779 | 2.516 | 0.287 |
|  | Teacher | 32 | 3.750 | 2.627 | 0.464 |
| Diff. Cookbook | Student | 77 | -0.312 | 1.948 | 0.222 |
|  | Teacher | 32 | -0.344 | 3.525 | 0.623 |
| Diff. Int. Exp. | Student | 77 | 0.130 | 2.582 | 0.294 |
|  | Teacher | 32 | 2.438 | 1.999 | 0.353 |
| Diff. Dev own theory | Student | 77 | 0.494 | 2.537 | 0.289 |
|  | Teacher | 32 | 4.000 | 2.502 | 0.442 |
| Diff. Num Sim. | Student | 77 | 0.623 | 2.254 | 0.257 |
|  | Teacher | 32 | 4.281 | 2.932 | 0.518 |
| Diff. Research | Student | 77 | 0.987 | 2.526 | 0.288 |
|  | Teacher | 32 | 4.219 | 2.485 | 0.439 |
| Diff. Crit. Ass. | Student | 77 | 0.494 | 2.718 | 0.310 |
|  | Teacher | 32 | 4.375 | 2.537 | 0.448 |

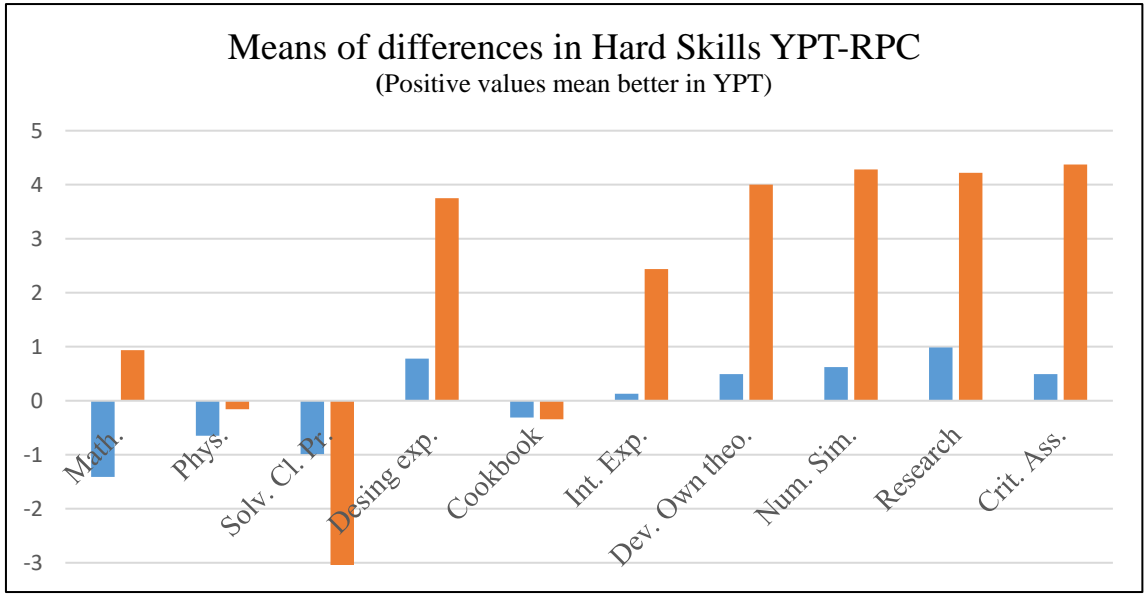

*Figure 2.: Differences in hard skills YPT-RPC. Blue: students, Orange: teachers*

There is no significant difference in "High School Physics" and "Cookbook Experiments", even though the very different typical scoring of the two groups. In the following, we show the results and comparison of the differences of Hard Skills in RPC and YPT.

**Comparison: Differences of Hard Skills between YPT and RPC**

|  | Test | Statistic | p |
|---|---|---|---|
| **Diff. Math.** | **Mann-Whitney** | **278.000** | **< .001** |
| Diff. Phys. | Mann-Whitney | 983.500 | 0.081 |
| **Diff. Solv. Cl. Pr.** | **Mann-Whitney** | **1741.500** | **< .001** |
| **Diff. Des. Exp.** | **Mann-Whitney** | **515.500** | **< .001** |
| Diff. Cookbook | Mann-Whitney | 1210.500 | 0.885 |
| **Diff. Int. Exp.** | **Mann-Whitney** | **551.000** | **< .001** |
| **Diff. Num Sim.** | **Mann-Whitney** | **374.500** | **< .001** |
| **Diff. Research** | **Mann-Whitney** | **407.000** | **< .001** |
| **Diff. Crit. Ass.** | **Mann-Whitney** | **355.500** | **< .001** |
| **Diff. Dev own theo.** | **Mann-Whitney** | **401.500** | **< .001** |

*Note: with p ≤ ,05 highlighted bold*

Differences with the *same sign* of teachers and students:

- "High school physics"
- "Solving close-ended problems in physics"
- "Cookbook experiments / Conducting experiment (based on clear manual)"
- "Interpreting experimental data, data analysis"
- "Developing own theoretical model"
- "Numerical simulations"
- "Interdependent research in scientific literature"

- "Critical assessment of others' results"

Differences with *opposite sign* by teachers and students:

- "High school mathematics"

By differences with same sign, we can state, that teachers and students see the effects quite similar, but in the case of "High school mathematics" it seems that the teachers tend to overestimate the effect of YPT – or underestimate the effect of the RPC.

For teachers, it means that they can mostly rely on their instincts, because with the only exception of "High school mathematics" students see the usefulness of YPT compared to RPC in the same way as teachers do. And yet, the relatively big differences are not negligible. Reason for that could be, that teachers might tend to think, that students in YPT learn faster or at least in the same speed as in the RPC. It is important not to forget, that YPT activities are often full with new situations and challenges for the students, what makes their development more diverse, but therefore it needs often more time and patience.

We recommend the teachers, that beside the motivation effect of YPT activities, they always consider novelty and complexity of problems in order to select the optimal working and learning speed for their students.

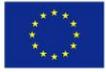
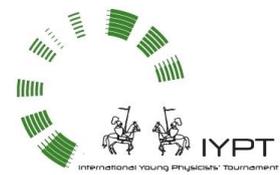

# The relationship between inquiry-based learning in YPT and the development of hard skills

*IO3 Dibali: 2019-1-SK01-KA201-060798*

## SUPPLEMENTARY MATERIALS

In this document, we provide supplementary materials that offer further details on the condensed guidelines presented in our report. These supplementary materials consist of three sections. The first section shows survey results on students' assessment of hard-skill development through regular physics classes, YPT-related activities, and other extracurricular activities. The second section present results from a survey of teachers' assessment of hard-skill development through these three types of activities. In section three, we present the comparison between students' and teachers' results in RPC and YPT. In the fourth section we present the research question and hypotheses.

## 1. Supplement: Students' Assessment of Hard-Skill Development

### 1.1 Participants

In total, 308 students from nine countries participated in the survey. The largest share of students was from Slovakia (54%), followed by Hungary (23%), the Czech Republic (7%), and Bulgaria (7%). While gender was not included in some in surveys, for the remainder the female-male split was about one third to two thirds. In some countries the share of male participants in the survey was even 70% and beyond (Czech Republic, Hungary). In only one country (Slovenia), the share of females exceeded that of male participants.

Students were classified based on the school years until they would write their final exams. Overall, for this categorization the split was even: 19% of students were in their final school year, 26% had one and 28% had two years until completion. About one fifth of the participants still had to complete three or more years until their final exams. Slovenia constitutes somewhat of an outlier with 22% of participants in their final year and 78% of participants in their second to last school year. As part of the survey, students were asked about their regular weekly physics classes. About half of participants took four hours of weekly physics classes. In the case of students from Slovakia and Slovenia, this share is even higher at 65% and 78%, respectively. 25% of participants from Bulgaria and 19% of participants from Hungary took 5 hours or more of weekly physics classes. Students also reported the time they spent on physics-related extracurricular activities. 28% of participants reported that they spent more than 20 hours per month on physics-related extracurricular activities, another 22% answered that they spent between 10 and 20 hours per month on these activities. Again, country differences seem to persist. 43% and 44% of students from Czech Republic and Slovenia, respectively, reported that they spend more than 20 hours per month on extra-curricular activities.

Participants indicated their most recent participation in YPT-related activities. Only in the case of "Work on problems" events, more than half (53%) of the students participated in YPT-related activities at least once. In the other events the majority of students had never participated. When asked about their participation in other physics competitions and events, students gave similar responses as for YPT-related events. With the exception of Ad hoc competitions (42%) and Other Science Olympiads (50%), more than half of the students had never participated in any events. Yet 46% of students had participated in a Physics Olympiad at least once. Only a part of the participants evaluated their overall experience with YPT. Yet for these 77 participants, the overall evaluation was very positive (median of 4). Although the results also show some outliers, the evaluation seems equally positive across all countries.

## 1.2. Detailed Data Analysis and Comparison of Hard Skills between RPC, YPT and Non-YPT

### 1.2.1 Basic Statistics of Students

**Gender of the Students**

| Country | Unknown | | Female | | Male | | Total | |
|---|---|---|---|---|---|---|---|---|
| | # | % | # | % | # | % | # | % |
| Austria | 13 | 100 | 0 | 0 | 0 | 0 | 13 | 4 |
| Bulgaria | 0 | 0 | 7 | 33 | 14 | 67 | 21 | 7 |
| Czech Rep. | 0 | 0 | 7 | 30 | 16 | 70 | 23 | 7 |
| Germany | 3 | 100 | 0 | 0 | 0 | 0 | 3 | 1 |
| Hungary | 1 | 1 | 19 | 26 | 52 | 72 | 72 | 23 |
| Iran | 1 | 100 | 0 | 0 | 0 | 0 | 1 | 0 |
| Russia | 1 | 100 | 0 | 0 | 0 | 0 | 1 | 0 |
| Slovakia | 0 | 0 | 63 | 38 | 102 | 62 | 165 | 54 |
| Slovenia | 0 | 0 | 5 | 56 | 4 | 44 | 9 | 3 |
| **Total** | **19** | **6** | **101** | **33** | **188** | **61** | **308** | **100** |

**Years to final exam**

| Country | Unknown | | 0 | | 1 | | 2 | | 3+ | | Total | |
|---|---|---|---|---|---|---|---|---|---|---|---|---|
| | # | % | # | % | # | % | # | % | # | % | # | % |
| Austria | 13 | 100 | 0 | 0 | 0 | 0 | 0 | 0 | 0 | 0 | 13 | 4 |
| Bulgaria | 0 | 0 | 7 | 33 | 6 | 29 | 4 | 19 | 4 | 19 | 21 | 7 |
| Czech Rep. | 0 | 0 | 9 | 39 | 5 | 22 | 6 | 26 | 3 | 13 | 23 | 7 |
| Germany | 3 | 100 | 0 | 0 | 0 | 0 | 0 | 0 | 0 | 0 | 3 | 1 |
| Hungary | 1 | 1 | 22 | 31 | 22 | 31 | 22 | 31 | 5 | 7 | 72 | 23 |
| Iran | 1 | 100 | 0 | 0 | 0 | 0 | 0 | 0 | 0 | 0 | 1 | 0 |
| Russia | 1 | 100 | 0 | 0 | 0 | 0 | 0 | 0 | 0 | 0 | 1 | 0 |
| Slovakia | 0 | 0 | 20 | 12 | 40 | 24 | 53 | 32 | 52 | 32 | 165 | 54 |
| Slovenia | 0 | 0 | 2 | 22 | 7 | 78 | 0 | 0 | 0 | 0 | 9 | 3 |
| **Total** | **19** | **6** | **60** | **19** | **80** | **26** | **85** | **28** | **64** | **21** | **308** | **100** |

**Regular physics classes per week**

| Country | Unknown | | 0 | | 1 | | 2 | | 3 | | 4 | | 5+ | | Total | |
|---|---|---|---|---|---|---|---|---|---|---|---|---|---|---|---|---|
| | # | % | # | % | # | % | # | % | # | % | # | % | # | % | # | % |
| Austria | 13 | 100 | 0 | 0 | 0 | 0 | 0 | 0 | 0 | 0 | 0 | 0 | 0 | 0 | 13 | 4 |

The European Commission's support for the production of this publication does not constitute an endorsement of the contents, which reflect the views only of the authors, and the Commission cannot be held responsible for any use which may be made of the information contained therein.

| | # | % | # | % | # | % | # | % | # | % | # | % | # | % | # | % |
|---|---|---|---|---|---|---|---|---|---|---|---|---|---|---|---|---|
| Bulgaria | 0 | 0 | 2 | 7 | 1 | 4 | 7 | 25 | 0 | 0 | 11 | 39 | 7 | 25 | 28 | 10 |
| Czech Rep. | 0 | 0 | 2 | 9 | 1 | 4 | 0 | 0 | 9 | 39 | 11 | 48 | 0 | 0 | 23 | 8 |
| Germany | 3 | 100 | 0 | 0 | 0 | 0 | 0 | 0 | 0 | 0 | 0 | 0 | 0 | 0 | 3 | 1 |
| Hungary | 1 | 1 | 1 | 1 | 2 | 3 | 14 | 19 | 21 | 29 | 20 | 27 | 14 | 19 | 73 | 25 |
| Iran | 1 | 100 | 0 | 0 | 0 | 0 | 0 | 0 | 0 | 0 | 0 | 0 | 0 | 0 | 1 | 0 |
| Russia | 1 | 100 | 0 | 0 | 0 | 0 | 0 | 0 | 0 | 0 | 0 | 0 | 0 | 0 | 1 | 0 |
| Slovakia | 0 | 0 | 3 | 2 | 5 | 4 | 2 | 1 | 37 | 27 | 90 | 65 | 2 | 1 | 139 | 48 |
| Slovenia | 0 | 0 | 0 | 0 | 0 | 0 | 0 | 0 | 2 | 22 | 7 | 78 | 0 | 0 | 9 | 3 |
| **Total** | **19** | **7** | **8** | **3** | **9** | **3** | **23** | **8** | **69** | **24** | **139** | **48** | **23** | **8** | **290** | **100** |

**Average hours spent on physics-related extracurricular activities per month**

| Country | Unknown | | ≤5 | | ≤10 | | ≤20 | | >20 | | Total | |
|---|---|---|---|---|---|---|---|---|---|---|---|---|
| | # | % | # | % | # | % | # | % | # | % | # | % |
| Austria | 0 | 0 | 13 | 100 | 0 | 0 | 0 | 0 | 0 | 0 | 13 | 4 |
| Bulgaria | 4 | 19 | 2 | 10 | 3 | 14 | 5 | 24 | 7 | 33 | 21 | 7 |
| Czech Rep. | 3 | 13 | 1 | 4 | 1 | 4 | 8 | 35 | 10 | 43 | 23 | 7 |
| Germany | 0 | 0 | 3 | 100 | 0 | 0 | 0 | 0 | 0 | 0 | 3 | 1 |
| Hungary | 12 | 17 | 4 | 6 | 11 | 15 | 24 | 33 | 21 | 29 | 72 | 23 |
| Iran | 0 | 0 | 1 | 100 | 0 | 0 | 0 | 0 | 0 | 0 | 1 | 0 |
| Russia | 0 | 0 | 1 | 100 | 0 | 0 | 0 | 0 | 0 | 0 | 1 | 0 |
| Slovakia | 56 | 34 | 1 | 1 | 38 | 23 | 27 | 16 | 43 | 26 | 165 | 54 |
| Slovenia | 1 | 11 | 1 | 11 | 0 | 0 | 3 | 33 | 4 | 44 | 9 | 3 |
| **Total** | **76** | **25** | **27** | **9** | **53** | **17** | **67** | **22** | **85** | **28** | **308** | **100** |

**Most recent participation in YPT-related activities**

| Event | This year | | Last year | | Earlier | | Never | | Total |
|---|---|---|---|---|---|---|---|---|---|
| | # | % | # | % | # | % | # | % | # |
| Preparatory seminar | 42 | 19 | 27 | 12 | 12 | 5 | 139 | 63 | 220 |
| Work on problems | 84 | 35 | 29 | 12 | 13 | 5 | 112 | 47 | 238 |
| Regional YPT event | 47 | 22 | 21 | 10 | 16 | 8 | 125 | 60 | 209 |
| National YPT event | 50 | 25 | 9 | 4 | 15 | 7 | 129 | 64 | 203 |
| AYPT or similar international event | 9 | 5 | 5 | 3 | 13 | 7 | 161 | 86 | 188 |
| IYPT | 21 | 11 | 6 | 3 | 14 | 7 | 151 | 79 | 192 |

**Participation in other physics competitions or preparation for them**

| Event | This year | | Last year | | Earlier | | Never | | Total |
|---|---|---|---|---|---|---|---|---|---|
| | # | % | # | % | # | % | # | % | # |
| Physics Olympiad | 46 | 19 | 29 | 12 | 36 | 15 | 128 | 54 | 239 |
| IJSO or EUSO | 2 | 1 | 9 | 5 | 3 | 2 | 173 | 93 | 187 |
| IYNT | 2 | 1 | 2 | 1 | 5 | 3 | 176 | 95 | 185 |
| Other Science Olympiad | 60 | 26 | 24 | 10 | 32 | 14 | 117 | 50 | 233 |

The European Commission's support for the production of this publication does not constitute an endorsement of the contents, which reflect the views only of the authors, and the Commission cannot be held responsible for any use which may be made of the information contained therein.

| | | | | | | | | |
|---|---|---|---|---|---|---|---|---|
| Project Science Competition | 18 | 15 | 9 | 8 | 8 | 7 | 83 | 70 | 118 |
| Seminar or correspondence | 42 | 20 | 16 | 7 | 28 | 13 | 129 | 60 | 215 |
| Ad hoc competitions | 73 | 32 | 38 | 17 | 29 | 13 | 87 | 38 | 227 |
| Debate club or similar | 18 | 9 | 14 | 7 | 17 | 8 | 156 | 76 | 205 |

**Overall experience with YPT**

| Valid | Missing | Mean | Median | SD | Min. | Max. |
|---|---|---|---|---|---|---|
| 73 | 235 | 3,82 | 4 | 0,96 | 1 | 5 |

## 1.2.2 Data with Correlations of Students for Self-evaluation, RPC, YPT and Non-YPT

**Self-evaluation by student**

| Hard Skills | Valid | Missing | Mean | Median | SD | Min. | Max. |
|---|---|---|---|---|---|---|---|
| High school mathematics | 195 | 113 | 4,18 | 4 | 0,83 | 1 | 5 |
| High school physics | 266 | 42 | 3,70 | 4 | 0,96 | 1 | 5 |
| Solve close-ended pro | 274 | 34 | 4,05 | 4 | 0,84 | 1 | 5 |
| Designing experiments | 278 | 30 | 3,95 | 4 | 0,87 | 1 | 5 |
| Conducting experiment | 264 | 44 | 4,12 | 4 | 0,86 | 1 | 5 |
| Interpreting experimental data, data analysis | 265 | 43 | 3,72 | 4 | 0,93 | 1 | 5 |
| Developing own theoretical model | 266 | 42 | 3,70 | 4 | 0,96 | 1 | 5 |
| Numerical simulations | 278 | 30 | 3,95 | 4 | 0,87 | 1 | 5 |
| Independent research in scientific literature | 195 | 113 | 4,18 | 4 | 0,83 | 1 | 5 |
| Critical assessment of others' results | 244 | 64 | 3,31 | 3 | 1,15 | 1 | 5 |

**Usefulness of RPC (regular physics classes)**

| Hard Skills | Valid | Missing | Mean | Median | SD | Min. | Max. |
|---|---|---|---|---|---|---|---|
| High school mathematics | 198 | 110 | 4,04 | 4 | 0,88 | 1 | 5 |
| High school physics | 263 | 45 | 3,35 | 3 | 1,14 | 1 | 5 |
| Solve close-ended pro | 267 | 41 | 4,07 | 4 | 0,92 | 1 | 5 |
| Designing experiments | 259 | 49 | 3,64 | 4 | 1,09 | 1 | 5 |
| Conducting experiment | 262 | 46 | 3,94 | 4 | 1,08 | 1 | 5 |
| Interpreting experimental data, data analysis | 258 | 50 | 3,35 | 3 | 1,19 | 1 | 5 |
| Developing own theoretical model | 263 | 45 | 3,35 | 3 | 1,14 | 1 | 5 |
| Numerical simulations | 259 | 49 | 3,64 | 4 | 1,09 | 1 | 5 |
| Independent research in scientific literature | 198 | 110 | 4,04 | 4 | 0,88 | 1 | 5 |
| Critical assessment of others' results | 239 | 69 | 2,89 | 3 | 1,20 | 1 | 5 |

**Usefulness of YPT activities**

| Hard Skills | Valid | Missing | Mean | Median | SD | Min. | Max. |
|---|---|---|---|---|---|---|---|
| High school mathematics | 140 | 168 | 3,63 | 4 | 1,00 | 1 | 5 |

The European Commission's support for the production of this publication does not constitute an endorsement of the contents, which reflect the views only of the authors, and the Commission cannot be held responsible for any use which may be made of the information contained therein.

| High school physics | 192 | 116 | 4,01 | 4 | 1,00 | 1 | 5 |
|---|---|---|---|---|---|---|---|
| Solve close-ended pro | 193 | 115 | 3,58 | 4 | 0,99 | 1 | 5 |
| Designing experiments | 204 | 104 | 3,81 | 4 | 1,04 | 1 | 5 |
| Conducting experiment | 184 | 124 | 4,04 | 4 | 0,97 | 1 | 5 |
| Interpreting experimental data, data analysis | 182 | 126 | 3,85 | 4 | 1,00 | 1 | 5 |
| Developing own theoretical model | 192 | 116 | 4,01 | 4 | 1,00 | 1 | 5 |
| Numerical simulations | 204 | 104 | 3,81 | 4 | 1,04 | 1 | 5 |
| Independent research in scientific literature | 140 | 168 | 3,63 | 4 | 1,00 | 1 | 5 |
| Critical assessment of others' results | 181 | 127 | 3,43 | 3 | 1,12 | 1 | 5 |

## Usefulness of Non-YPT activities

| Hard Skills | Valid | Missing | Mean | Median | SD | Min. | Max. |
|---|---|---|---|---|---|---|---|
| High school mathematics | 189 | 119 | 4,11 | 4 | 1,00 | 1 | 5 |
| High school physics | 246 | 62 | 3,72 | 4 | 1,01 | 1 | 5 |
| Solve close-ended pro | 246 | 62 | 3,97 | 4 | 0,99 | 1 | 5 |
| Designing experiments | 245 | 63 | 3,81 | 4 | 0,95 | 1 | 5 |
| Conducting experiment | 239 | 69 | 3,85 | 4 | 1,01 | 1 | 5 |
| Interpreting experimental data, data analysis | 240 | 68 | 3,70 | 4 | 1,04 | 1 | 5 |
| Developing own theoretical model | 246 | 62 | 3,72 | 4 | 1,01 | 1 | 5 |
| Numerical simulations | 245 | 63 | 3,81 | 4 | 0,95 | 1 | 5 |
| Independent research in scientific literature | 189 | 119 | 4,11 | 4 | 1,00 | 1 | 5 |
| Critical assessment of others' results | 225 | 83 | 3,45 | 3 | 1,14 | 1 | 5 |

## Correlations in Self Evaluations of Hard Skills

| | Hard Skills | 1 | 2 | 3 | 4 | 5 | 6 | 7 | 8 | 9 | 10 |
|---|---|---|---|---|---|---|---|---|---|---|---|
| 1 | High school mathematics | 1,00 | | | | | | | | | |
| 2 | High school physics | 0,43 | 1,00 | | | | | | | | |
| 3 | Solve close-ended pro | 0,47 | 0,41 | 1,00 | | | | | | | |
| 4 | Designing experiments | 0,37 | 0,29 | 0,51 | 1,00 | | | | | | |
| 5 | Conducting experiment | 0,56 | 0,39 | 0,66 | 0,63 | 1,00 | | | | | |
| 6 | Interpreting experimental data, data analysis | 0,61 | 0,33 | 0,39 | 0,24 | 0,45 | 1,00 | | | | |
| 7 | Developing own theoretical model | 0,55 | 0,34 | 0,25 | 0,28 | 0,25 | 0,59 | 1,00 | | | |
| 8 | Numerical simulations | 0,52 | 0,42 | 0,52 | 0,38 | 0,46 | 0,43 | 0,18 | 1,00 | | |
| 9 | Independent research in scientific literature | 0,47 | 0,35 | 0,31 | 0,29 | 0,33 | 0,38 | 0,19 | 0,56 | 1,00 | |
| 10 | Critical assessment of others' results | 0,47 | 0,47 | 0,23 | 0,27 | 0,30 | 0,41 | 0,41 | 0,41 | 0,40 | 1,00 |

Note: Pearson correlation coefficients.

## Correlation in Usefulness of RPC

| | Hard Skills | 1 | 2 | 3 | 4 | 5 | 6 | 7 | 8 | 9 | 10 |
|---|---|---|---|---|---|---|---|---|---|---|---|
| 1 | High school mathematics | 1,00 | | | | | | | | | |
| 2 | High school physics | 0,54 | 1,00 | | | | | | | | |

| | | 1 | 2 | 3 | 4 | 5 | 6 | 7 | 8 | 9 | 10 |
|---|---|---|---|---|---|---|---|---|---|---|---|
| **3** | Solve close-ended pro | 0,44 | 0,75 | 1,00 | | | | | | | |
| **4** | Designing experiments | 0,38 | 0,44 | 0,50 | 1,00 | | | | | | |
| **5** | Conducting experiment | 0,48 | 0,39 | 0,44 | 0,67 | 1,00 | | | | | |
| **6** | Interpreting experimental data, data analysis | 0,43 | 0,41 | 0,50 | 0,67 | 0,69 | 1,00 | | | | |
| **7** | Developing own theoretical model | 0,37 | 0,45 | 0,48 | 0,59 | 0,46 | 0,61 | 1,00 | | | |
| **8** | Numerical simulations | 0,25 | 0,20 | 0,23 | 0,54 | 0,47 | 0,54 | 0,58 | 1,00 | | |
| **9** | Independent research in scientific literature | 0,31 | 0,37 | 0,31 | 0,59 | 0,50 | 0,53 | 0,60 | 0,59 | 1,00 | |
| **10** | Critical assessment of others' results | 0,32 | 0,29 | 0,29 | 0,55 | 0,48 | 0,54 | 0,55 | 0,56 | 0,75 | 1,00 |

Note: Pearson correlation coefficients.

**Correlations in Usefulness of YPT activities**

| | **Hard Skills** | 1 | 2 | 3 | 4 | 5 | 6 | 7 | 8 | 9 | 10 |
|---|---|---|---|---|---|---|---|---|---|---|---|
| **1** | High school mathematics | 1,00 | | | | | | | | | |
| **2** | High school physics | 0,65 | 1,00 | | | | | | | | |
| **3** | Solve close-ended pro | 0,53 | 0,43 | 1,00 | | | | | | | |
| **4** | Designing experiments | 0,65 | 0,50 | 0,61 | 1,00 | | | | | | |
| **5** | Conducting experiment | 0,78 | 0,62 | 0,71 | 0,74 | 1,00 | | | | | |
| **6** | Interpreting experimental data, data analysis | 0,86 | 0,63 | 0,57 | 0,66 | 0,79 | 1,00 | | | | |
| **7** | Developing own theoretical model | 0,74 | 0,56 | 0,63 | 0,67 | 0,69 | 0,75 | 1,00 | | | |
| **8** | Numerical simulations | 0,80 | 0,57 | 0,52 | 0,67 | 0,77 | 0,74 | 0,63 | 1,00 | | |
| **9** | Independent research in scientific literature | 0,54 | 0,45 | 0,43 | 0,50 | 0,53 | 0,53 | 0,39 | 0,60 | 1,00 | |
| **10** | Critical assessment of others' results | 0,71 | 0,60 | 0,41 | 0,51 | 0,60 | 0,66 | 0,59 | 0,65 | 0,57 | 1,00 |

Note: Pearson correlation coefficients.

## 1.2.3 Differences in usefulness of RPC, YPT and other activities for Hard Skills

To verify the descriptive statistics from above, we use t-tests to test differences between the perceived usefulness of regular physics classes, YPT-related activities, and other activities. The results give a highly differentiated picture. While regular physics classes seem to be more useful to "Solve close-ended problems" ($p = 0,000$) than YPT-related activities, we find that YPT-related activities and other activities are more useful than regular physics classes for "Designing experiments", "Interpreting experimental data, data analysis", "Developing own theoretical model", "Numerical simulations", "Independent research in scientific literature", and "Critical assessment of others' results". We also observe that YPT-related activities are perceived as more useful than other activities to develop skills for "Interpreting experimental data, data analysis" ($p = 0,003$), "Developing own theoretical model" ($p = 0,000$), "Numerical simulations" ($p = 0,000$), and "Critical assessment of others' results" ($p = 0,039$). On the other hand, other activities appear more useful than YPT-related activities to develop abilities to "Solve close-ended problems" ($p = 0,010$) and to conduct "Independent research in scientific literature" ($p = 0,000$).

**Usefulness of RPC vs. YPT activities**

| **Hard Skills** | **t** | **df** | **p** |
|---|---|---|---|
| High school mathematics | 0,288 | 136 | 0,774 |
| High school physics | 0,524 | 184 | 0,601 |
| Solve close-ended problems | **4,409** | **178** | **0,000** |
| Designing experiments | **-3,157** | **131** | **0,002** |
| Conducting experiment | -0,095 | 176 | 0,924 |
| Interpreting experimental data, data analysis | **-3,593** | **180** | **0,000** |
| Developing own theoretical model | **-8,185** | **173** | **0,000** |
| Numerical simulations | **-7,447** | **170** | **0,000** |
| Independent research in scientific literature | **-1,760** | **169** | **0,080** |
| Critical assessment of others' results | **-4,323** | **173** | **0,000** |

Note: Student's t-Test, coefficients with $p \leq 0,10$ highlighted bold.

## Usefulness of RPC vs. Non-YPT activities

| Hard Skills | t | df | p |
|---|---|---|---|
| High school mathematics | -1,160 | 185 | 0,248 |
| High school physics | 0,419 | 262 | 0,676 |
| Solve close-ended problems | 1,425 | 240 | 0,156 |
| Designing experiments | **-4,715** | **240** | **0,000** |
| Conducting experiment | 1,108 | 232 | 0,269 |
| Interpreting experimental data, data analysis | **-2,156** | **238** | **0,032** |
| Developing own theoretical model | **-5,971** | **228** | **0,000** |
| Numerical simulations | **-6,490** | **216** | **0,000** |
| Independent research in scientific literature | **-8,060** | **238** | **0,000** |
| Critical assessment of others' results | **-4,315** | **233** | **0,000** |

Note: Student's t-Test, coefficients with $p \leq 0,10$ highlighted bold.

## Usefulness of YPT activities vs. other activities

| Hard Skills | t | df | p |
|---|---|---|---|
| High school mathematics | -1,000 | 128 | 0,319 |
| High school physics | -0,495 | 178 | 0,621 |
| Solve close-ended problems | **-2,588** | **169** | **0,010** |
| Designing experiments | -0,076 | 127 | 0,939 |
| Conducting experiment | 0,648 | 168 | 0,518 |
| Interpreting experimental data, data analysis | **2,970** | **175** | **0,003** |
| Developing own theoretical model | **4,345** | **162** | **0,000** |
| Numerical simulations | **3,765** | **166** | **0,000** |
| Independent research in scientific literature | **-4,069** | **171** | **0,000** |
| Critical assessment of others' results | **2,079** | **168** | **0,039** |

Note: Student's t-Test, coefficients with $p \leq 0,10$ highlighted bold.

The European Commission's support for the production of this publication does not constitute an endorsement of the contents, which reflect the views only of the authors, and the Commission cannot be held responsible for any use which may be made of the information contained therein.

## 1.3 Impact of years to final exam on usefulness of RPC, YPT and other activities

For regular physics classes, we find lower perceived usefulness for "High school mathematics" (p = 0,046) for students who still had two years until their final exam. Students that were three or more years away from their final exam indicated lower usefulness to develop skills for "High school physics" (p = 0,064) and to "Solve close-ended problems" (p = 0,052). At the same time, students who had only one or two years left until the final exam considered regular physics classes more useful for "Developing own theoretical model", "Numerical simulations", "Independent research in scientific literature", and "Critical assessment of others' results". With few exceptions, students that were three or more years away from their final exam considered YPT-related activities as less useful to develop their hard skills than students who were closer to their final exams. With the exception of "High school mathematics", "Conducting experiment", and "Critical assessment of others' results", we found no differences in the perceived usefulness of participation in other activities based on time to final exam.

**Differences in usefulness of regular classes based on years to final exam**

| Hard Skills – RPC | 1 | 2 | 3+ | R² |
|---|---|---|---|---|
| High school mathematics | -0,205 | **-0,357** | -0,155 | 0,021 |
| Std. Error | 0,177 | **0,178** | 0,200 | |
| p-value | 0,246 | **0,046** | 0,440 | |
| High school physics | -0,087 | -0,164 | **-0,315** | 0,014 |
| Std. Error | 0,159 | 0,158 | **0,169** | |
| p-value | 0,584 | 0,301 | **0,064** | |
| Solve close-ended problems | -0,042 | -0,136 | **-0,338** | 0,018 |
| Std. Error | 0,163 | 0,162 | **0,173** | |
| p-value | 0,796 | 0,402 | **0,052** | |
| Designing experiments | 0,069 | 0,319 | 0,018 | 0,014 |
| Std. Error | 0,202 | 0,199 | 0,215 | |
| p-value | 0,732 | 0,110 | 0,934 | |
| Conducting experiment | 0,120 | 0,311 | 0,310 | 0,014 |
| Std. Error | 0,193 | 0,191 | 0,202 | |
| p-value | 0,534 | 0,104 | 0,126 | |
| Interpreting experimental data, data analysis | 0,209 | 0,316 | **0,396** | 0,016 |
| Std. Error | 0,195 | 0,194 | **0,208** | |
| p-value | 0,286 | 0,104 | **0,058** | |
| Developing own theoretical model | **0,356** | **0,561** | 0,187 | 0,036 |
| Std. Error | **0,196** | **0,194** | 0,210 | |
| p-value | **0,071** | **0,004** | 0,373 | |
| Numerical simulations | 0,185 | **0,831** | **0,667** | 0,080 |
| Std. Error | 0,217 | **0,213** | **0,230** | |
| p-value | 0,393 | **0,000** | **0,004** | |
| Independent research in scientific literature | **0,428** | **0,715** | **0,428** | 0,043 |
| Std. Error | **0,217** | **0,213** | **0,225** | |
| p-value | **0,049** | **0,001** | **0,059** | |
| Critical assessment of others' results | **0,434** | **0,606** | **0,591** | 0,038 |

The European Commission's support for the production of this publication does not constitute an endorsement of the contents, which reflect the views only of the authors, and the Commission cannot be held responsible for any use which may be made of the information contained therein.

| | | | | |
|---|---|---|---|---|
| Std. Error | **0,210** | **0,208** | **0,223** | |
| p-value | **0,040** | **0,004** | **0,008** | |

Note: Linear regression, baseline: year of final exam, coefficients with $p \leq 0,10$ highlighted bold.

## Differences in usefulness of YPT activities based on years to final exam

| Hard Skills – YPT | 1 | 2 | 3+ | R² |
|---|---|---|---|---|
| High school mathematics | -0,188 | **-0,851** | **-0,877** | 0,152 |
| Std. Error | 0,184 | **0,188** | **0,206** | |
| p-value | 0,310 | **0,000** | **0,000** | |
| High school physics | -0,108 | **-0,386** | **-0,690** | 0,069 |
| Std. Error | 0,187 | **0,190** | **0,210** | |
| p-value | 0,566 | **0,044** | **0,001** | |
| Solve close-ended problems | 0,094 | -0,223 | -0,230 | 0,021 |
| Std. Error | 0,194 | 0,200 | 0,216 | |
| p-value | 0,628 | 0,266 | 0,288 | |
| Designing experiments | -0,100 | **-0,394** | -0,311 | 0,025 |
| Std. Error | 0,221 | **0,233** | 0,268 | |
| p-value | 0,653 | **0,094** | 0,249 | |
| Conducting experiment | -0,157 | **-0,645** | **-0,775** | 0,094 |
| Std. Error | 0,193 | **0,198** | **0,220** | |
| p-value | 0,417 | **0,001** | **0,001** | |
| Interpreting experimental data, data analysis | -0,222 | **-0,580** | **-0,862** | 0,095 |
| Std. Error | 0,188 | **0,193** | **0,216** | |
| p-value | 0,240 | **0,003** | **0,000** | |
| Developing own theoretical model | -0,159 | **-0,536** | **-0,659** | 0,071 |
| Std. Error | 0,188 | **0,193** | **0,216** | |
| p-value | 0,400 | **0,006** | **0,003** | |
| Numerical simulations | -0,133 | **-0,673** | **-0,790** | 0,087 |
| Std. Error | 0,216 | **0,222** | **0,243** | |
| p-value | 0,537 | **0,003** | **0,001** | |
| Independent research in scientific literature | -0,160 | -0,290 | -0,395 | 0,016 |
| Std. Error | 0,228 | 0,229 | 0,254 | |
| p-value | 0,484 | 0,207 | 0,121 | |
| Critical assessment of others' results | -0,163 | **-0,432** | **-0,769** | 0,072 |
| Std. Error | 0,199 | **0,200** | **0,225** | |
| p-value | 0,413 | **0,032** | **0,001** | |

Note: Linear regression, baseline: year of final exam, coefficients with $p \leq 0,10$ highlighted bold.

## Differences in usefulness of other Non-YPT activities based on years to final exam

| Hard Skills – Non-YPT | 1 | 2 | 3+ | R² |
|---|---|---|---|---|
| High school mathematics | 0,165 | **0,348** | 0,047 | 0,018 |
| Std. Error | 0,201 | **0,205** | 0,230 | |
| p-value | 0,411 | **0,091** | 0,840 | |
| High school physics | 0,047 | -0,018 | 0,000 | 0,001 |
| Std. Error | 0,166 | 0,168 | 0,180 | |
| p-value | 0,780 | 0,917 | 0,999 | |
| Solve close-ended problems | 0,185 | 0,130 | -0,033 | 0,008 |
| Std. Error | 0,180 | 0,182 | 0,192 | |
| p-value | 0,305 | 0,475 | 0,864 | |
| Designing experiments | 0,072 | 0,101 | 0,019 | 0,002 |
| Std. Error | 0,185 | 0,185 | 0,196 | |
| p-value | 0,696 | 0,584 | 0,925 | |
| Conducting experiment | **0,335** | **0,337** | 0,176 | 0,018 |
| Std. Error | **0,188** | **0,186** | 0,197 | |
| p-value | **0,076** | **0,072** | 0,371 | |
| Interpreting experimental data, data analysis | 0,225 | 0,086 | -0,046 | 0,012 |
| Std. Error | 0,175 | 0,173 | 0,185 | |
| p-value | 0,199 | 0,620 | 0,805 | |
| Developing own theoretical model | 0,159 | 0,257 | 0,020 | 0,011 |
| Std. Error | 0,192 | 0,191 | 0,204 | |
| p-value | 0,410 | 0,180 | 0,923 | |
| Numerical simulations | 0,214 | 0,199 | 0,236 | 0,006 |
| Std. Error | 0,217 | 0,216 | 0,235 | |
| p-value | 0,325 | 0,358 | 0,317 | |
| Independent research in scientific literature | 0,212 | -0,060 | 0,082 | 0,013 |
| Std. Error | 0,176 | 0,177 | 0,187 | |
| p-value | 0,229 | 0,733 | 0,662 | |
| Critical assessment of others' results | **0,357** | **0,357** | 0,224 | 0,018 |
| Std. Error | **0,192** | **0,195** | 0,206 | |
| p-value | **0,064** | **0,068** | 0,278 | |

Note: Linear regression, baseline: year of final exam, coefficients with $p \leq 0,10$ highlighted bold.

## 1.4 Impact of physics classes on usefulness of RPC, YPT and other activities

We test the hypothesis that the perceived usefulness of regular physics classes, YPT-related activities, and other activities depends on the students' weekly physics classes. Below, we show regression results for the perceived usefulness with the responses of students without weekly physics classes as baseline.

Contrary to our expectations, we observe that students perceive their regular physics classes as more useful to develop skills for "Numerical simulations" when they attend only few ($\leq 3$ hours) weekly physics classes. At the same time, except for "Critical assessment of others' results", we find no differences in the perceived usefulness of YPT-related activities contingent on the number of weekly physics classes. However, we observe lower perceived usefulness of participation in other activities for students who take only little (1 hour) weekly physics classes.

**Differences in usefulness of regular classes based on regular physics classes per week**

| Hard Skills – RPC | 1 | 2 | 3 | 4 | 5+ | R² |
|---|---|---|---|---|---|---|
| High school mathematics | -0,310 | -0,210 | -0,143 | 0,012 | 0,106 | 0,018 |
| Std. Error | 0,493 | 0,373 | 0,387 | 0,399 | 0,408 | |
| p-value | 0,531 | 0,575 | 0,712 | 0,976 | 0,795 | |
| High school physics | 0,268 | 0,106 | 0,219 | 0,443 | 0,325 | 0,018 |
| Std. Error | 0,472 | 0,353 | 0,362 | 0,373 | 0,396 | |
| p-value | 0,571 | 0,764 | 0,547 | 0,237 | 0,412 | |
| Solve close-ended problems | 0,643 | 0,089 | 0,299 | 0,327 | 0,506 | 0,028 |
| Std. Error | 0,511 | 0,356 | 0,366 | 0,378 | 0,399 | |
| p-value | 0,210 | 0,803 | 0,414 | 0,387 | 0,205 | |
| Designing experiments | 0,667 | 0,508 | 0,548 | 0,693 | 0,258 | 0,013 |
| Std. Error | 0,617 | 0,477 | 0,488 | 0,502 | 0,526 | |
| p-value | 0,281 | 0,288 | 0,263 | 0,168 | 0,625 | |
| Conducting experiment | 0,738 | 0,547 | 0,587 | 0,436 | 0,253 | 0,014 |
| Std. Error | 0,600 | 0,419 | 0,429 | 0,444 | 0,468 | |
| p-value | 0,220 | 0,192 | 0,172 | 0,327 | 0,589 | |
| Interpreting experimental data, data analysis | 0,625 | 0,303 | 0,140 | 0,403 | 0,244 | 0,011 |
| Std. Error | 0,567 | 0,400 | 0,410 | 0,428 | 0,455 | |
| p-value | 0,272 | 0,450 | 0,733 | 0,348 | 0,593 | |
| Developing own theoretical model | 0,881 | 0,332 | 0,527 | 0,444 | 0,262 | 0,015 |
| Std. Error | 0,612 | 0,460 | 0,471 | 0,485 | 0,510 | |
| p-value | 0,152 | 0,472 | 0,264 | 0,361 | 0,608 | |
| Numerical simulations | **1,381** | 0,580 | **0,856** | 0,273 | -0,083 | 0,066 |
| Std. Error | **0,654** | 0,493 | **0,504** | 0,522 | 0,548 | |
| p-value | **0,036** | 0,240 | **0,091** | 0,602 | 0,879 | |
| Independent research in scientific literature | -0,167 | -0,154 | -0,214 | -0,375 | -0,600 | 0,012 |
| Std. Error | 0,633 | 0,502 | 0,513 | 0,535 | 0,559 | |
| p-value | 0,793 | 0,760 | 0,677 | 0,484 | 0,284 | |
| Critical assessment of others' results | -0,107 | 0,160 | 0,125 | 0,128 | -0,250 | 0,009 |
| Std. Error | 0,618 | 0,436 | 0,448 | 0,466 | 0,500 | |
| p-value | 0,863 | 0,714 | 0,780 | 0,783 | 0,617 | |

Note: Linear regression, baseline: no weekly physics classes, coefficients with p ≤ 0,10 highlighted bold.

**Differences in usefulness of YPT activities based on regular physics classes per week**

| Hard Skills - RPC | 1 | 2 | 3 | 4 | 5+ | R² |
|---|---|---|---|---|---|---|
| High school mathematics | 0,917 | 0,545 | 0,652 | **1,105** | 0,883 | 0,055 |
| Std. Error | 0,630 | 0,499 | 0,511 | **0,518** | 0,549 | |
| p-value | 0,147 | 0,276 | 0,203 | **0,034** | 0,109 | |
| High school physics | 0,143 | -0,230 | 0,023 | 0,167 | 0,400 | 0,046 |
| Std. Error | 0,589 | 0,481 | 0,491 | 0,500 | 0,529 | |
| p-value | 0,809 | 0,633 | 0,962 | 0,739 | 0,450 | |
| Solve close-ended problems | 0,433 | 0,072 | 0,325 | 0,406 | 0,529 | 0,030 |
| Std. Error | 0,590 | 0,448 | 0,462 | 0,470 | 0,508 | |
| p-value | 0,464 | 0,873 | 0,483 | 0,388 | 0,299 | |
| Designing experiments | 0,500 | -0,323 | 0,161 | 0,290 | 0,067 | 0,077 |
| Std. Error | 0,691 | 0,577 | 0,593 | 0,600 | 0,618 | |
| p-value | 0,471 | 0,577 | 0,786 | 0,630 | 0,914 | |
| Conducting experiment | 0,393 | -0,098 | 0,250 | 0,485 | 0,450 | 0,059 |
| Std. Error | 0,629 | 0,513 | 0,525 | 0,530 | 0,565 | |
| p-value | 0,533 | 0,848 | 0,635 | 0,361 | 0,427 | |
| Interpreting experimental data, data analysis | 0,250 | 0,008 | 0,440 | 0,656 | 0,500 | 0,070 |
| Std. Error | 0,611 | 0,498 | 0,510 | 0,517 | 0,545 | |
| p-value | 0,683 | 0,987 | 0,389 | 0,206 | 0,360 | |
| Developing own theoretical model | 0,200 | -0,092 | 0,122 | 0,323 | 0,000 | 0,026 |
| Std. Error | 0,613 | 0,446 | 0,459 | 0,467 | 0,501 | |
| p-value | 0,745 | 0,837 | 0,791 | 0,491 | 1,000 | |
| Numerical simulations | 0,417 | -0,272 | 0,250 | 0,350 | 0,250 | 0,064 |
| Std. Error | 0,708 | 0,561 | 0,575 | 0,584 | 0,617 | |
| p-value | 0,557 | 0,628 | 0,664 | 0,550 | 0,686 | |
| Independent research in scientific literature | 1,083 | 0,506 | 0,957 | 0,650 | 1,036 | 0,046 |
| Std. Error | 0,717 | 0,568 | 0,582 | 0,591 | 0,630 | |
| p-value | 0,133 | 0,375 | 0,102 | 0,273 | 0,102 | |
| Critical assessment of others' results | 0,417 | 0,394 | **0,869** | 0,853 | 0,821 | 0,060 |
| Std. Error | 0,635 | 0,503 | **0,514** | 0,524 | 0,557 | |
| p-value | 0,512 | 0,435 | **0,093** | 0,105 | 0,142 | |

Note: Linear regression, baseline: no weekly physics classes, coefficients with $p \leq 0,10$ highlighted bold.

**Differences in usefulness of other Non-YPT activities based on RPC per week**

| Hard Skills – Non-YPT | 1 | 2 | 3 | 4 | 5+ | R² |
|---|---|---|---|---|---|---|
| High school mathematics | -1,429 | -0,247 | -0,403 | -0,021 | -0,571 | 0,073 |
| Std. Error | 0,523 | 0,384 | 0,401 | 0,415 | 0,427 | |
| p-value | 0,007 | 0,521 | 0,317 | 0,959 | 0,182 | |
| High school physics | **-0,875** | -0,328 | -0,153 | 0,039 | -0,114 | 0,035 |
| Std. Error | **0,471** | 0,343 | 0,354 | 0,366 | 0,389 | |
| p-value | **0,064** | 0,340 | 0,665 | 0,914 | 0,770 | |
| Solve close-ended problems | **-0,917** | -0,326 | -0,302 | -0,132 | -0,159 | 0,019 |
| Std. Error | **0,533** | 0,360 | 0,372 | 0,387 | 0,407 | |
| p-value | **0,086** | 0,366 | 0,418 | 0,733 | 0,696 | |
| Designing experiments | -0,196 | 0,044 | 0,276 | 0,343 | 0,229 | 0,019 |
| Std. Error | 0,525 | 0,395 | 0,405 | 0,420 | 0,445 | |
| p-value | 0,709 | 0,911 | 0,497 | 0,415 | 0,608 | |
| Conducting experiment | -0,250 | 0,052 | 0,232 | 0,187 | 0,000 | 0,011 |
| Std. Error | 0,546 | 0,369 | 0,382 | 0,400 | 0,423 | |
| p-value | 0,647 | 0,889 | 0,543 | 0,639 | 1,000 | |
| Interpreting experimental data, data analysis | -0,321 | -0,006 | 0,017 | 0,393 | 0,139 | 0,026 |

The European Commission's support for the production of this publication does not constitute an endorsement of the contents, which reflect the views only of the authors, and the Commission cannot be held responsible for any use which may be made of the information contained therein.

| | | | | | | |
|---|---|---|---|---|---|---|
| Std. Error | 0,490 | 0,346 | 0,356 | 0,371 | 0,402 | |
| p-value | 0,513 | 0,985 | 0,963 | 0,291 | 0,730 | |
| Developing own theoretical model | -0,571 | -0,620 | -0,357 | -0,371 | -0,254 | 0,025 |
| Std. Error | 0,548 | 0,400 | 0,411 | 0,424 | 0,457 | |
| p-value | 0,298 | 0,122 | 0,386 | 0,382 | 0,579 | |
| Numerical simulations | 0,429 | 0,272 | 0,660 | 0,667 | 0,600 | 0,031 |
| Std. Error | 0,632 | 0,477 | 0,489 | 0,501 | 0,529 | |
| p-value | 0,499 | 0,570 | 0,179 | 0,185 | 0,258 | |
| Independent research in scientific literature | -0,619 | -0,401 | -0,253 | -0,369 | -0,186 | 0,012 |
| Std. Error | 0,490 | 0,378 | 0,388 | 0,402 | 0,427 | |
| p-value | 0,207 | 0,289 | 0,515 | 0,359 | 0,664 | |
| Critical assessment of others' results | -0,857 | -0,105 | -0,047 | -0,218 | -0,457 | 0,023 |
| Std. Error | 0,577 | 0,404 | 0,415 | 0,429 | 0,456 | |
| p-value | 0,139 | 0,795 | 0,910 | 0,611 | 0,317 | |

Note: Linear regression, baseline: no weekly physics classes, coefficients with $p \leq 0,10$ highlighted bold.

## 1.5 Impact of participation in YPT on usefulness of RPC, YPT and other activities

For some types of hard skills, we observe that students that participated in YPT-related activities consider RPC and other Non-YPT activities as less useful to develop these hard skills. What is interesting is that we observe these effects only for students that participated in YPT-related activities some time ago but not for students that recently participated in these activities. This may suggest that synergies between the YPT-related activities and regular physics classes as well as other activities are limited. We observe no differences in the perceived usefulness of YPT-related activities based on the most recent participation.

**Differences in usefulness of RPC based on most recent participation in YPT activities**

| Hard Skills - RPC | Earlier | This year | $R^2$ |
|---|---|---|---|
| High school mathematics | **-0,900** | 0,481 | 0,135 |
| Std. Error | **0,264** | 0,531 | |
| p-value | **0,001** | 0,368 | |
| High school physics | **-0,747** | 0,279 | 0,070 |
| Std. Error | **0,217** | 0,451 | |
| p-value | **0,001** | 0,537 | |
| Solve close-ended problems | **-0,979** | 0,493 | 0,119 |
| Std. Error | **0,218** | 0,442 | |
| p-value | **0,000** | 0,266 | |
| Designing experiments | **-1,310** | -0,250 | 0,130 |
| Std. Error | **0,266** | 0,576 | |
| p-value | **0,000** | 0,665 | |
| Conducting experiment | **-1,219** | -0,528 | 0,125 |
| Std. Error | **0,255** | 0,553 | |
| p-value | **0,000** | 0,341 | |
| Interpreting experimental data, data analysis | **-1,621** | -0,771 | 0,206 |

The European Commission's support for the production of this publication does not constitute an endorsement of the contents, which reflect the views only of the authors, and the Commission cannot be held responsible for any use which may be made of the information contained therein.

| | | | |
|---|---|---|---|
| Std. Error | **0,254** | 0,621 | |
| p-value | **0,000** | 0,216 | |
| Developing own theoretical model | **-1,191** | -0,341 | 0,123 |
| Std. Error | **0,256** | 0,624 | |
| p-value | **0,000** | 0,586 | |
| Numerical simulations | **-1,262** | -1,962 | 0,134 |
| Std. Error | **0,279** | 1,164 | |
| p-value | **0,000** | 0,094 | |
| Independent research in scientific literature | **-1,405** | -0,355 | 0,144 |
| Std. Error | **0,271** | 0,806 | |
| p-value | **0,000** | 0,661 | |
| Critical assessment of others' results | **-1,349** | 0,051 | 0,116 |
| Std. Error | **0,297** | 0,630 | |
| p-value | **0,000** | 0,936 | |

Note: Linear regression, baseline: no participation, coefficients with $p \leq 0,10$ highlighted bold.

**Differences in usefulness of YPT activities based on most recent participation in YPT activities**

| **Hard Skills - YPT** | **Earlier** | **This year** | **$R^2$** |
|---|---|---|---|
| High school mathematics | 0,297 | 0,547 | 0,022 |
| Std. Error | 0,263 | 0,493 | |
| p-value | 0,261 | 0,271 | |
| High school physics | -0,005 | 0,307 | 0,004 |
| Std. Error | 0,256 | 0,482 | |
| p-value | 0,984 | 0,525 | |
| Solve close-ended problems | -0,120 | -0,517 | 0,011 |
| Std. Error | 0,265 | 0,511 | |
| p-value | 0,651 | 0,314 | |
| Designing experiments | 0,286 | 0,548 | 0,028 |
| Std. Error | 0,299 | 0,580 | |
| p-value | 0,344 | 0,349 | |
| Conducting experiment | 0,211 | 0,785 | 0,028 |
| Std. Error | 0,258 | 0,497 | |
| p-value | 0,414 | 0,117 | |
| Interpreting experimental data, data analysis | 0,238 | **0,944** | 0,043 |
| Std. Error | 0,240 | **0,464** | |
| p-value | 0,324 | **0,044** | |
| Developing own theoretical model | 0,221 | 0,560 | 0,023 |
| Std. Error | 0,226 | 0,435 | |
| p-value | 0,330 | 0,202 | |
| Numerical simulations | 0,383 | -0,055 | 0,014 |
| Std. Error | 0,314 | 0,591 | |
| p-value | 0,226 | 0,927 | |
| Independent research in scientific literature | 0,297 | 0,297 | 0,010 |

The European Commission's support for the production of this publication does not constitute an endorsement of the contents, which reflect the views only of the authors, and the Commission cannot be held responsible for any use which may be made of the information contained therein.

| | | | |
|---|---|---|---|
| Std. Error | 0,314 | 0,590 | |
| p-value | 0,347 | 0,616 | |
| Critical assessment of others' results | 0,208 | 0,641 | 0,021 |
| Std. Error | 0,271 | 0,494 | |
| p-value | 0,444 | 0,197 | |

Note: Linear regression, baseline: no participation, coefficients with p ≤ 0,10 highlighted bold.

**Differences in usefulness of other Non-YPT activities based on most recent participation in YPT activities**

| Hard Skills – Non-YPT | Earlier | This year | R² |
|---|---|---|---|
| High school mathematics | -0,484 | -0,217 | 0,031 |
| Std. Error | 0,298 | 0,750 | |
| p-value | 0,108 | 0,773 | |
| High school physics | -0,241 | -0,386 | 0,011 |
| Std. Error | 0,219 | 0,455 | |
| p-value | 0,274 | 0,398 | |
| Solve close-ended problems | -0,075 | -0,075 | 0,001 |
| Std. Error | 0,239 | 0,569 | |
| p-value | 0,755 | 0,896 | |
| Designing experiments | **-0,563** | 0,508 | 0,036 |
| Std. Error | **0,258** | 0,615 | |
| p-value | **0,031** | 0,410 | |
| Conducting experiment | **-0,895** | 0,000 | 0,088 |
| Std. Error | **0,237** | 0,688 | |
| p-value | **0,000** | 1,000 | |
| Interpreting experimental data, data analysis | **-0,939** | 0,394 | 0,099 |
| Std. Error | **0,239** | 0,554 | |
| p-value | **0,000** | 0,478 | |
| Developing own theoretical model | **-0,470** | -1,119 | 0,042 |
| Std. Error | **0,259** | 0,614 | |
| p-value | **0,071** | 0,070 | |
| Numerical simulations | **-0,668** | -0,563 | 0,040 |
| Std. Error | **0,289** | 0,684 | |
| p-value | **0,022** | 0,412 | |
| Independent research in scientific literature | **-1,053** | 0,667 | 0,131 |
| Std. Error | **0,227** | 0,542 | |
| p-value | **0,000** | 0,221 | |
| Critical assessment of others' results | **-1,199** | 0,503 | 0,141 |
| Std. Error | **0,249** | 0,592 | |
| p-value | **0,000** | 0,398 | |

Note: Linear regression, baseline: no participation, coefficients with p ≤ 0,10 highlighted bold.

The European Commission's support for the production of this publication does not constitute an endorsement of the contents, which reflect the views only of the authors, and the Commission cannot be held responsible for any use which may be made of the information contained therein.

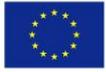
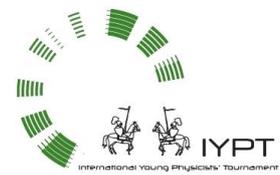

*DEVELOPMENT OF INQUIRY-BASED LEARNING VIA IYPT*

## 1.6 Impact of participation in Non-YPT competitions on usefulness of RPC, YPT and other activities

We test the hypothesis that the perceived usefulness of regular physics classes, YPT-related activities, and other activities depends on the students' most recent participation in other, non-YPT activities. Below, we show regression results for the perceived usefulness with the responses of students who never participated in other activities as baseline. Depending on the year of the survey, the year of reference— "This year"—is either 2021 or 2020.

For some types of hard skills, we observe that students that participated in Non-YPT activities consider regular physics classes as less useful—particularly in case of "High school physics" and the ability to "Solve close-ended problems". We also find that students that participated in other activities considered YPT-related activities as more useful to develop the skills for "Developing own theoretical model" and "Independent research in scientific literature". This, in contrast to the results above, suggests that there may be synergies between the YPT-related activities and Non-YPT activities. Only in the case of "Independent research in scientific literature", we observe differences in the perceived usefulness of Non-YPT activities contingent on the most recent participation in these activities.

**Differences in usefulness of RPC based on most recent participation in other Non-YPT activities**

| Hard Skills - RPC | Earlier | This year | $R^2$ |
|---|---|---|---|
| High school mathematics | -0,277 | **-2,194** | 0,076 |
| Std. Error | 0,211 | **0,895** | |
| p-value | 0,193 | **0,016** | |
| High school physics | -0,384 | **-2,134** | 0,074 |
| Std. Error | 0,232 | **0,980** | |
| p-value | 0,100 | **0,032** | |
| Solve close-ended problems | **-0,454** | **-2,246** | 0,090 |
| Std. Error | **0,232** | 0,978 | |
| p-value | **0,053** | 0,024 | |
| Designing experiments | -0,302 | -1,222 | 0,021 |
| Std. Error | 0,300 | 1,280 | |
| p-value | 0,317 | 0,342 | |
| Conducting experiment | -0,139 | -1,619 | 0,018 |
| Std. Error | 0,317 | 1,354 | |
| p-value | 0,662 | 0,235 | |
| Interpreting experimental data, data analysis | **-0,797** | -1,547 | 0,084 |
| Std. Error | **0,306** | 1,287 | |
| p-value | **0,011** | 0,233 | |
| Developing own theoretical model | **-0,649** | -1,190 | 0,069 |
| Std. Error | **0,275** | 1,155 | |
| p-value | **0,021** | 0,306 | |
| Numerical simulations | -0,295 | -0,339 | 0,017 |
| Std. Error | 0,255 | 1,055 | |
| p-value | 0,251 | 0,749 | |
| Independent research in scientific literature | -0,378 | -0,937 | 0,022 |

The European Commission's support for the production of this publication does not constitute an endorsement of the contents, which reflect the views only of the authors, and the Commission cannot be held responsible for any use which may be made of the information contained therein.

| | | | |
|---|---|---|---|
| Std. Error | 0,301 | 1,287 | |
| p-value | 0,213 | 0,468 | |
| Critical assessment of others' results | -0,086 | -0,726 | 0,004 |
| Std. Error | 0,313 | 1,331 | |
| p-value | 0,784 | 0,587 | |

Note: Linear regression, baseline: no participation, coefficients with p ≤ 0,10 highlighted bold.

**Differences in usefulness of YPT activities based on most recent participation in other Non-YPT activities**

| Hard Skills - YPT | Earlier | This year | $R^2$ |
|---|---|---|---|
| High school mathematics | 0,190 | -1,190 | 0,040 |
| Std. Error | 0,241 | 0,911 | |
| p-value | 0,432 | 0,196 | |
| High school physics | -0,306 | -1,163 | 0,046 |
| Std. Error | 0,240 | 0,912 | |
| p-value | 0,208 | 0,207 | |
| Solve close-ended problems | 0,198 | -0,756 | 0,021 |
| Std. Error | 0,259 | 0,991 | |
| p-value | 0,446 | 0,449 | |
| Designing experiments | 0,388 | -0,707 | 0,046 |
| Std. Error | 0,265 | 0,998 | |
| p-value | 0,148 | 0,481 | |
| Conducting experiment | 0,156 | -1,225 | 0,041 |
| Std. Error | 0,238 | 0,893 | |
| p-value | 0,514 | 0,175 | |
| Interpreting experimental data, data analysis | 0,159 | -1,250 | 0,041 |
| Std. Error | 0,225 | 0,872 | |
| p-value | 0,482 | 0,157 | |
| **Developing own theoretical model** | **0,425** | -1,146 | 0,103 |
| Std. Error | **0,209** | 0,787 | |
| p-value | **0,046** | 0,150 | |
| Numerical simulations | 0,190 | -1,048 | 0,026 |
| Std. Error | 0,277 | 1,050 | |
| p-value | 0,495 | 0,322 | |
| **Independent research in scientific literature** | **0,643** | -0,357 | 0,063 |
| Std. Error | **0,334** | 1,200 | |
| p-value | **0,059** | 0,767 | |
| Critical assessment of others' results | 0,048 | -1,000 | 0,015 |
| Std. Error | 0,283 | 1,069 | |
| p-value | 0,867 | 0,353 | |

Note: Linear regression, baseline: no participation, coefficients with p ≤ 0,10 highlighted bold.

The European Commission's support for the production of this publication does not constitute an endorsement of the contents, which reflect the views only of the authors, and the Commission cannot be held responsible for any use which may be made of the information contained therein.

**Differences in usefulness of Non-YPT activities based on most recent participation in  Non-YPT activities**

| Hard Skills - Other | Earlier | This year | R² |
|---|---|---|---|
| High school mathematics | 0,163 | -0,045 | 0,005 |
| Std. Error | 0,255 | 1,078 | |
| p-value | 0,525 | 0,966 | |
| High school physics | 0,039 | -0,121 | 0,001 |
| Std. Error | 0,215 | 0,921 | |
| p-value | 0,857 | 0,896 | |
| Solve close-ended problems | 0,247 | 0,081 | 0,011 |
| Std. Error | 0,254 | 1,066 | |
| p-value | 0,333 | 0,940 | |
| Designing experiments | -0,063 | 0,328 | 0,002 |
| Std. Error | 0,287 | 1,183 | |
| p-value | 0,826 | 0,782 | |
| Conducting experiment | -0,069 | 0,322 | 0,002 |
| Std. Error | 0,267 | 1,095 | |
| p-value | 0,796 | 0,769 | |
| Interpreting experimental data, data analysis | -0,132 | 0,172 | 0,004 |
| Std. Error | 0,247 | 1,010 | |
| p-value | 0,594 | 0,865 | |
| Developing own theoretical model | 0,250 | 0,386 | 0,013 |
| Std. Error | 0,259 | 1,042 | |
| p-value | 0,339 | 0,712 | |
| Numerical simulations | -0,354 | 0,596 | 0,021 |
| Std. Error | 0,310 | 1,203 | |
| p-value | 0,258 | 0,621 | |
| Independent research in scientific literature | **-0,578** | 0,031 | 0,058 |
| Std. Error | **0,253** | 1,050 | |
| p-value | **0,025** | 0,977 | |
| Critical assessment of others' results | -0,359 | 0,459 | 0,020 |
| Std. Error | 0,298 | 1,209 | |
| p-value | 0,232 | 0,705 | |

Note: Linear regression, baseline: no participation, coefficients with $p \leq 0,10$ highlighted bold.

## 1.7 Country differences

### 1.7.1 Across-country differences

To test the impact of country differences on our results, we use ANOVA to test for differences in self-evaluation and perceived usefulness of regular physics classes, YPT-related activities, and other activities contingent on the student's home country. We observe that students' self-evaluations for most types of hard skills differ by country. We find across-country differences in the perceived usefulness of regular physics classes for eight out of ten hard skills. In the case of YPT-related activities, however, we observe that the perceived usefulness for all types of hard skills depends on students' home countries. We observe country differences for five out of ten types of hard skills for the perceived usefulness of participation in other activities.

**Differences in self-evaluation based on country**

| Hard Skills – self-evaluation | df | F | p |
|---|---|---|---|
| High school mathematics | **12,899** | **2,231** | **0,026** |
| High school physics | 2,230 | 0,322 | 0,957 |
| Solve close-ended problems | 5,077 | 1,031 | 0,410 |
| Designing experiments | **12,170** | **4,770** | **0,001** |
| Conducting experiment | **13,233** | **4,432** | **0,002** |
| Interpreting experimental data, data analysis | **13,632** | **3,829** | **0,005** |
| Developing own theoretical model | 1,349 | 0,453 | 0,770 |
| Numerical simulations | **23,109** | **5,440** | **0,000** |
| Independent research in scientific literature | 6,652 | 1,267 | 0,284 |
| Critical assessment of others' results | 2,050 | 0,592 | 0,669 |

Note: ANOVA (Value ~ Country), coefficients with p ≤ 0,10 highlighted bold.

**Differences in usefulness of regular physics classes based on country**

| Hard Skills – RPC | df | F | p |
|---|---|---|---|
| High school mathematics | **8,264** | **2,742** | **0,030** |
| High school physics | 1,226 | 0,365 | 0,833 |
| Solve close-ended problems | 2,683 | 0,785 | 0,536 |
| Designing experiments | **13,066** | **2,579** | **0,038** |
| Conducting experiment | **38,296** | **9,334** | **0,000** |
| Interpreting experimental data, data analysis | **21,258** | **4,719** | **0,001** |
| Developing own theoretical model | 5,550 | 1,154 | 0,332 |
| Numerical simulations | **48,752** | **9,621** | **0,000** |
| Independent research in scientific literature | **27,770** | **5,158** | **0,001** |
| Critical assessment of others' results | **56,740** | **11,722** | **0,000** |

Note: ANOVA (Value ~ Country), coefficients with p ≤ 0,10 highlighted bold.

**Differences in usefulness of YPT activities based on country**

| Hard Skills – YPT | df | F | p |
|---|---|---|---|
| High school mathematics | 52,205 | 7,542 | 0,000 |
| High school physics | 53,342 | 7,649 | 0,000 |
| Solve close-ended problems | 17,245 | 2,685 | 0,011 |
| Designing experiments | 18,976 | 5,350 | 0,000 |
| Conducting experiment | 24,766 | 6,651 | 0,000 |
| Interpreting experimental data, data analysis | 24,654 | 6,972 | 0,000 |
| Developing own theoretical model | 11,134 | 3,104 | 0,017 |
| Numerical simulations | 35,661 | 8,236 | 0,000 |
| Independent research in scientific literature | 14,385 | 2,985 | 0,020 |
| Critical assessment of others' results | 13,743 | 3,636 | 0,007 |

Note: ANOVA (Value ~ Country), coefficients with $p \leq 0,10$ highlighted bold.

**Differences in usefulness of other activities classes based on country**

| Hard Skills – Other | df | F | p |
|---|---|---|---|
| High school mathematics | **8,691** | **2,221** | **0,068** |
| High school physics | 4,938 | 1,378 | 0,242 |
| Solve close-ended problems | 3,788 | 0,976 | 0,422 |
| Designing experiments | 1,866 | 0,450 | 0,772 |
| Conducting experiment | **8,101** | **2,039** | **0,090** |
| Interpreting experimental data, data analysis | 6,305 | 1,770 | 0,135 |
| Developing own theoretical model | 4,499 | 1,068 | 0,373 |
| Numerical simulations | 0,763 | 0,144 | 0,965 |
| Independent research in scientific literature | 2,472 | 0,655 | 0,624 |
| Critical assessment of others' results | 6,371 | 1,488 | 0,207 |

Note: ANOVA (Value ~ Country), coefficients with $p \leq 0,10$ highlighted bold.

## 1.7.2 Within-country differences

To further investigate the results from above, we provide country-level summary statistics for students' self-evaluation and the usefulness of regular physics classes, YPT-related activities, and other activities for each hard skill separately. Note: No data available for Austria, Germany, Iran, and Russia.

### High school mathematics

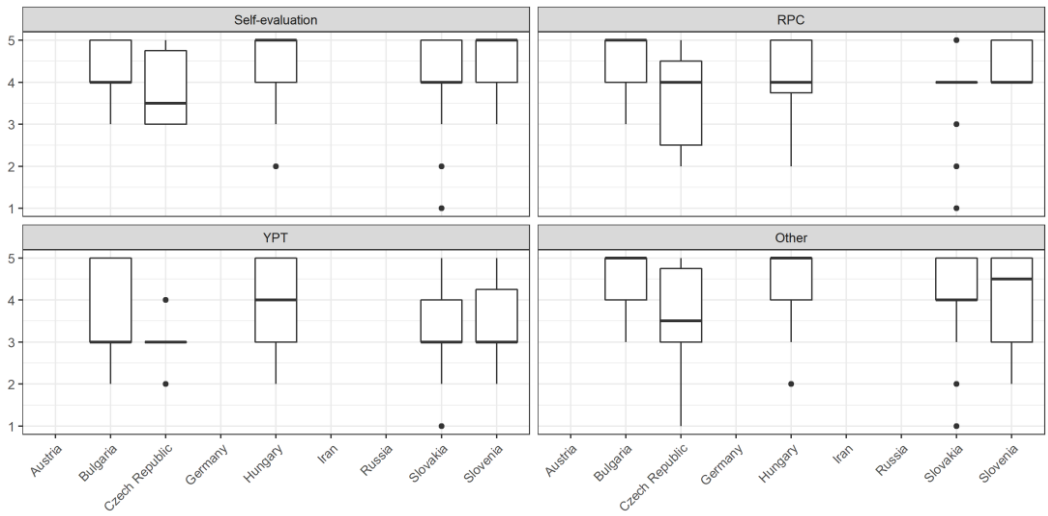

| Country | Type | Valid | Missing | Mean | Median | SD | Min. | Max. |
|---|---|---|---|---|---|---|---|---|
| Austria | Self-evaluation | 0 | 13 | 0,00 | 0 | 0,00 | 0 | 0 |
| | RPC | 0 | 13 | 0,00 | 0 | 0,00 | 0 | 0 |
| | YPT | 0 | 13 | 0,00 | 0 | 0,00 | 0 | 0 |
| | Other | 0 | 13 | 0,00 | 0 | 0,00 | 0 | 0 |
| Bulgaria | Self-evaluation | 21 | 0 | 4,48 | 5 | 0,60 | 3 | 5 |
| | RPC | 17 | 4 | 3,76 | 3 | 1,03 | 2 | 5 |
| | YPT | 21 | 0 | 4,43 | 5 | 0,81 | 3 | 5 |
| | Other | 21 | 0 | 4,38 | 4 | 0,67 | 3 | 5 |
| Czech Rep. | Self-evaluation | 7 | 16 | 3,57 | 4 | 1,27 | 2 | 5 |
| | RPC | 6 | 17 | 3,00 | 3 | 0,63 | 2 | 4 |
| | YPT | 10 | 13 | 3,40 | 3,5 | 1,51 | 1 | 5 |
| | Other | 6 | 17 | 3,83 | 3,5 | 0,98 | 3 | 5 |
| Germany | Self-evaluation | 0 | 3 | 0,00 | 0 | 0,00 | 0 | 0 |
| | RPC | 0 | 3 | 0,00 | 0 | 0,00 | 0 | 0 |
| | YPT | 0 | 3 | 0,00 | 0 | 0,00 | 0 | 0 |
| | Other | 0 | 3 | 0,00 | 0 | 0,00 | 0 | 0 |
| Hungary | Self-evaluation | 68 | 4 | 4,06 | 4 | 0,93 | 2 | 5 |
| | RPC | 41 | 31 | 4,15 | 4 | 0,88 | 2 | 5 |
| | YPT | 65 | 7 | 4,23 | 5 | 0,95 | 2 | 5 |
| | Other | 67 | 5 | 4,43 | 5 | 0,70 | 2 | 5 |
| Iran | Self-evaluation | 0 | 1 | 0,00 | 0 | 0,00 | 0 | 0 |
| | RPC | 0 | 1 | 0,00 | 0 | 0,00 | 0 | 0 |
| | YPT | 0 | 1 | 0,00 | 0 | 0,00 | 0 | 0 |
| | Other | 0 | 1 | 0,00 | 0 | 0,00 | 0 | 0 |
| Russia | Self-evaluation | 0 | 1 | 0,00 | 0 | 0,00 | 0 | 0 |
| | RPC | 0 | 1 | 0,00 | 0 | 0,00 | 0 | 0 |
| | YPT | 0 | 1 | 0,00 | 0 | 0,00 | 0 | 0 |
| | Other | 0 | 1 | 0,00 | 0 | 0,00 | 0 | 0 |
| Slovakia | Self-evaluation | 93 | 72 | 3,92 | 4 | 0,86 | 1 | 5 |
| | RPC | 68 | 97 | 3,35 | 3 | 0,96 | 1 | 5 |
| | YPT | 85 | 80 | 4,04 | 4 | 0,97 | 1 | 5 |
| | Other | 92 | 73 | 3,95 | 4 | 0,88 | 1 | 5 |
| Slovenia | Self-evaluation | 9 | 0 | 4,44 | 4 | 0,53 | 4 | 5 |
| | RPC | 8 | 1 | 3,50 | 3 | 1,07 | 2 | 5 |
| | YPT | 8 | 1 | 4,00 | 4,5 | 1,20 | 2 | 5 |
| | Other | 9 | 0 | 4,56 | 5 | 0,73 | 3 | 5 |

**High school physics**

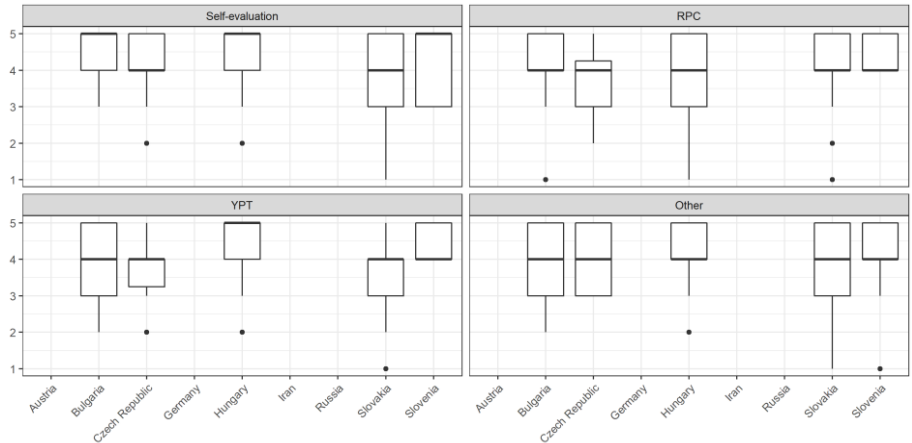

| Country | Type | Valid | Missing | Mean | Median | SD | Min. | Max. |
|---|---|---|---|---|---|---|---|---|
| Austria | Self-evaluation | 0 | 13 | 0,00 | 0 | 0,00 | 0 | 0 |
| | RPC | 0 | 13 | 0,00 | 0 | 0,00 | 0 | 0 |
| | YPT | 0 | 13 | 0,00 | 0 | 0,00 | 0 | 0 |
| | Other | 0 | 13 | 0,00 | 0 | 0,00 | 0 | 0 |
| Bulgaria | Self-evaluation | 21 | 0 | 4,10 | 4 | 0,94 | 1 | 5 |
| | RPC | 17 | 4 | 3,88 | 4 | 0,99 | 2 | 5 |
| | YPT | 21 | 0 | 4,00 | 4 | 1,05 | 2 | 5 |
| | Other | 21 | 0 | 4,48 | 5 | 0,75 | 3 | 5 |
| Czech Rep. | Self-evaluation | 20 | 3 | 3,90 | 4 | 0,85 | 2 | 5 |
| | RPC | 18 | 5 | 3,83 | 4 | 0,79 | 2 | 5 |
| | YPT | 23 | 0 | 4,00 | 4 | 0,80 | 3 | 5 |
| | Other | 18 | 5 | 4,17 | 4 | 0,86 | 2 | 5 |
| Germany | Self-evaluation | 0 | 3 | 0,00 | 0 | 0,00 | 0 | 0 |
| | RPC | 0 | 3 | 0,00 | 0 | 0,00 | 0 | 0 |
| | YPT | 0 | 3 | 0,00 | 0 | 0,00 | 0 | 0 |
| | Other | 0 | 3 | 0,00 | 0 | 0,00 | 0 | 0 |
| Hungary | Self-evaluation | 68 | 4 | 4,04 | 4 | 1,07 | 1 | 5 |
| | RPC | 41 | 31 | 4,51 | 5 | 0,87 | 2 | 5 |
| | YPT | 65 | 7 | 4,26 | 4 | 0,82 | 2 | 5 |

The European Commission's support for the production of this publication does not constitute an endorsement of the contents, which reflect the views only of the authors, and the Commission cannot be held responsible for any use which may be made of the information contained therein.

| | | | | | | | | |
|---|---|---|---|---|---|---|---|---|
| | Other | 67 | 5 | 4,43 | 5 | 0,70 | 2 | 5 |
| Iran | Self-evaluation | 0 | 1 | 0,00 | 0 | 0,00 | 0 | 0 |
| | RPC | 0 | 1 | 0,00 | 0 | 0,00 | 0 | 0 |
| | YPT | 0 | 1 | 0,00 | 0 | 0,00 | 0 | 0 |
| | Other | 0 | 1 | 0,00 | 0 | 0,00 | 0 | 0 |
| Russia | Self-evaluation | 0 | 1 | 0,00 | 0 | 0,00 | 0 | 0 |
| | RPC | 0 | 1 | 0,00 | 0 | 0,00 | 0 | 0 |
| | YPT | 0 | 1 | 0,00 | 0 | 0,00 | 0 | 0 |
| | Other | 0 | 1 | 0,00 | 0 | 0,00 | 0 | 0 |
| Slovakia | Self-evaluation | 161 | 4 | 4,06 | 4 | 0,86 | 1 | 5 |
| | RPC | 108 | 57 | 3,64 | 4 | 1,04 | 1 | 5 |
| | YPT | 148 | 17 | 3,93 | 4 | 0,98 | 1 | 5 |
| | Other | 160 | 5 | 3,96 | 4 | 0,93 | 1 | 5 |
| Slovenia | Self-evaluation | 9 | 0 | 4,33 | 4 | 0,50 | 4 | 5 |
| | RPC | 7 | 2 | 4,43 | 4 | 0,53 | 4 | 5 |
| | YPT | 9 | 0 | 4,00 | 4 | 1,32 | 1 | 5 |
| | Other | 9 | 0 | 4,22 | 5 | 0,97 | 3 | 5 |

**Solve close-ended problems**

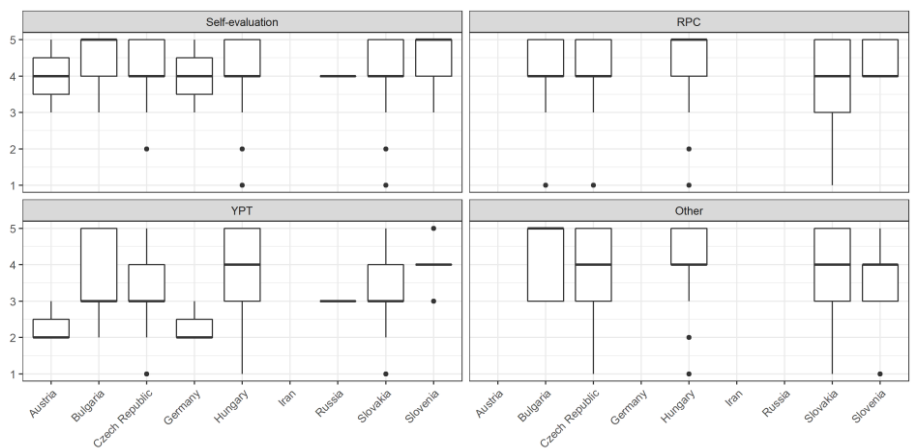

| Country | Type | Valid | Missing | Mean | Median | SD | Min. | Max. |
|---|---|---|---|---|---|---|---|---|
| Austria | Self-evaluation | 0 | 13 | 0,00 | 0 | 0,00 | 0 | 0 |
| | RPC | 3 | 10 | 2,33 | 2 | 0,58 | 2 | 3 |
| | YPT | 0 | 13 | 0,00 | 0 | 0,00 | 0 | 0 |
| | Other | 3 | 10 | 4,00 | 4 | 1,00 | 3 | 5 |
| Bulgaria | Self-evaluation | 21 | 0 | 4,05 | 4 | 0,97 | 1 | 5 |
| | RPC | 17 | 4 | 3,88 | 3 | 1,11 | 2 | 5 |
| | YPT | 20 | 1 | 4,20 | 5 | 0,95 | 3 | 5 |
| | Other | 21 | 0 | 4,38 | 5 | 0,86 | 3 | 5 |
| Czech Rep. | Self-evaluation | 20 | 3 | 4,10 | 4 | 0,97 | 1 | 5 |
| | RPC | 20 | 3 | 3,35 | 3 | 1,14 | 1 | 5 |
| | YPT | 21 | 2 | 3,81 | 4 | 1,03 | 1 | 5 |

| Country | Type | Valid | Missing | Mean | Median | SD | Min. | Max. |
|---|---|---|---|---|---|---|---|---|
| | Other | 18 | 5 | 4,06 | 4 | 0,87 | 2 | 5 |
| Germany | Self-evaluation | 0 | 3 | 0,00 | 0 | 0,00 | 0 | 0 |
| | RPC | 3 | 0 | 2,33 | 2 | 0,58 | 2 | 3 |
| | YPT | 0 | 3 | 0,00 | 0 | 0,00 | 0 | 0 |
| | Other | 3 | 0 | 4,00 | 4 | 1,00 | 3 | 5 |
| Hungary | Self-evaluation | 67 | 5 | 4,18 | 5 | 1,06 | 1 | 5 |
| | RPC | 41 | 31 | 3,85 | 4 | 1,04 | 1 | 5 |
| | YPT | 61 | 11 | 4,10 | 4 | 1,00 | 1 | 5 |
| | Other | 68 | 4 | 4,15 | 4 | 0,82 | 1 | 5 |
| Iran | Self-evaluation | 0 | 1 | 0,00 | 0 | 0,00 | 0 | 0 |
| | RPC | 0 | 1 | 0,00 | 0 | 0,00 | 0 | 0 |
| | YPT | 0 | 1 | 0,00 | 0 | 0,00 | 0 | 0 |
| | Other | 0 | 1 | 0,00 | 0 | 0,00 | 0 | 0 |
| Russia | Self-evaluation | 0 | 1 | 0,00 | 0 | 0,00 | 0 | 0 |
| | RPC | 1 | 0 | 3,00 | 3 | 0,00 | 3 | 3 |
| | YPT | 0 | 1 | 0,00 | 0 | 0,00 | 0 | 0 |
| | Other | 1 | 0 | 4,00 | 4 | 0,00 | 4 | 4 |
| Slovakia | Self-evaluation | 150 | 15 | 4,01 | 4 | 0,86 | 1 | 5 |
| | RPC | 100 | 65 | 3,51 | 3 | 0,89 | 1 | 5 |
| | YPT | 135 | 30 | 3,92 | 4 | 0,96 | 1 | 5 |
| | Other | 151 | 14 | 3,95 | 4 | 0,84 | 1 | 5 |
| Slovenia | Self-evaluation | 9 | 0 | 4,44 | 4 | 0,53 | 4 | 5 |
| | RPC | 8 | 1 | 4,00 | 4 | 0,53 | 3 | 5 |
| | YPT | 9 | 0 | 3,67 | 4 | 1,22 | 1 | 5 |
| | Other | 9 | 0 | 4,33 | 5 | 0,87 | 3 | 5 |

**Designing experiments**

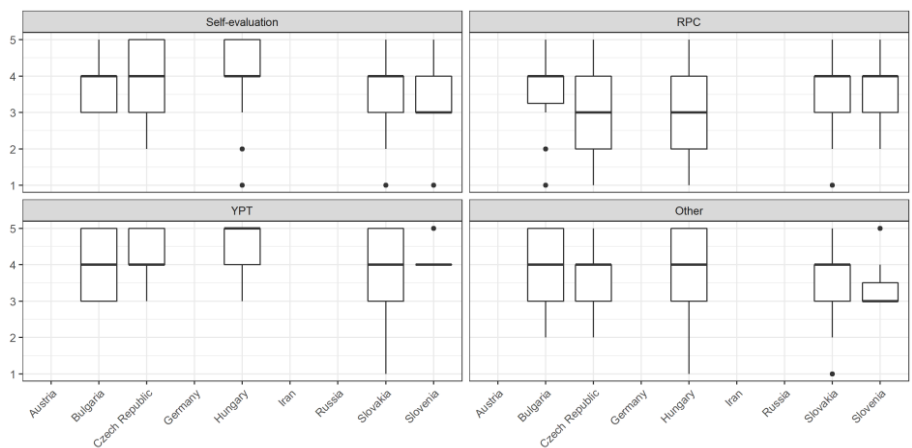

| Country | Type | Valid | Missing | Mean | Median | SD | Min. | Max. |
|---|---|---|---|---|---|---|---|---|
| Austria | Self-evaluation | 0 | 13 | 0,00 | 0 | 0,00 | 0 | 0 |
| | RPC | 0 | 13 | 0,00 | 0 | 0,00 | 0 | 0 |
| | YPT | 0 | 13 | 0,00 | 0 | 0,00 | 0 | 0 |
| | Other | 0 | 13 | 0,00 | 0 | 0,00 | 0 | 0 |
| Bulgaria | Self-evaluation | 18 | 3 | 3,78 | 4 | 1,06 | 1 | 5 |
| | RPC | 17 | 4 | 4,00 | 4 | 0,94 | 3 | 5 |
| | YPT | 19 | 2 | 3,89 | 4 | 0,99 | 2 | 5 |
| | Other | 19 | 2 | 3,89 | 4 | 0,74 | 3 | 5 |
| Czech Rep. | Self-evaluation | 22 | 1 | 2,91 | 3 | 1,48 | 1 | 5 |
| | RPC | 18 | 5 | 4,17 | 4 | 0,71 | 3 | 5 |
| | YPT | 21 | 2 | 3,67 | 4 | 1,02 | 2 | 5 |
| | Other | 19 | 4 | 3,84 | 4 | 1,07 | 2 | 5 |
| Germany | Self-evaluation | 0 | 3 | 0,00 | 0 | 0,00 | 0 | 0 |
| | RPC | 0 | 3 | 0,00 | 0 | 0,00 | 0 | 0 |
| | YPT | 0 | 3 | 0,00 | 0 | 0,00 | 0 | 0 |
| | Other | 0 | 3 | 0,00 | 0 | 0,00 | 0 | 0 |
| Hungary | Self-evaluation | 63 | 9 | 3,11 | 3 | 1,38 | 1 | 5 |
| | RPC | 43 | 29 | 4,60 | 5 | 0,69 | 3 | 5 |
| | YPT | 61 | 11 | 3,80 | 4 | 1,24 | 1 | 5 |
| | Other | 66 | 6 | 4,03 | 4 | 0,94 | 1 | 5 |
| Iran | Self-evaluation | 0 | 1 | 0,00 | 0 | 0,00 | 0 | 0 |
| | RPC | 0 | 1 | 0,00 | 0 | 0,00 | 0 | 0 |
| | YPT | 0 | 1 | 0,00 | 0 | 0,00 | 0 | 0 |
| | Other | 0 | 1 | 0,00 | 0 | 0,00 | 0 | 0 |
| Russia | Self-evaluation | 0 | 1 | 0,00 | 0 | 0,00 | 0 | 0 |
| | RPC | 0 | 1 | 0,00 | 0 | 0,00 | 0 | 0 |
| | YPT | 0 | 1 | 0,00 | 0 | 0,00 | 0 | 0 |
| | Other | 0 | 1 | 0,00 | 0 | 0,00 | 0 | 0 |
| Slovakia | Self-evaluation | 151 | 14 | 3,45 | 4 | 0,96 | 1 | 5 |
| | RPC | 105 | 60 | 3,72 | 4 | 1,08 | 1 | 5 |
| | YPT | 138 | 27 | 3,68 | 4 | 0,92 | 1 | 5 |
| | Other | 153 | 12 | 3,54 | 4 | 0,94 | 1 | 5 |
| Slovenia | Self-evaluation | 9 | 0 | 3,56 | 4 | 0,88 | 2 | 5 |
| | RPC | 9 | 0 | 4,22 | 4 | 0,44 | 4 | 5 |
| | YPT | 7 | 2 | 3,43 | 3 | 0,79 | 3 | 5 |
| | Other | 9 | 0 | 3,33 | 3 | 1,12 | 1 | 5 |

The European Commission's support for the production of this publication does not constitute an endorsement of the contents, which reflect the views only of the authors, and the Commission cannot be held responsible for any use which may be made of the information contained therein.

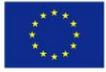

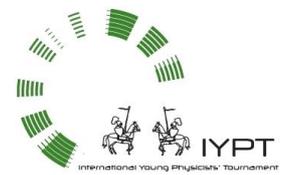

*DEVELOPMENT OF INQUIRY-BASED
LEARNING VIA IYPT*

**Conducting experiment**

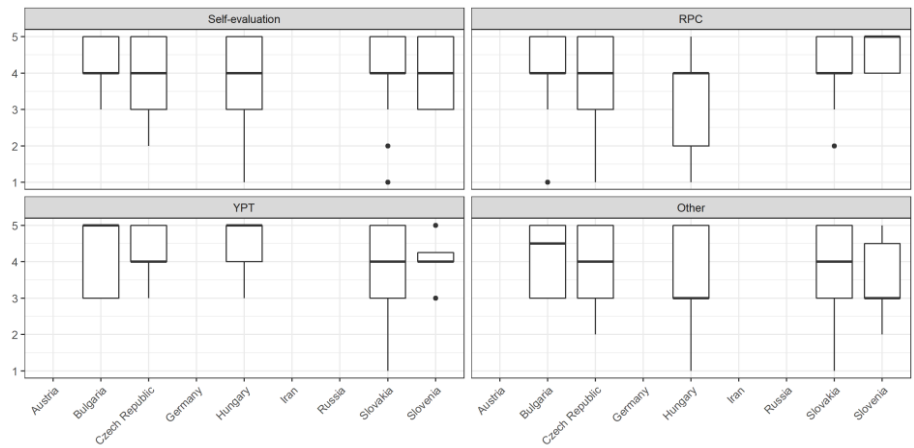

| Country | Type | Valid | Missing | Mean | Median | SD | Min. | Max. |
|---------|------|-------|---------|------|--------|-----|------|------|
| Austria | Self-evaluation | 0 | 13 | 0,00 | 0 | 0,00 | 0 | 0 |
| | RPC | 0 | 13 | 0,00 | 0 | 0,00 | 0 | 0 |
| | YPT | 0 | 13 | 0,00 | 0 | 0,00 | 0 | 0 |
| | Other | 0 | 13 | 0,00 | 0 | 0,00 | 0 | 0 |
| Bulgaria | Self-evaluation | 20 | 1 | 4,05 | 4 | 1,00 | 1 | 5 |
| | RPC | 18 | 3 | 4,17 | 5 | 0,99 | 3 | 5 |
| | YPT | 20 | 1 | 4,20 | 4,5 | 0,89 | 3 | 5 |
| | Other | 20 | 1 | 4,30 | 4 | 0,66 | 3 | 5 |
| Czech Rep. | Self-evaluation | 22 | 1 | 3,82 | 4 | 1,26 | 1 | 5 |
| | RPC | 18 | 5 | 4,22 | 4 | 0,65 | 3 | 5 |
| | YPT | 21 | 2 | 3,76 | 4 | 1,04 | 2 | 5 |
| | Other | 19 | 4 | 3,95 | 4 | 1,08 | 2 | 5 |
| Germany | Self-evaluation | 0 | 3 | 0,00 | 0 | 0,00 | 0 | 0 |
| | RPC | 0 | 3 | 0,00 | 0 | 0,00 | 0 | 0 |
| | YPT | 0 | 3 | 0,00 | 0 | 0,00 | 0 | 0 |
| | Other | 0 | 3 | 0,00 | 0 | 0,00 | 0 | 0 |
| Hungary | Self-evaluation | 63 | 9 | 3,30 | 4 | 1,36 | 1 | 5 |
| | RPC | 41 | 31 | 4,41 | 5 | 0,81 | 3 | 5 |
| | YPT | 59 | 13 | 3,59 | 3 | 1,10 | 1 | 5 |
| | Other | 65 | 7 | 4,12 | 4 | 0,93 | 1 | 5 |
| Iran | Self-evaluation | 0 | 1 | 0,00 | 0 | 0,00 | 0 | 0 |
| | RPC | 0 | 1 | 0,00 | 0 | 0,00 | 0 | 0 |
| | YPT | 0 | 1 | 0,00 | 0 | 0,00 | 0 | 0 |
| | Other | 0 | 1 | 0,00 | 0 | 0,00 | 0 | 0 |
| Russia | Self-evaluation | 0 | 1 | 0,00 | 0 | 0,00 | 0 | 0 |
| | RPC | 0 | 1 | 0,00 | 0 | 0,00 | 0 | 0 |
| | YPT | 0 | 1 | 0,00 | 0 | 0,00 | 0 | 0 |
| | Other | 0 | 1 | 0,00 | 0 | 0,00 | 0 | 0 |
| Slovakia | Self-evaluation | 148 | 17 | 4,16 | 4 | 0,80 | 2 | 5 |

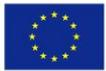

Co-funded by the
Erasmus+ Programme
of the European Union

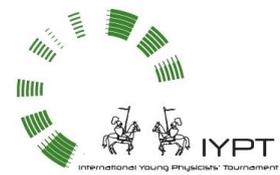

|          | RPC            | 99  | 66 | 3,83 | 4 | 1,05 | 1 | 5 |
|          | YPT            | 132 | 33 | 3,94 | 4 | 0,95 | 1 | 5 |
|          | Other          | 151 | 14 | 4,13 | 4 | 0,83 | 1 | 5 |
| Slovenia | Self-evaluation | 9  | 0  | 4,67 | 5 | 0,50 | 4 | 5 |
|          | RPC            | 8   | 1  | 4,13 | 4 | 0,64 | 3 | 5 |
|          | YPT            | 7   | 2  | 3,57 | 3 | 1,13 | 2 | 5 |
|          | Other          | 9   | 0  | 4,00 | 4 | 0,87 | 3 | 5 |

**Interpreting experimental data, data analysis**

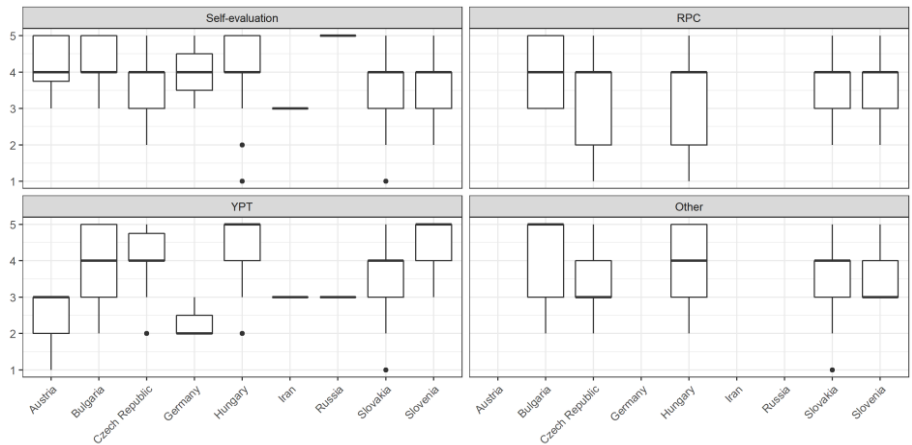

| Country    | Type            | Valid | Missing | Mean | Median | SD   | Min. | Max. |
|------------|-----------------|-------|---------|------|--------|------|------|------|
| Austria    | Self-evaluation | 0     | 13      | 0,00 | 0      | 0,00 | 0    | 0    |
|            | RPC             | 13    | 0       | 2,54 | 3      | 0,78 | 1    | 3    |
|            | YPT             | 0     | 13      | 0,00 | 0      | 0,00 | 0    | 0    |
|            | Other           | 12    | 1       | 4,17 | 4      | 0,83 | 3    | 5    |
| Bulgaria   | Self-evaluation | 20    | 1       | 4,00 | 4      | 0,79 | 3    | 5    |
|            | RPC             | 16    | 5       | 3,94 | 4      | 1,00 | 2    | 5    |
|            | YPT             | 19    | 2       | 4,16 | 5      | 1,01 | 2    | 5    |
|            | Other           | 20    | 1       | 4,10 | 4      | 0,72 | 3    | 5    |
| Czech Rep. | Self-evaluation | 21    | 2       | 3,10 | 4      | 1,41 | 1    | 5    |
|            | RPC             | 18    | 5       | 3,94 | 4      | 0,94 | 2    | 5    |
|            | YPT             | 22    | 1       | 3,45 | 3      | 1,01 | 2    | 5    |
|            | Other           | 19    | 4       | 3,79 | 4      | 0,85 | 2    | 5    |
| Germany    | Self-evaluation | 0     | 3       | 0,00 | 0      | 0,00 | 0    | 0    |
|            | RPC             | 3     | 0       | 2,33 | 2      | 0,58 | 2    | 3    |
|            | YPT             | 0     | 3       | 0,00 | 0      | 0,00 | 0    | 0    |
|            | Other           | 3     | 0       | 4,00 | 4      | 1,00 | 3    | 5    |
| Hungary    | Self-evaluation | 64    | 8       | 3,30 | 4      | 1,36 | 1    | 5    |
|            | RPC             | 43    | 29      | 4,44 | 5      | 0,85 | 2    | 5    |
|            | YPT             | 58    | 14      | 3,90 | 4      | 0,97 | 2    | 5    |
|            | Other           | 67    | 5       | 4,24 | 4      | 0,82 | 1    | 5    |
| Iran       | Self-evaluation | 0     | 1       | 0,00 | 0      | 0,00 | 0    | 0    |
|  | RPC | 1 | 0 | 3,00 | 3 | 0,00 | 3 | 3 |
|  | YPT | 0 | 1 | 0,00 | 0 | 0,00 | 0 | 0 |
|  | Other | 1 | 0 | 3,00 | 3 | 0,00 | 3 | 3 |
| Russia | Self-evaluation | 0 | 1 | 0,00 | 0 | 0,00 | 0 | 0 |
|  | RPC | 1 | 0 | 3,00 | 3 | 0,00 | 3 | 3 |
|  | YPT | 0 | 1 | 0,00 | 0 | 0,00 | 0 | 0 |
|  | Other | 1 | 0 | 5,00 | 5 | 0,00 | 5 | 5 |
| Slovakia | Self-evaluation | 145 | 20 | 3,82 | 4 | 0,87 | 2 | 5 |
|  | RPC | 100 | 65 | 3,67 | 4 | 0,99 | 1 | 5 |
|  | YPT | 138 | 27 | 3,80 | 4 | 0,92 | 1 | 5 |
|  | Other | 146 | 19 | 3,82 | 4 | 0,86 | 1 | 5 |
| Slovenia | Self-evaluation | 9 | 0 | 3,78 | 4 | 0,97 | 2 | 5 |
|  | RPC | 9 | 0 | 4,44 | 5 | 0,73 | 3 | 5 |
|  | YPT | 8 | 1 | 3,50 | 3 | 0,76 | 3 | 5 |
|  | Other | 9 | 0 | 3,56 | 4 | 1,13 | 2 | 5 |

## Developing own theoretical model

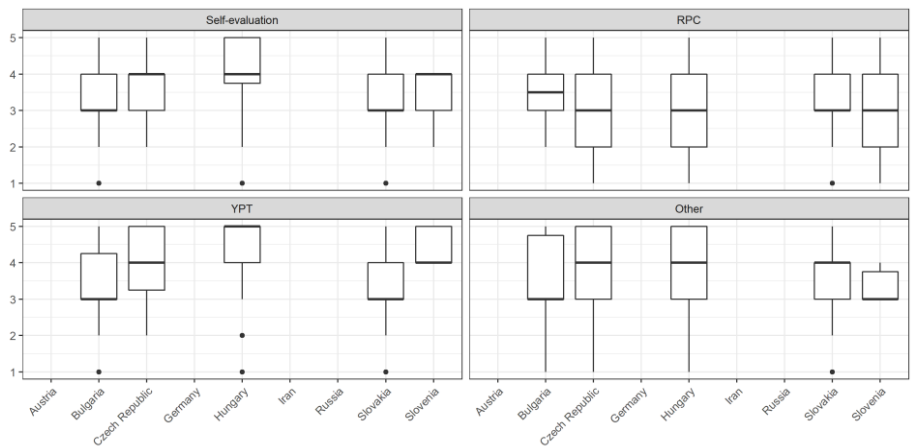

| Country | Type | Valid | Missing | Mean | Median | SD | Min. | Max. |
|---|---|---|---|---|---|---|---|---|
| Austria | Self-evaluation | 0 | 13 | 0,00 | 0 | 0,00 | 0 | 0 |
|  | RPC | 0 | 13 | 0,00 | 0 | 0,00 | 0 | 0 |
|  | YPT | 0 | 13 | 0,00 | 0 | 0,00 | 0 | 0 |
|  | Other | 0 | 13 | 0,00 | 0 | 0,00 | 0 | 0 |
| Bulgaria | Self-evaluation | 18 | 3 | 3,50 | 3,5 | 0,99 | 2 | 5 |
|  | RPC | 16 | 5 | 3,38 | 3 | 1,20 | 1 | 5 |
|  | YPT | 18 | 3 | 3,50 | 3 | 1,20 | 1 | 5 |
|  | Other | 19 | 2 | 3,37 | 3 | 1,07 | 1 | 5 |
| Czech Rep. | Self-evaluation | 21 | 2 | 2,90 | 3 | 1,18 | 1 | 5 |
|  | RPC | 18 | 5 | 4,06 | 4 | 1,06 | 2 | 5 |
|  | YPT | 21 | 2 | 3,67 | 4 | 1,24 | 1 | 5 |
|  | Other | 19 | 4 | 3,68 | 4 | 0,95 | 2 | 5 |
| Germany | Self-evaluation | 0 | 3 | 0,00 | 0 | 0,00 | 0 | 0 |

The European Commission's support for the production of this publication does not constitute an endorsement of the contents, which reflect the views only of the authors, and the Commission cannot be held responsible for any use which may be made of the information contained therein.

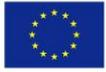
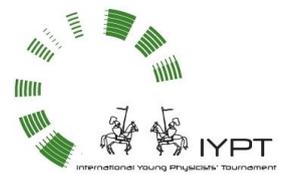

| | | Valid | Missing | Mean | Median | SD | Min. | Max. |
|---|---|---|---|---|---|---|---|---|
| | RPC | 0 | 3 | 0,00 | 0 | 0,00 | 0 | 0 |
| | YPT | 0 | 3 | 0,00 | 0 | 0,00 | 0 | 0 |
| | Other | 0 | 3 | 0,00 | 0 | 0,00 | 0 | 0 |
| Hungary | Self-evaluation | 63 | 9 | 3,16 | 3 | 1,26 | 1 | 5 |
| | RPC | 42 | 30 | 4,43 | 5 | 0,99 | 1 | 5 |
| | YPT | 57 | 15 | 3,89 | 4 | 1,03 | 1 | 5 |
| | Other | 64 | 8 | 4,03 | 4 | 1,01 | 1 | 5 |
| Iran | Self-evaluation | 0 | 1 | 0,00 | 0 | 0,00 | 0 | 0 |
| | RPC | 0 | 1 | 0,00 | 0 | 0,00 | 0 | 0 |
| | YPT | 0 | 1 | 0,00 | 0 | 0,00 | 0 | 0 |
| | Other | 0 | 1 | 0,00 | 0 | 0,00 | 0 | 0 |
| Russia | Self-evaluation | 0 | 1 | 0,00 | 0 | 0,00 | 0 | 0 |
| | RPC | 0 | 1 | 0,00 | 0 | 0,00 | 0 | 0 |
| | YPT | 0 | 1 | 0,00 | 0 | 0,00 | 0 | 0 |
| | Other | 0 | 1 | 0,00 | 0 | 0,00 | 0 | 0 |
| Slovakia | Self-evaluation | 141 | 24 | 3,30 | 3 | 1,00 | 1 | 5 |
| | RPC | 101 | 64 | 3,45 | 3 | 1,06 | 1 | 5 |
| | YPT | 132 | 33 | 3,61 | 4 | 0,98 | 1 | 5 |
| | Other | 145 | 20 | 3,33 | 3 | 1,05 | 1 | 5 |
| Slovenia | Self-evaluation | 9 | 0 | 2,89 | 3 | 1,36 | 1 | 5 |
| | RPC | 9 | 0 | 4,33 | 4 | 0,50 | 4 | 5 |
| | YPT | 6 | 3 | 3,33 | 3 | 0,52 | 3 | 4 |
| | Other | 9 | 0 | 3,33 | 4 | 0,87 | 2 | 4 |

**Numerical simulations**

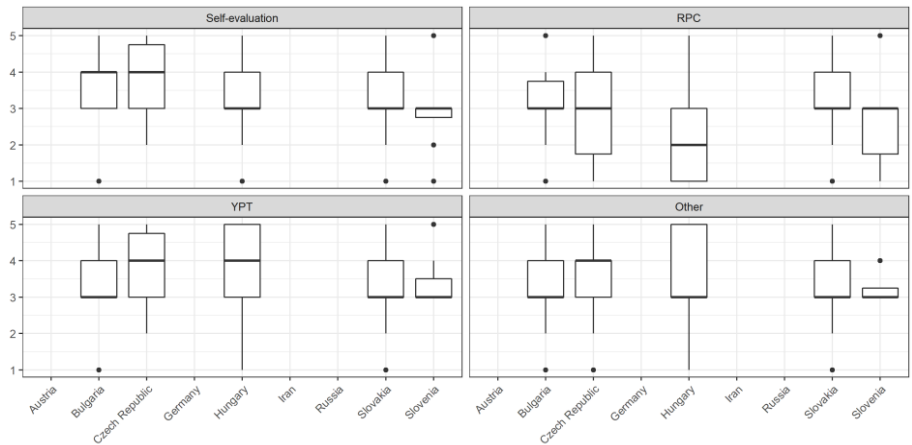

| Country | Type | Valid | Missing | Mean | Median | SD | Min. | Max. |
|---|---|---|---|---|---|---|---|---|
| Austria | Self-evaluation | 0 | 13 | 0,00 | 0 | 0,00 | 0 | 0 |
| | RPC | 0 | 13 | 0,00 | 0 | 0,00 | 0 | 0 |
| | YPT | 0 | 13 | 0,00 | 0 | 0,00 | 0 | 0 |
| | Other | 0 | 13 | 0,00 | 0 | 0,00 | 0 | 0 |
| Bulgaria | Self-evaluation | 18 | 3 | 3,11 | 3 | 1,18 | 1 | 5 |

The European Commission's support for the production of this publication does not constitute an endorsement of the contents, which reflect the views only of the authors, and the Commission cannot be held responsible for any use which may be made of the information contained therein.

|  |  |  |  |  |  |  |  |  |
|---|---|---|---|---|---|---|---|---|
|  | RPC | 16 | 5 | 3,31 | 3 | 1,20 | 1 | 5 |
|  | YPT | 17 | 4 | 3,29 | 3 | 1,26 | 1 | 5 |
|  | Other | 19 | 2 | 3,53 | 4 | 1,17 | 1 | 5 |
| Czech Rep. | Self-evaluation | 20 | 3 | 2,90 | 3 | 1,41 | 1 | 5 |
|  | RPC | 18 | 5 | 3,83 | 4 | 0,99 | 2 | 5 |
|  | YPT | 20 | 3 | 3,45 | 4 | 1,19 | 1 | 5 |
|  | Other | 18 | 5 | 3,72 | 4 | 1,07 | 2 | 5 |
| Germany | Self-evaluation | 0 | 3 | 0,00 | 0 | 0,00 | 0 | 0 |
|  | RPC | 0 | 3 | 0,00 | 0 | 0,00 | 0 | 0 |
|  | YPT | 0 | 3 | 0,00 | 0 | 0,00 | 0 | 0 |
|  | Other | 0 | 3 | 0,00 | 0 | 0,00 | 0 | 0 |
| Hungary | Self-evaluation | 62 | 10 | 2,16 | 2 | 1,09 | 1 | 5 |
|  | RPC | 41 | 31 | 3,83 | 4 | 1,28 | 1 | 5 |
|  | YPT | 57 | 15 | 3,44 | 3 | 1,27 | 1 | 5 |
|  | Other | 62 | 10 | 3,37 | 3 | 1,24 | 1 | 5 |
| Iran | Self-evaluation | 0 | 1 | 0,00 | 0 | 0,00 | 0 | 0 |
|  | RPC | 0 | 1 | 0,00 | 0 | 0,00 | 0 | 0 |
|  | YPT | 0 | 1 | 0,00 | 0 | 0,00 | 0 | 0 |
|  | Other | 0 | 1 | 0,00 | 0 | 0,00 | 0 | 0 |
| Russia | Self-evaluation | 0 | 1 | 0,00 | 0 | 0,00 | 0 | 0 |
|  | RPC | 0 | 1 | 0,00 | 0 | 0,00 | 0 | 0 |
|  | YPT | 0 | 1 | 0,00 | 0 | 0,00 | 0 | 0 |
|  | Other | 0 | 1 | 0,00 | 0 | 0,00 | 0 | 0 |
| Slovakia | Self-evaluation | 131 | 34 | 3,22 | 3 | 1,08 | 1 | 5 |
|  | RPC | 99 | 66 | 3,21 | 3 | 1,03 | 1 | 5 |
|  | YPT | 127 | 38 | 3,49 | 3 | 1,08 | 1 | 5 |
|  | Other | 137 | 28 | 3,23 | 3 | 1,10 | 1 | 5 |
| Slovenia | Self-evaluation | 8 | 1 | 2,63 | 3 | 1,30 | 1 | 5 |
|  | RPC | 7 | 2 | 3,43 | 3 | 0,79 | 3 | 5 |
|  | YPT | 4 | 5 | 3,25 | 3 | 0,50 | 3 | 4 |
|  | Other | 8 | 1 | 2,88 | 3 | 1,13 | 1 | 5 |

**Independent research in scientific literature**

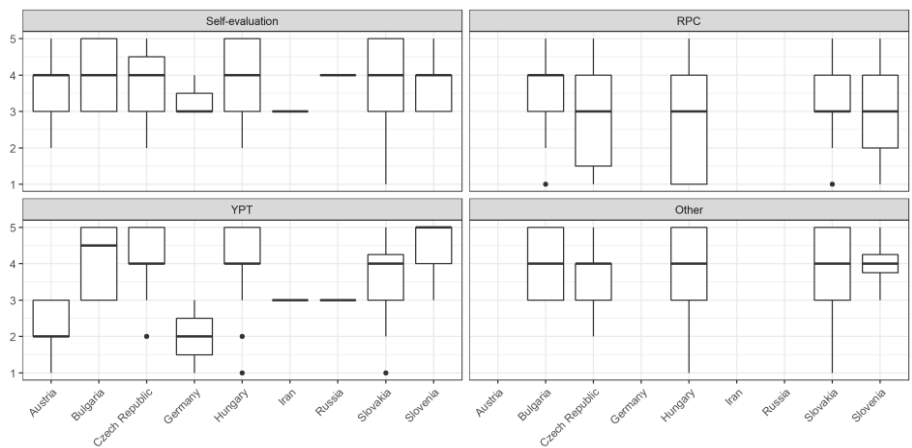

| Country | Type | Valid | Missing | Mean | Median | SD | Min. | Max. |
|---|---|---|---|---|---|---|---|---|
| Austria | Self-evaluation | 0 | 13 | 0,00 | 0 | 0,00 | 0 | 0 |
| | RPC | 13 | 0 | 2,23 | 2 | 0,60 | 1 | 3 |
| | YPT | 0 | 13 | 0,00 | 0 | 0,00 | 0 | 0 |
| | Other | 13 | 0 | 3,77 | 4 | 0,83 | 2 | 5 |
| Bulgaria | Self-evaluation | 17 | 4 | 3,59 | 4 | 1,06 | 1 | 5 |
| | RPC | 16 | 5 | 4,13 | 4,5 | 0,96 | 3 | 5 |
| | YPT | 18 | 3 | 4,11 | 4 | 0,90 | 3 | 5 |
| | Other | 20 | 1 | 3,95 | 4 | 0,89 | 3 | 5 |
| Czech Rep. | Self-evaluation | 23 | 0 | 2,91 | 3 | 1,41 | 1 | 5 |
| | RPC | 18 | 5 | 4,11 | 4 | 0,90 | 2 | 5 |
| | YPT | 22 | 1 | 3,68 | 4 | 0,89 | 2 | 5 |
| | Other | 19 | 4 | 3,79 | 4 | 0,92 | 2 | 5 |
| Germany | Self-evaluation | 0 | 3 | 0,00 | 0 | 0,00 | 0 | 0 |
| | RPC | 3 | 0 | 2,00 | 2 | 1,00 | 1 | 3 |
| | YPT | 0 | 3 | 0,00 | 0 | 0,00 | 0 | 0 |
| | Other | 3 | 0 | 3,33 | 3 | 0,58 | 3 | 4 |
| Hungary | Self-evaluation | 64 | 8 | 2,80 | 3 | 1,37 | 1 | 5 |
| | RPC | 44 | 28 | 4,07 | 4 | 0,97 | 1 | 5 |
| | YPT | 64 | 8 | 3,89 | 4 | 1,11 | 1 | 5 |
| | Other | 67 | 5 | 3,93 | 4 | 0,96 | 2 | 5 |
| Iran | Self-evaluation | 0 | 1 | 0,00 | 0 | 0,00 | 0 | 0 |
| | RPC | 1 | 0 | 3,00 | 3 | 0,00 | 3 | 3 |
| | YPT | 0 | 1 | 0,00 | 0 | 0,00 | 0 | 0 |
| | Other | 1 | 0 | 3,00 | 3 | 0,00 | 3 | 3 |
| Russia | Self-evaluation | 0 | 1 | 0,00 | 0 | 0,00 | 0 | 0 |
| | RPC | 1 | 0 | 3,00 | 3 | 0,00 | 3 | 3 |
| | YPT | 0 | 1 | 0,00 | 0 | 0,00 | 0 | 0 |
| | Other | 1 | 0 | 4,00 | 4 | 0,00 | 4 | 4 |
| Slovakia | Self-evaluation | 144 | 21 | 3,51 | 3 | 0,99 | 1 | 5 |
| | RPC | 100 | 65 | 3,80 | 4 | 0,96 | 1 | 5 |
| | YPT | 142 | 23 | 3,99 | 4 | 0,93 | 1 | 5 |
| | Other | 153 | 12 | 3,87 | 4 | 0,94 | 1 | 5 |
| Slovenia | Self-evaluation | 9 | 0 | 3,11 | 3 | 1,54 | 1 | 5 |
| | RPC | 9 | 0 | 4,44 | 5 | 0,73 | 3 | 5 |
| | YPT | 8 | 1 | 4,00 | 4 | 0,76 | 3 | 5 |
| | Other | 9 | 0 | 3,89 | 4 | 0,78 | 3 | 5 |

**Critical assessment of others' results**

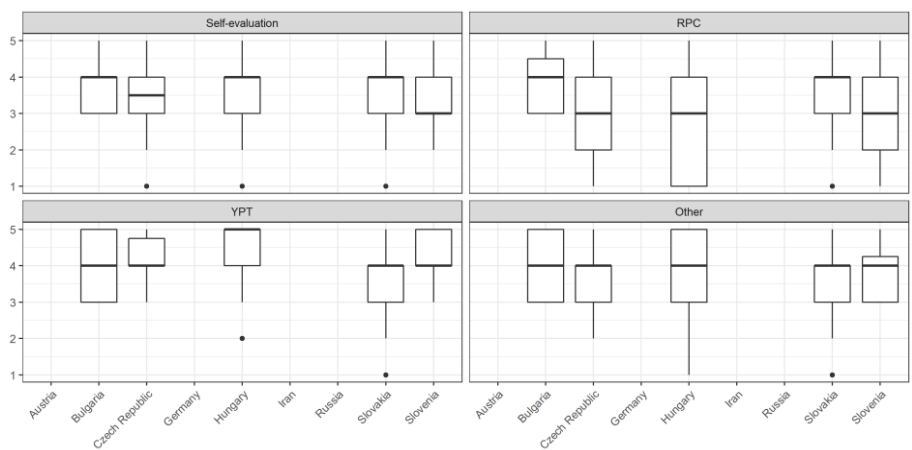

| Country | Type | Valid | Missing | Mean | Median | SD | Min. | Max. |
|---|---|---|---|---|---|---|---|---|
| Austria | Self-evaluation | 0 | 13 | 0,00 | 0 | 0,00 | 0 | 0 |
| | RPC | 0 | 13 | 0,00 | 0 | 0,00 | 0 | 0 |
| | YPT | 0 | 13 | 0,00 | 0 | 0,00 | 0 | 0 |
| | Other | 0 | 13 | 0,00 | 0 | 0,00 | 0 | 0 |
| Bulgaria | Self-evaluation | 19 | 2 | 3,79 | 4 | 0,85 | 3 | 5 |
| | RPC | 16 | 5 | 4,00 | 4 | 0,97 | 3 | 5 |
| | YPT | 18 | 3 | 4,00 | 4 | 0,84 | 3 | 5 |
| | Other | 20 | 1 | 3,80 | 4 | 0,62 | 3 | 5 |
| Czech Rep. | Self-evaluation | 23 | 0 | 3,00 | 3 | 1,38 | 1 | 5 |
| | RPC | 18 | 5 | 4,11 | 4 | 0,68 | 3 | 5 |
| | YPT | 21 | 2 | 3,67 | 4 | 1,02 | 2 | 5 |
| | Other | 20 | 3 | 3,50 | 3,5 | 1,10 | 1 | 5 |
| Germany | Self-evaluation | 0 | 3 | 0,00 | 0 | 0,00 | 0 | 0 |
| | RPC | 0 | 3 | 0,00 | 0 | 0,00 | 0 | 0 |
| | YPT | 0 | 3 | 0,00 | 0 | 0,00 | 0 | 0 |
| | Other | 0 | 3 | 0,00 | 0 | 0,00 | 0 | 0 |
| Hungary | Self-evaluation | 62 | 10 | 2,61 | 3 | 1,42 | 1 | 5 |
| | RPC | 41 | 31 | 4,20 | 5 | 1,03 | 2 | 5 |
| | YPT | 60 | 12 | 3,45 | 4 | 1,32 | 1 | 5 |
| | Other | 65 | 7 | 3,71 | 4 | 1,03 | 1 | 5 |
| Iran | Self-evaluation | 0 | 1 | 0,00 | 0 | 0,00 | 0 | 0 |
| | RPC | 0 | 1 | 0,00 | 0 | 0,00 | 0 | 0 |
| | YPT | 0 | 1 | 0,00 | 0 | 0,00 | 0 | 0 |
| | Other | 0 | 1 | 0,00 | 0 | 0,00 | 0 | 0 |
| Russia | Self-evaluation | 0 | 1 | 0,00 | 0 | 0,00 | 0 | 0 |
| | RPC | 0 | 1 | 0,00 | 0 | 0,00 | 0 | 0 |
| | YPT | 0 | 1 | 0,00 | 0 | 0,00 | 0 | 0 |
| | Other | 0 | 1 | 0,00 | 0 | 0,00 | 0 | 0 |
| Slovakia | Self-evaluation | 145 | 20 | 3,68 | 4 | 0,90 | 1 | 5 |
| | RPC | 98 | 67 | 3,60 | 4 | 1,01 | 1 | 5 |
| | YPT | 133 | 32 | 3,77 | 4 | 0,92 | 1 | 5 |
| | Other | 151 | 14 | 3,76 | 4 | 0,90 | 1 | 5 |
| Slovenia | Self-evaluation | 9 | 0 | 3,11 | 3 | 1,27 | 1 | 5 |
| | RPC | 9 | 0 | 4,22 | 4 | 0,67 | 3 | 5 |
| | YPT | 8 | 1 | 3,88 | 4 | 0,83 | 3 | 5 |
| | Other | 9 | 0 | 3,44 | 3 | 0,88 | 2 | 5 |

## 2. Supplement: Teachers' Assessment of Hard-Skill Development

### 2.1 Data characteristics

In this project, we have conducted a survey among 11 Slovakian, 9 Bulgarian, 6 Hungarian, 4 Czech and 3 Slovenian teachers, who are involved in preparing high school students for IYPT or any local organized YPT competitions. We have mapped they observed or assumed effect on soft (e.g. teamwork, creativity) and physical hard skills (e.g. high school physics, data analysis) in different teaching forms (RCP, YPT and Non-YPT competitions). Given the COVID situation, teachers carried out their preparatory work in 2020/2021 mainly online. This is why it is important to mention that most colleagues have been involved in preparing for YPT-type competitions for several years. Teachers had to fill in a questionnaire and answer 16x3 quantitative and 15 qualitative questions about the impact and characteristics of RCP, YPT and Non-YPT competitions.

**Descriptives - Num other competititons**

| Country | Mean | SD | N |
|---------|-------|-------|----|
| BG | 1.889 | 1.537 | 9 |
| CZ | 2.750 | 1.708 | 4 |
| HU | 3.167 | 1.602 | 6 |
| SK | 2.091 | 2.700 | 11 |
| SLO | 3.000 | 2.000 | 3 |

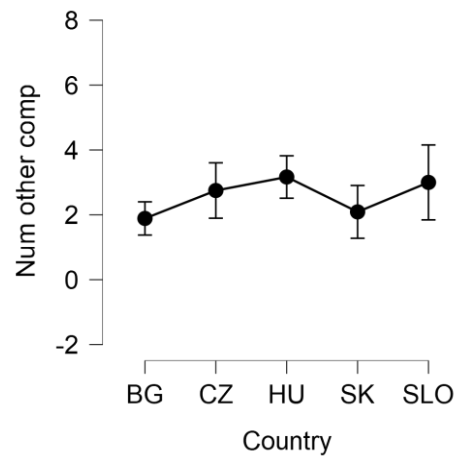

Data on teachers' answers form the questionnaire were provided in Excel format. For carrying out the empirical analysis, the software JASP[9] was used. First, descriptive analyses on skills as well as a correlation matrix using all variables. Secondly, for testing the hypotheses paired and independent t-tests (and Wilcoxon or Mann-Whitney-tests where needed) were computed. If Wilcoxon or Mann-Whitney-test was used, it is always the relevant result, t-tests are in these cases not relevant.

### 2.2 Results of Teachers Survey

The study investigates hard skills in the context of RPC, YPT, and non-YPT. Based on the teachers' evaluation, we can determine which effect their colleagues see in different educational settings. In addition, we can explore relationships between effects that provide indirect insights into teachers' work.

---

## 2.2.1 Regular physics classroom: RPC results

**Descriptive Statistics: Results of Hard Skills in RPC by teachers**

| | RPC [High school mathematics] | RPC [High school physics] | RPC [Solving close-ended problems] | RPC [Designing experiments] | RPC [Cookbook experiments] | RPC [Interp. exp. data, data analysis] | RPC [Dev. own theor. model] | RPC [Numerical simulations] | RPC [Indep. research in scientific litr.] | RPC [Crit. asses. of other's res.] |
|---|---|---|---|---|---|---|---|---|---|---|
| Valid | 32 | 33 | 33 | 33 | 33 | 33 | 33 | 33 | 33 | 33 |
| Mean | 6.719 | 8.242 | 7.758 | 4.758 | 6.121 | 6.212 | 3.364 | 2.909 | 3.606 | 4.212 |
| Std. Deviation | 2.466 | 1.937 | 1.969 | 2.513 | 2.274 | 2.315 | 2.560 | 2.638 | 2.423 | 2.408 |

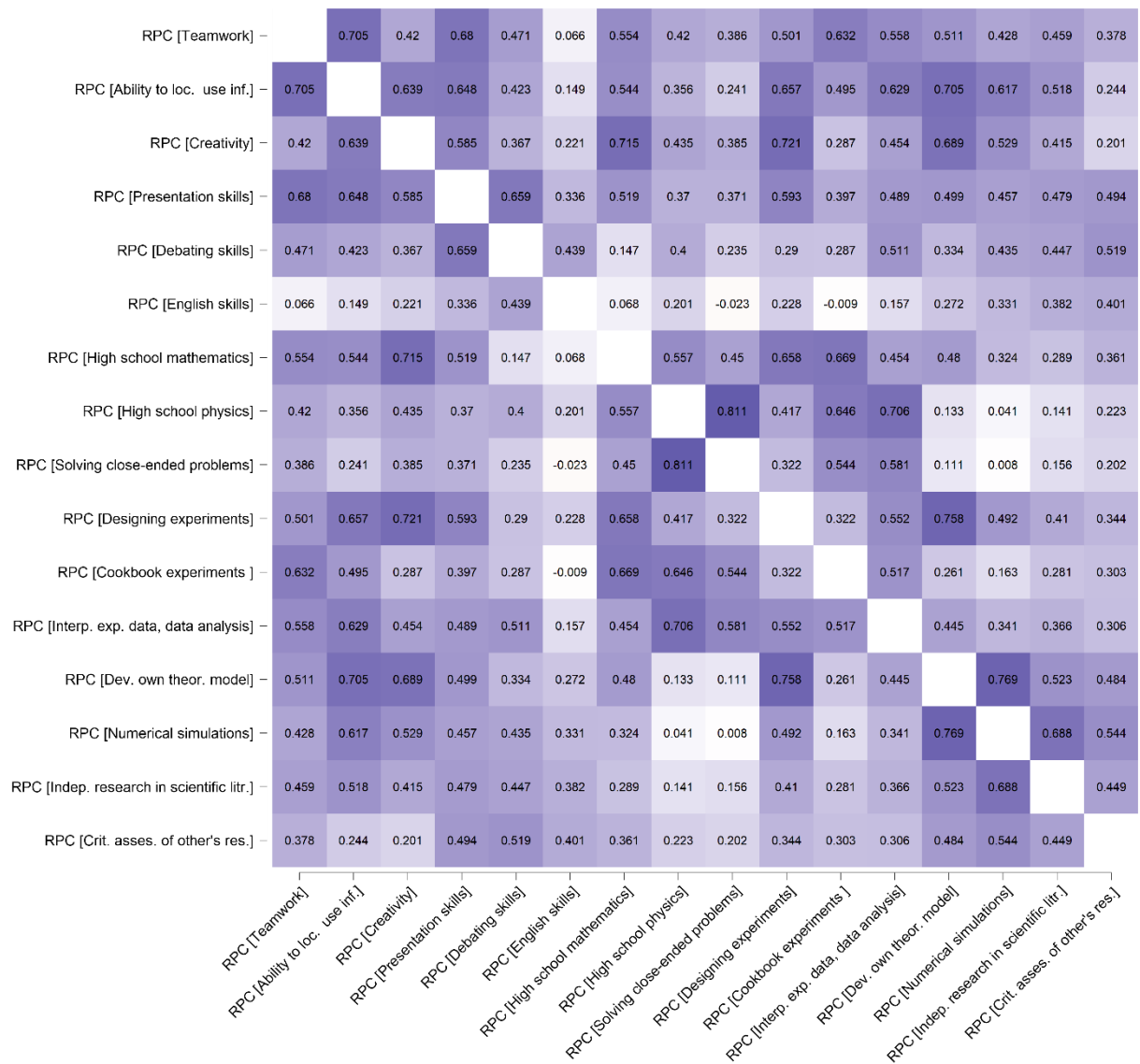

**Note: Correlations between Soft and Hard Skills in RPC (Pearsons' r, * p<.05, ** p<.01,  *** p<.001)**

As it can be seen, the strongest correlations in RPC are

- *Solving close ended problems* and *High School physics* r = 0.811***

- *Numerical simulations* and *Developing own theoretical models* r = 0.769***

- *Developing own theoretical models* and *Designing experiments* r = 0.758***

- *Designing experiments* and *Creativity* r = 0.721***

- *High school mathematics* and *Creativity* r = 0,715***.

The European Commission's support for the production of this publication does not constitute an endorsement of the contents, which reflect the views only of the authors, and the Commission cannot be held responsible for any use which may be made of the information contained therein.

## 2.2.2. YPT results

**Descriptive Statistics: Results of Hard Skills in YPT by teachers**

| | YPT [High school mathematics] | YPT [High school physics] | YPT [Solving close-ended problems] | YPT [Designing experiments] | YPT [Cookbook experiments] | YPT [Interp. exp. data, data analysis] | YPT [Dev. own theor. model] | YPT [Numerical simulations] | YPT [Indep. research in scientific litr.] | YPT [Crit. asses. of other's res.] |
|---|---|---|---|---|---|---|---|---|---|---|
| Valid | 33 | 33 | 33 | 33 | 33 | 33 | 33 | 33 | 33 | 33 |
| Mean | 7.455 | 8.121 | 4.788 | 8.485 | 5.939 | 8.697 | 7.364 | 7.182 | 7.818 | 8.455 |
| Std. Deviation | 2.320 | 2.147 | 3.029 | 1.734 | 3.082 | 1.723 | 2.485 | 2.493 | 2.455 | 1.889 |

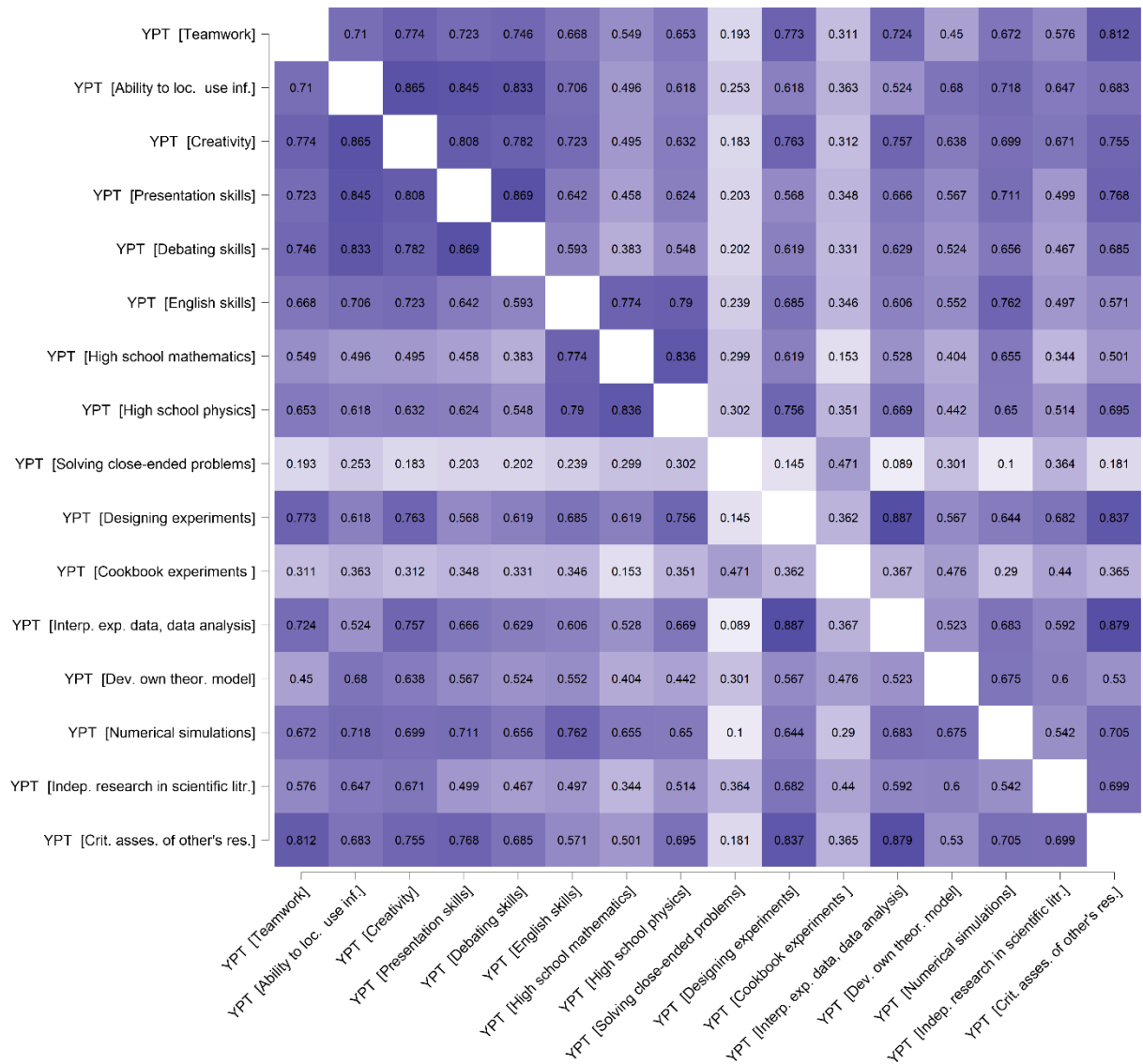

**Note: Correlations between Soft and Hard Skills in YPT (Pearsons' r, * p<.05, ** p<.01, *** p<.001)**

As it can be seen, the strongest correlations in YPT are

The European Commission's support for the production of this publication does not constitute an endorsement of the contents, which reflect the views only of the authors, and the Commission cannot be held responsible for any use which may be made of the information contained therein.

- *Interpreting experimental data, data analysis* and *Designing experiments*  r = 0.887***

- *Interpreting experimental data, data analysis* and *Critical assessment of others' results* r = 0.879***

- *Designing experiments* and *Critical assessment of others' results* r = 0.837***

- *High school mathematics* and *High school physics*  r = 0,836***.

### 2.2.3 Non-YPT results

**Descriptive Statistics: Results of Hard Skills in Non-YPT by teachers**

| | Non YPT [High school mathematics] | Non YPT [High school physics] | Non YPT [Solving close-ended problems] | Non YPT [Designing experiments] | Non YPT [Cookbook experiments] | Non YPT [Interp. exp. data, data analysis] | Non YPT [Dev. own theor. model] | Non YPT [Numerical simulations] | Non YPT [Indep. research in scientific litr.] | Non YPT [Crit. asses. of other's res.] |
|---|---|---|---|---|---|---|---|---|---|---|
| Valid | 29 | 28 | 29 | 29 | 29 | 28 | 29 | 29 | 28 | 29 |
| Mean | 7.414 | 8.107 | 7.862 | 4.310 | 4.862 | 5.286 | 3.793 | 3.241 | 5.214 | 3.207 |
| Std. Deviation | 2.719 | 2.409 | 2.722 | 3.037 | 3.148 | 3.253 | 3.245 | 2.923 | 3.665 | 2.969 |

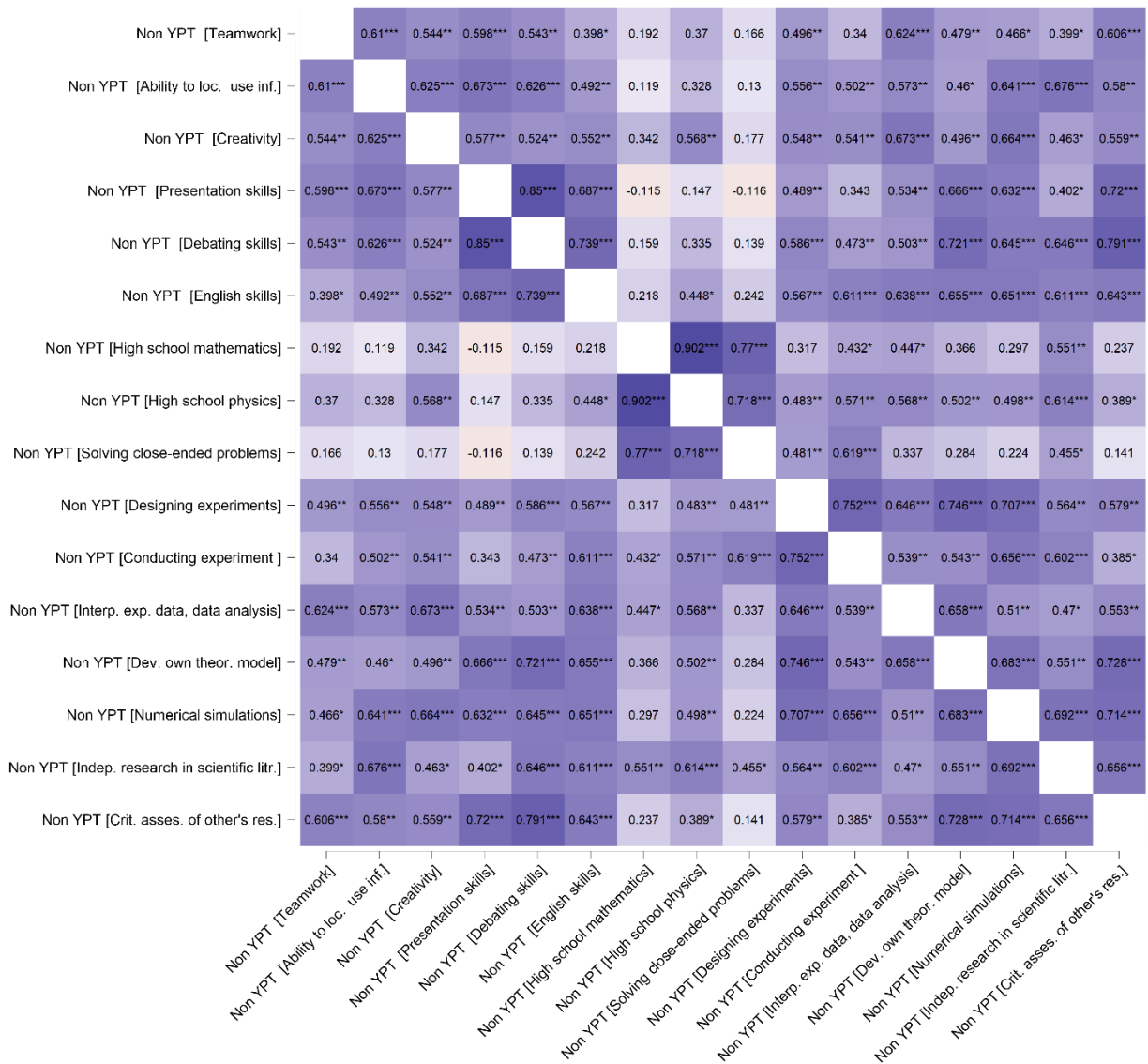

**Note: Correlations between Soft and Hard Skills in Non-YPT (Pearsons' r, * p<.05, ** p<.01,  *** p<.001)**

The European Commission's support for the production of this publication does not constitute an endorsement of the contents, which reflect the views only of the authors, and the Commission cannot be held responsible for any use which may be made of the information contained therein.

As it can be seen, (some of) the strongest correlations in Non-YPT are

- *High school mathematics* and *High School physics* r = 0.902***

- *Debating skills* and *Critical assessment of others' results* r = 0.791***

- *Solving close ended problems* and *High School mathematics* r = 0.77***

- *Designing experiments* and *Conducting (cookbook) experiments* r = 0.752***

- *Designing experiments* and *Developing own theoretical models* r = 0.746***

- *Solving close ended problems* and *High School physics* r = 0.718***

## 2.2.4 Teachers Paired T-Test on Hard Skills (RPC vs. YPT)

The most important step in answering the research questions is to compare the evaluations given for different platforms. After examining the distributions, we compare the values obtained first for RCP and YPT, then for YPT and non-YPT competitions using an appropriate procedure (Student's t-test or Mann-Whitney test), and we also present the results using graphs.

### Test of Normality (Shapiro-Wilk): Hard Skills in RPC vs. YPT

|  |  | W | p |
|---|---|---|---|
| RPC [High school mathematics] | - YPT [High school mathematics] | 0.912 | 0.013 |
| RPC [Solving close-ended problems] | - YPT [Solving close-ended problems] | 0.967 | 0.394 |
| RPC [Cookbook experiments ] | - YPT [Cookbook experiments ] | 0.978 | 0.725 |
| RPC [Dev. own theor. model] | - YPT [Dev. own theor. model] | 0.958 | 0.227 |
| RPC [Indep. research in scientific litr.] | - YPT [Indep. research in scientific litr.] | 0.957 | 0.214 |
| RPC [High school physics] | - YPT [High school physics] | 0.891 | 0.003 |
| RPC [Designing experiments] | - YPT [Designing experiments] | 0.974 | 0.598 |
| RPC [Interp. exp. data, data analysis] | - YPT [Interp. exp. data, data analysis] | 0.963 | 0.312 |
| RPC [Numerical simulations] | - YPT [Numerical simulations] | 0.970 | 0.481 |
| RPC [Crit. asses. of other's res.] | - YPT [Crit. asses. of other's res.] | 0.965 | 0.361 |

Note. Significant results suggest a deviation from normality.

### Paired Samples T-Test : Hard Skills RPC vs. YPT

| RPC | YPT | Test | Statistic | df | p |
|---|---|---|---|---|---|
| **RPC [High school mathematics]** | - YPT [High school mathematics] | Wilcoxon | 93.500 | | 0.037 |
| **RPC [Solving close-ended problems]** | - YPT [Solving close-ended problems] | Student | 5.010 | 32 | < .001 |
| RPC [Cookbook experiments ] | - YPT [Cookbook experiments ] | Student | 0.291 | 32 | 0.773 |
| RPC [Dev. own theor. model] | - **YPT [Dev. own theor. model]** | Student | -9.332 | 32 | < .001 |
| RPC [Indep. research in scientific litr.] | - **YPT [Indep. research in scientific litr.]** | Student | -9.891 | 32 | < .001 |
| RPC [High school physics] | - YPT [High school physics] | Student | 0.333 | 32 | 0.741 |
|  |  | Wilcoxon | 116.000 | | 1.000 |
| RPC [Designing experiments] | - **YPT [Designing experiments]** | Student | -8.269 | 32 | < .001 |
| RPC [Interp. exp. data, data analysis] | - **YPT [Interp. exp. data, data analysis]** | Student | -7.187 | 32 | < .001 |
| RPC [Numerical simulations] | - **YPT [Numerical simulations]** | Student | -8.505 | 32 | < .001 |
| RPC [Crit. asses. of other's res.] | - **YPT [Crit. asses. of other's res.]** | Student | -9.336 | 32 | < .001 |

## 2.2.5 Teachers Paired T-Test on Hard Skill (YPT vs. Non-YPT, without CZ)

As RPC is a form of education developed for all high school students, we get much more useful and more information, especially for hard skills, from comparing YPT and non-YPT type competitions. As the competitions are already open to interested and / or talented students, the result of comparing them can be useful for teachers, as we want to turn as many interested students with different backgrounds to physics and research activities in general. The results presented below show well what additional opportunities YPT-type competitions have for interested and talented students compared to traditional ones.

**Descriptive Statistics: Hard Skills in YPT**

|  | YPT [High school mathematics] | YPT [High school physics] | YPT [Solving close-ended problems] | YPT [Designing experiments] | YPT [Cookbook experiments] | YPT [Interp. exp. data, data analysis] | YPT [Dev. own theor. model] | YPT [Numerical simulations] | YPT [Indep. research in scientific litr.] | YPT [Crit. asses. of other's res.] |
|---|---|---|---|---|---|---|---|---|---|---|
| Valid | 33 | 33 | 33 | 33 | 33 | 33 | 33 | 33 | 33 | 33 |
| Missing | 0 | 0 | 0 | 0 | 0 | 0 | 0 | 0 | 0 | 0 |
| Mean | 7.455 | 8.121 | 4.788 | 8.485 | 5.939 | 8.697 | 7.364 | 7.182 | 7.818 | 8.455 |
| Std. Deviation | 2.320 | 2.147 | 3.029 | 1.734 | 3.082 | 1.723 | 2.485 | 2.493 | 2.455 | 1.889 |
| Minimum | 2.000 | 2.000 | 0.000 | 2.000 | 1.000 | 2.000 | 1.000 | 0.000 | 2.000 | 2.000 |
| Maximum | 10.000 | 10.000 | 10.000 | 10.000 | 10.000 | 10.000 | 10.000 | 10.000 | 10.000 | 10.000 |

**Descriptive Statistics: Hard Skills in Non-Ypt (without CZ)**

|  | Non YPT [High school mathematics] | Non YPT [High school physics] | Non YPT [Solving close-ended problems] | Non YPT [Designing experiments] | Non YPT [Cookbook experiments] | Non YPT [Interp. exp. data, data analysis] | Non YPT [Dev. own theor. model] | Non YPT [Numerical simulations] | Non YPT [Indep. research in scientific litr.] | Non YPT [Crit. asses. of other's res.] |
|---|---|---|---|---|---|---|---|---|---|---|
| Valid | 29 | 28 | 29 | 29 | 29 | 28 | 29 | 29 | 28 | 29 |
| Missing | 4 | 5 | 4 | 4 | 4 | 5 | 4 | 4 | 5 | 4 |
| Mean | 7.414 | 8.107 | 7.862 | 4.310 | 4.862 | 5.286 | 3.793 | 3.241 | 5.214 | 3.207 |
| Std. Deviation | 2.719 | 2.409 | 2.722 | 3.037 | 3.148 | 3.253 | 3.245 | 2.923 | 3.665 | 2.969 |
| Minimum | 1.000 | 2.000 | 0.000 | 0.000 | 0.000 | 0.000 | 0.000 | 0.000 | 0.000 | 0.000 |
| Maximum | 10.000 | 10.000 | 10.000 | 10.000 | 10.000 | 10.000 | 9.000 | 9.000 | 10.000 | 10.000 |

**Test of Normality (Shapiro-Wilk): Hard Skills (YPT vs. Non-YPT)**

|  |  | W | p |
|---|---|---|---|
| YPT [High school mathematics] | - Non YPT [High school mathematics] | 0.890 | 0.006 |
| YPT [Solving close-ended problems] | - Non YPT [Solving close-ended problems] | 0.969 | 0.532 |
| YPT [Conducting experiment ] | - Non YPT [Conducting experiment ] | 0.954 | 0.238 |
| YPT [Dev. own theor. model] | - Non YPT [Dev. own theor. model] | 0.950 | 0.187 |
| YPT [Indep. research in scientific litr.] | - Non YPT [Indep. research in scientific litr.] | 0.914 | 0.024 |
| YPT [High school physics] | - Non YPT [High school physics] | 0.807 | < .001 |
| YPT [Designing experiments] | - Non YPT [Designing experiments] | 0.950 | 0.186 |
| YPT [Interp. exp. data, data analysis] | - Non YPT [Interp. exp. data, data analysis] | 0.837 | < .001 |
| YPT [Numerical simulations] | - Non YPT [Numerical simulations] | 0.967 | 0.478 |
| YPT [Crit. asses. of other's res.] | - Non YPT [Crit. asses. of other's res.] | 0.947 | 0.157 |

*Note.* Significant results suggest a deviation from normality.

#### Paired Samples T-Test: Hard Skills (YPT vs. Non-YPT)

| YPT | Non-YPT | Test | Statistic | df | p |
|---|---|---|---|---|---|
| YPT [High school mathematics] | - Non YPT [High school mathematics] | Wilcoxon | 81.500 | | 0.828 |
| YPT [Solving close-ended problems] | - **Non YPT [Solving close-ended problems]** | Student | -3.841 | 28 | < .001 |
| YPT [Conducting experiment ] | - Non YPT [Conducting experiment ] | Student | 1.629 | 28 | 0.115 |
| **YPT [Dev. own theor. model]** | - Non YPT [Dev. own theor. model] | Student | 5.554 | 28 | < .001 |
| **YPT [Indep. research in sci. litr.]** | - Non YPT [Indep. research in sci. litr.] | Student | 4.400 | 27 | < .001 |
| | | Wilcoxon | 259.500 | | < .001 |
| YPT [High school physics] | - Non YPT [High school physics] | Wilcoxon | 35.500 | | 0.855 |
| **YPT [Designing experiments]** | - Non YPT [Designing experiments] | Student | 8.267 | 28 | < .001 |
| **YPT [Interp. exp. data, data analysis]** | - Non YPT [Interp. exp. data, data analysis] | Student | 5.953 | 27 | < .001 |
| | | Wilcoxon | 325.000 | | < .001 |
| **YPT [Numerical simulations]** | - Non YPT [Numerical simulations] | Student | 6.841 | 28 | < .001 |
| **YPT [Crit. asses. of other's res.]** | - Non YPT [Crit. asses. of other's res.] | Student | 9.374 | 28 | < .001 |

There is no difference between "High school mathematics", "High school physics" development, and "Conducting experiments (based on clear manual)". Non-YPT is significantly better in "Solving close-ended problems in physics", and in all other hard skill, the developmental impact of YPT is seen as more serious by the teachers interviewed.

## 2.2.6 Teachers Paired T-Test on Hard Skills (RPC vs. Non-YPT, without CZ)

Based on our research hypothesis, we do not expect many differences, but of course a few differences may have a good chance. Where the normality test is not met, a Wilcoxon test is performed. After presenting the results of Hard Skills in RPC and Non-YPT, the comparison of Hard Skills in RPC vs. Non-YPT can be seen in the following table the significantly different values, where the higher value was marked with bold letters (which is the standard symbol system for the results presented).

The comparison of Hard Skills in RPC vs. Non-YPT (without CZ) can be seen in the following table – with colored background

#### Test of Normality (Shapiro-Wilk): Hard Skills (RPC vs. Non-YPT)

| | | W | p |
|---|---|---|---|
| RPC [High school mathematics] | - Non YPT [High school mathematics] | 0.927 | 0.046 |
| RPC [Solving close-ended problems] | - Non YPT [Solving close-ended problems] | 0.774 | < .001 |
| RPC [Conducting experiment ] | - Non YPT [Conducting experiment ] | 0.977 | 0.756 |
| RPC [Dev. own theor. model] | - Non YPT [Dev. own theor. model] | 0.970 | 0.549 |
| RPC [Indep. research in scientific litr.] | - Non YPT [Indep. research in scientific litr.] | 0.956 | 0.282 |
| RPC [High school physics] | - Non YPT [High school physics] | 0.853 | 0.001 |
| RPC [Designing experiments] | - Non YPT [Designing experiments] | 0.967 | 0.472 |
| RPC [Interp. exp. data, data analysis] | - Non YPT [Interp. exp. data, data analysis] | 0.953 | 0.230 |
| RPC [Numerical simulations] | - Non YPT [Numerical simulations] | 0.958 | 0.290 |
| RPC [Crit. asses. of other's res.] | - Non YPT [Crit. asses. of other's res.] | 0.921 | 0.033 |

Note. Significant results suggest a deviation from normality.

#### Paired Samples T-Test: Hard Skills (RPC vs. Non-YPT)

| RPC | Non-YPT | Test | Statistic | df | p |
|---|---|---|---|---|---|
| RPC [High school mathematics] | - Non YPT [High school mathematics] | Wilcoxon | 88.500 | | 0.130 |
| RPC [Solving close-ended problems] | - Non YPT [Solving close-ended problems] | Wilcoxon | 77.500 | | 0.308 |
| **RPC [Conducting experiment ]** | - Non YPT [Conducting experiment ] | Student | 2.292 | 28 | 0.030 |
| RPC [Dev. own theor. model] | - Non YPT [Dev. own theor. model] | Student | -0.484 | 28 | 0.632 |
| RPC [Indep. research in sci. litr.] | - **Non YPT [Indep. research in sci. litr.]** | Student | -2.097 | 27 | 0.045 |
| RPC [High school physics] | - Non YPT [High school physics] | Wilcoxon | 47.500 | | 0.916 |

The European Commission's support for the production of this publication does not constitute an endorsement of the contents, which reflect the views only of the authors, and the Commission cannot be held responsible for any use which may be made of the information contained therein.

**Paired Samples T-Test: Hard Skills (RPC vs. Non-YPT)**

| RPC | Non-YPT | Test | Statistic | df | p |
|---|---|---|---|---|---|
| RPC [Designing experiments] | - Non YPT [Designing experiments] | Student | 0.580 | 28 | 0.566 |
| RPC [Interp. exp. data, data analysis] | - Non YPT [Interp. exp. data, data analysis] | Student | 1.537 | 27 | 0.136 |
| RPC [Numerical simulations] | - Non YPT [Numerical simulations] | Student | -0.533 | 28 | 0.598 |
| **RPC [Crit. asses. of other's res.]** | - Non YPT [Crit. asses. of other's res.] | Wilcoxon | 189.000 | | 0.011 |

The results show practically minimal discrepancy for most of the hard skills tested. There is only one strongly significant difference in favor of RPC over Non-YPT: the "Critical assessment of others' results" within RPC is significantly better W = 189 p = .011. Beside this "Conducting experiments (based on clear manual)" t = 2.292 p =.03 seems to be better in RPC than in Non-YPT competitions, and "Independent research in scientific literature" t = - 2.097 p = .045 seems to be better in Non-YPT than in RPC.

One of the most striking questions in our research is whether we see these significant differences between YPT and RPC or Non-YPT in their impact on Hard Skills. Because the comparison in 5.1 shows, that with the only exception of the "Independent research in scientific literature", RPC supposed to have the same or significantly better effect as Non-YPT competitions on students Hard Skills, it is reasonable to limit the following comparison to YPT vs. RPC. To do this, we perform paired t-tests – or Wilcoxon-test, if needed.

## 2.2.7 Summary of the Results by Teachers in Hard Skills

Investigations on the Hard Skills show that

-YPT has an overall <u>significantly higher</u> positive influence than RPC and Non-YPT competitions:

> - *Designing experiments*
> - *Interpreting experimental data, data analysis*
> - *Developing own theoretical model*
> - *Numerical simulations*
> - *Independent research in scientific literature*
> - *Critical assessment of others' results*

- YPT has an overall the <u>same influence</u> as RPC and Non-YPT competitions:

> - *High school mathematics* (slightly better in RPC p=.037, but the same in Non-YPT)
> - *High school physics*
> - *Conducting experiment (based on clear manual)/Cookbook experiments*

- YPT has a <u>significantly lower</u> positive influence than RPC and Non-YPT competitions:

> - *Solving close-ended problems in physics*

The European Commission's support for the production of this publication does not constitute an endorsement of the contents, which reflect the views only of the authors, and the Commission cannot be held responsible for any use which may be made of the information contained therein.

## 2.3 Effect of the Country on Hard Skills

The scores given for hard skills – in RPC, YPT and Non-YPT too - seem to be very similar in all investigated countries too. As samples we show first the distributions of *High school mathematics, High school physics and Solving close-ended problems in physics* in the 5 countries: they are basically the same – what also the ANOVA tests are suggesting too.

**ANOVA - RPC [High school mathematics]**

| Cases | Sum of Squares | df | Mean Square | F | p | η² |
|---|---|---|---|---|---|---|
| Country | 13.757 | 4 | 3.439 | 0.531 | 0.714 | 0.073 |
| Residuals | 174.712 | 27 | 6.471 | | | |

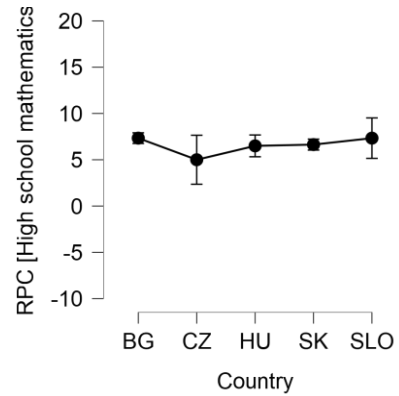

**Descriptives - RPC [High school mathematics]**

| Country | Mean | SD | N |
|---|---|---|---|
| BG | 7.333 | 1.732 | 9 |
| CZ | 5.000 | 4.583 | 3 |
| HU | 6.500 | 2.881 | 6 |
| SK | 6.636 | 1.963 | 11 |
| SLO | 7.333 | 3.786 | 3 |

**ANOVA - RPC [High school physics]**

| Cases | Sum of Squares | df | Mean Square | F | p | η² |
|---|---|---|---|---|---|---|
| Country | 15.833 | 4 | 3.958 | 1.063 | 0.393 | 0.132 |
| Residuals | 104.227 | 28 | 3.722 | | | |

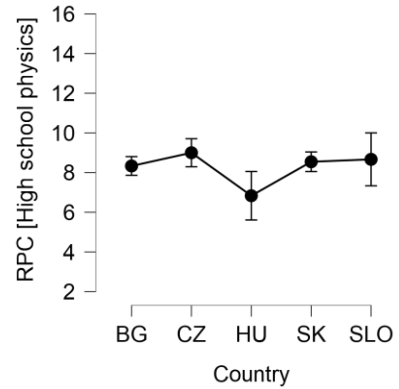

**Descriptives - RPC [High school physics]**

| Country | Mean | SD | N |
|---|---|---|---|
| BG | 8.333 | 1.414 | 9 |
| CZ | 9.000 | 1.414 | 4 |
| HU | 6.833 | 2.994 | 6 |
| SK | 8.545 | 1.635 | 11 |
| SLO | 8.667 | 2.309 | 3 |

**ANOVA - RPC [Solving close-ended problems]**

| Cases | Sum of Squares | df | Mean Square | F | p | η² |
|---|---|---|---|---|---|---|
| Country | 11.061 | 4 | 2.765 | 0.685 | 0.608 | 0.089 |
| Residuals | 113.000 | 28 | 4.036 | | | |

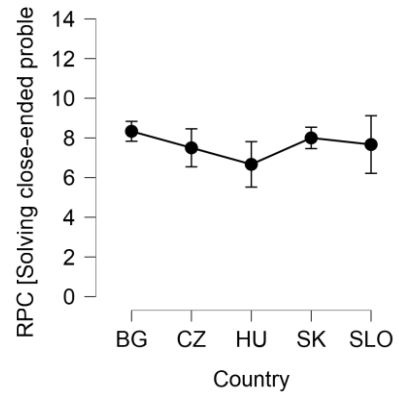

**Descriptives - RPC [Solving close-ended problems]**

| Country | Mean | SD | N |
|---|---|---|---|
| BG | 8.333 | 1.500 | 9 |
| CZ | 7.500 | 1.915 | 4 |
| HU | 6.667 | 2.805 | 6 |
| SK | 8.000 | 1.789 | 11 |
| SLO | 7.667 | 2.517 | 3 |

The European Commission's support for the production of this publication does not constitute an endorsement of the contents, which reflect the views only of the authors, and the Commission cannot be held responsible for any use which may be made of the information contained therein.

In the following you can see the significant effect of the country in hard skills:

**ANOVA - RPC [Dev. own theor. model]**

| Cases | Sum of Squares | df | Mean Square | F | p | $\eta^2_p$ |
|---|---|---|---|---|---|---|
| Country | 97.104 | 4 | 24.276 | 6.040 | 0.001 | 0.463 |
| Residuals | 112.533 | 28 | 4.019 | | | |

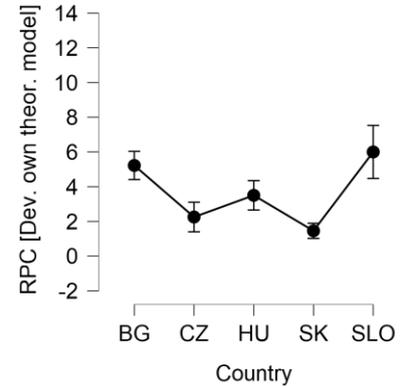

**Descriptives - RPC [Dev. own theor. model]**

| Country | Mean | SD | N |
|---|---|---|---|
| BG | 5.222 | 2.438 | 9 |
| CZ | 2.250 | 1.708 | 4 |
| HU | 3.500 | 2.074 | 6 |
| SK | 1.455 | 1.440 | 11 |
| SLO | 6.000 | 2.646 | 3 |

**ANOVA - RPC [Numerical simulations]**

| Cases | Sum of Squares | df | Mean Square | F | p | $\eta^2_p$ |
|---|---|---|---|---|---|---|
| Country | 68.611 | 4 | 17.153 | 3.116 | 0.031 | 0.308 |
| Residuals | 154.116 | 28 | 5.504 | | | |

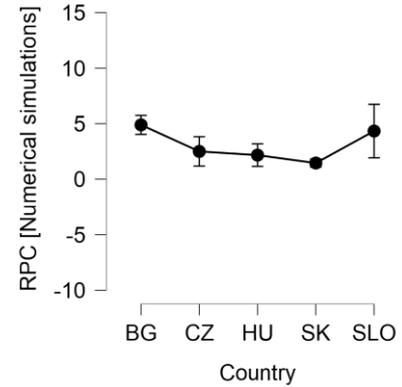

**Descriptives - RPC [Numerical simulations]**

| Country | Mean | SD | N |
|---|---|---|---|
| BG | 4.889 | 2.571 | 9 |
| CZ | 2.500 | 2.646 | 4 |
| HU | 2.167 | 2.483 | 6 |
| SK | 1.455 | 1.214 | 11 |
| SLO | 4.333 | 4.163 | 3 |

The European Commission's support for the production of this publication does not constitute an endorsement of the contents, which reflect the views only of the authors, and the Commission cannot be held responsible for any use which may be made of the information contained therein.

**ANOVA - YPT [Cookbook experiments ]**

| Cases | Sum of Squares | df | Mean Square | F | p | $\eta^2_p$ |
|---|---|---|---|---|---|---|
| Country | 94.187 | 4 | 23.547 | 3.144 | 0.030 | 0.310 |
| Residuals | 209.692 | 28 | 7.489 | | | |

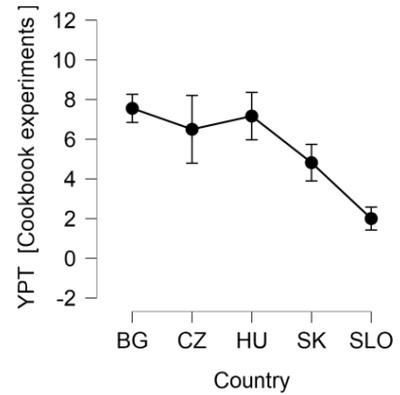

**Descriptives - YPT [Cookbook experiments ]**

| Country | Mean | SD | N |
|---|---|---|---|
| BG | 7.556 | 2.128 | 9 |
| CZ | 6.500 | 3.416 | 4 |
| HU | 7.167 | 2.927 | 6 |
| SK | 4.818 | 3.060 | 11 |
| SLO | 2.000 | 1.000 | 3 |

**ANOVA - YPT [Dev. own theor. model]**

| Cases | Sum of Squares | df | Mean Square | F | p | $\eta^2_p$ |
|---|---|---|---|---|---|---|
| Country | 72.702 | 4 | 18.176 | 4.073 | 0.010 | 0.368 |
| Residuals | 124.934 | 28 | 4.462 | | | |

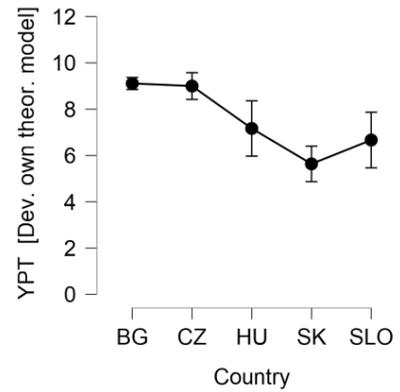

**Descriptives - YPT [Dev. own theor. model]**

| Country | Mean | SD | N |
|---|---|---|---|
| BG | 9.111 | 0.782 | 9 |
| CZ | 9.000 | 1.155 | 4 |
| HU | 7.167 | 2.927 | 6 |
| SK | 5.636 | 2.541 | 11 |
| SLO | 6.667 | 2.082 | 3 |

**ANOVA - Non YPT [Conducting experiment ]**

| Cases | Sum of Squares | df | Mean Square | F | p | $\eta^2$ |
|---|---|---|---|---|---|---|
| Country | 87.665 | 3 | 29.222 | 3.849 | 0.022 | 0.316 |
| Residuals | 189.783 | 25 | 7.591 | | | |

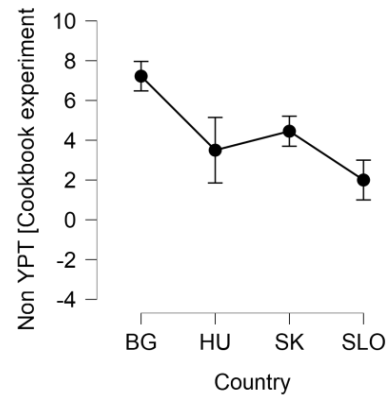

The European Commission's support for the production of this publication does not constitute an endorsement of the contents, which reflect the views only of the authors, and the Commission cannot be held responsible for any use which may be made of the information contained therein.

**Descriptives - Non YPT [Cookbook experiments ]**

| Country | Mean | SD | N |
|---------|------|------|----|
| BG | 7.222 | 2.224 | 9 |
| HU | 3.500 | 4.037 | 6 |
| SK | 4.455 | 2.505 | 11 |
| SLO | 2.000 | 1.732 | 3 |

The European Commission's support for the production of this publication does not constitute an endorsement of the contents, which reflect the views only of the authors, and the Commission cannot be held responsible for any use which may be made of the information contained therein.

# 3. Supplement: Comparison between Students' and Teachers' scores

## 3.1 Hard Skills Results of the n = 77 Students of the Comparison with Teachers

### Descriptives: 77 Students Scores in RPC and YPT

|  | N | Mean | SD | SE |
|---|---|---|---|---|
| High sch. math. - RPC | 34 | 8.294 | 1.567 | 0.269 |
| High sch. math. -YPT | 34 | 6.882 | 1.855 | 0.318 |
| High sch. phy.- RPC | 77 | 8.234 | 1.555 | 0.177 |
| High sch. phy.-YPT | 77 | 7.584 | 1.956 | 0.223 |
| Solv. clos-end. prob. - RPC | 77 | 8.338 | 1.501 | 0.171 |
| Solv. clos-end. prob. -YPT | 77 | 7.351 | 1.931 | 0.220 |
| Des. exp.- RPC | 77 | 7.091 | 2.141 | 0.244 |
| Des. exp.-YPT | 77 | 7.870 | 1.929 | 0.220 |
| Cookbook exp. - RPC | 77 | 8.260 | 1.787 | 0.204 |
| Cookbook exp. -YPT | 77 | 7.948 | 1.891 | 0.216 |
| Int. exp. data - RPC | 77 | 7.403 | 2.028 | 0.231 |
| Int. exp. data -YPT | 76 | 7.632 | 2.006 | 0.230 |
| Dev. own. th. mod. - RPC | 77 | 6.701 | 1.994 | 0.227 |
| Dev. own. th. mod. -YPT | 76 | 7.289 | 2.159 | 0.248 |
| Num. sim. - RPC | 77 | 6.364 | 2.194 | 0.250 |
| Num. sim. -YPT | 77 | 6.987 | 1.909 | 0.218 |
| Research in sci. lit. - RPC | 77 | 6.805 | 2.230 | 0.254 |
| Research in sci. lit. -YPT | 77 | 7.792 | 1.880 | 0.214 |
| Crit. ass. - RPC | 76 | 7.079 | 2.128 | 0.244 |
| Crit. ass. -YPT | 77 | 7.481 | 1.875 | 0.214 |

### Test of Normality (Shapiro-Wilk)

|  |  | W | p |
|---|---|---|---|
| High sch. math. - RPC | - High sch. math. -YPT | 0.879 | 0.001 |
| High sch. phy.- RPC | - High sch. phy.-YPT | 0.887 | < .001 |
| Solv. clos-end. prob. - RPC | - Solv. clos-end. prob. -YPT | 0.839 | < .001 |
| Des. exp.- RPC | - Des. exp.-YPT | 0.915 | < .001 |
| Cookbook exp. - RPC | - Cookbook exp. -YPT | 0.865 | < .001 |
| Int. exp. data - RPC | - Int. exp. data -YPT | 0.911 | < .001 |
| Dev. own. th. mod. - RPC | - Dev. own. th. mod. -YPT | 0.904 | < .001 |
| Num. sim. - RPC | - Num. sim. -YPT | 0.850 | < .001 |
| Research in sci. lit. - RPC | - Research in sci. lit. -YPT | 0.859 | < .001 |
| Crit. ass. - RPC | - Crit. ass. -YPT | 0.886 | < .001 |

*Note.* Significant results suggest a deviation from normality.

### Comparison (Wilcoxon): 77 Students RPC vs. YPT

| RPC | YPT | W | df | p |
|---|---|---|---|---|
| **High sch. math. - RPC** | - High sch. math. -YPT | **355.000** |  | **0.002** |
| **High sch. phy.- RPC** | - High sch. phy.-YPT | **619.000** |  | **0.003** |
| **Solv. clos-end. prob. - RPC** | - Solv. clos-end. prob. -YPT | **570.500** |  | **< .001** |
| Des. exp.- RPC | - **Des. exp.-YPT** | **270.000** |  | **0.012** |

The European Commission's support for the production of this publication does not constitute an endorsement of the contents, which reflect the views only of the authors, and the Commission cannot be held responsible for any use which may be made of the information contained therein.

**Comparison (Wilcoxon): 77 Students RPC vs. YPT**

| RPC | YPT | W | df | p |
|---|---|---|---|---|
| Cookbook exp. - RPC | - Cookbook exp. -YPT | 376.000 | | 0.163 |
| Int. exp. data - RPC | - Int. exp. data -YPT | 406.500 | | 0.410 |
| Dev. own. th. mod. - RPC | **- Dev. own. th. mod. -YPT** | **296.000** | | **0.029** |
| Num. sim. - RPC | **- Num. sim. -YPT** | **175.500** | | **0.019** |
| Research in sci. lit. - RPC | **- Research in sci. lit. -YPT** | **169.500** | | **0.002** |
| Crit. ass. - RPC | - Crit. ass. -YPT | 321.500 | | 0.222 |

*Note.* Wilcoxon signed-rank test.

### 3.2 Hard Skills in RPC and YPT: Students (n = 77) vs. Teachers (n = 32) (both on 1-10 scale).

**Descriptive Statistics: Hard Skills in YPT and RPC, Students and Teachers**

| | High sch. math. -YPT | | High sch. math. - RPC | | High sch. phy.-YPT | | High sch. phy.- RPC | |
|---|---|---|---|---|---|---|---|---|
| | Student | Teacher | Student | Teacher | Student | Teacher | Student | Teacher |
| Valid | 34 | 32 | 34 | 31 | 77 | 32 | 77 | 32 |
| Missing | 43 | 0 | 43 | 1 | 0 | 0 | 0 | 0 |
| Mean | 6.882 | 7.438 | 8.294 | 6.710 | 7.584 | 8.125 | 8.234 | 8.281 |
| Std. Deviation | 1.855 | 2.355 | 1.567 | 2.506 | 1.956 | 2.181 | 1.555 | 1.955 |
| Minimum | 4.000 | 2.000 | 4.000 | 0.000 | 2.000 | 2.000 | 4.000 | 2.000 |
| Maximum | 10.000 | 10.000 | 10.000 | 10.000 | 10.000 | 10.000 | 10.000 | 10.000 |

**Descriptive Statistics: Hard Skills in YPT and RPC, Students and Teachers**

| | Solv. clos-end. prob. -YPT | | Solv. clos-end. prob. - RPC | | Des. exp.-YPT | | Des. exp.- RPC | |
|---|---|---|---|---|---|---|---|---|
| | Student | Teacher | Student | Teacher | Student | Teacher | Student | Teacher |
| Valid | 77 | 32 | 77 | 32 | 77 | 32 | 77 | 32 |
| Missing | 0 | 0 | 0 | 0 | 0 | 0 | 0 | 0 |
| Mean | 7.351 | 4.688 | 8.338 | 7.781 | 7.870 | 8.531 | 7.091 | 4.781 |
| Std. Deviation | 1.931 | 3.021 | 1.501 | 1.996 | 1.929 | 1.741 | 2.141 | 2.549 |
| Minimum | 2.000 | 0.000 | 4.000 | 2.000 | 2.000 | 2.000 | 2.000 | 1.000 |
| Maximum | 10.000 | 10.000 | 10.000 | 10.000 | 10.000 | 10.000 | 10.000 | 9.000 |

**Descriptive Statistics: Hard Skills in YPT and RPC, Students and Teachers**

| | Cookbook exp. -YPT | | Cookbook exp. - RPC | | Int. exp. data -YPT | | Int. exp. data - RPC | |
|---|---|---|---|---|---|---|---|---|
| | Student | Teacher | Student | Teacher | Student | Teacher | Student | Teacher |
| Valid | 77 | 32 | 77 | 32 | 76 | 32 | 77 | 32 |
| Missing | 0 | 0 | 0 | 0 | 1 | 0 | 0 | 0 |
| Mean | 7.948 | 5.906 | 8.260 | 6.250 | 7.632 | 8.750 | 7.403 | 6.313 |
| Std. Deviation | 1.891 | 3.125 | 1.787 | 2.185 | 2.006 | 1.723 | 2.028 | 2.278 |
| Minimum | 2.000 | 1.000 | 2.000 | 1.000 | 2.000 | 2.000 | 2.000 | 1.000 |
| Maximum | 10.000 | 10.000 | 10.000 | 10.000 | 10.000 | 10.000 | 10.000 | 10.000 |

**Descriptive Statistics: Hard Skills in YPT and RPC, Students and Teachers**

| | Dev. own. th. mod. -YPT | | Dev. own. th. mod. - RPC | | Num. sim. -YPT | | Num. sim. - RPC | |
|---|---|---|---|---|---|---|---|---|
| | Student | Teacher | Student | Teacher | Student | Teacher | Student | Teacher |
| Valid | 76 | 32 | 77 | 32 | 77 | 32 | 77 | 32 |
| Missing | 1 | 0 | 0 | 0 | 0 | 0 | 0 | 0 |

**Descriptive Statistics: Hard Skills in YPT and RPC, Students and Teachers**

| | Dev. own. th. mod. -YPT | | Dev. own. th. mod. - RPC | | Num. sim. -YPT | | Num. sim. - RPC | |
|---|---|---|---|---|---|---|---|---|
| | **Student** | **Teacher** | **Student** | **Teacher** | **Student** | **Teacher** | **Student** | **Teacher** |
| Mean | 7.289 | 7.375 | 6.701 | 3.375 | 6.987 | 7.219 | 6.364 | 2.938 |
| Std. Deviation | 2.159 | 2.524 | 1.994 | 2.600 | 1.909 | 2.524 | 2.194 | 2.675 |
| Minimum | 2.000 | 1.000 | 2.000 | 0.000 | 2.000 | 0.000 | 2.000 | 0.000 |
| Maximum | 10.000 | 10.000 | 10.000 | 10.000 | 10.000 | 10.000 | 10.000 | 9.000 |

**Descriptive Statistics: Hard Skills in YPT and RPC, Students and Teachers**

| | Research in sci. lit. -YPT | | Research in sci. lit. - RPC | | Crit. ass. -YPT | | Crit. ass. - RPC | |
|---|---|---|---|---|---|---|---|---|
| | **Student** | **Teacher** | **Student** | **Teacher** | **Student** | **Teacher** | **Student** | **Teacher** |
| Valid | 77 | 32 | 77 | 32 | 77 | 32 | 76 | 32 |
| Missing | 0 | 0 | 0 | 0 | 0 | 0 | 1 | 0 |
| Mean | 7.792 | 7.844 | 6.805 | 3.625 | 7.481 | 8.469 | 7.079 | 4.094 |
| Std. Deviation | 1.880 | 2.490 | 2.230 | 2.459 | 1.875 | 1.917 | 2.128 | 2.347 |
| Minimum | 2.000 | 2.000 | 2.000 | 0.000 | 2.000 | 2.000 | 2.000 | 0.000 |
| Maximum | 10.000 | 10.000 | 10.000 | 9.000 | 10.000 | 10.000 | 10.000 | 10.000 |

**Test of Normality (Shapiro-Wilk)**

| | | **W** | **p** |
|---|---|---|---|
| High sch. math. - RPC | Student | 0.768 | < .001 |
| | Teacher | 0.927 | 0.037 |
| High sch. math. -YPT | Student | 0.778 | < .001 |
| | Teacher | 0.878 | 0.002 |
| High sch. phy.- RPC | Student | 0.825 | < .001 |
| | Teacher | 0.811 | < .001 |
| High sch. phy.-YPT | Student | 0.871 | < .001 |
| | Teacher | 0.820 | < .001 |
| Solv. clos-end. prob. - RPC | Student | 0.816 | < .001 |
| | Teacher | 0.891 | 0.004 |
| Solv. clos-end. prob. -YPT | Student | 0.862 | < .001 |
| | Teacher | 0.924 | 0.027 |
| Des. exp.- RPC | Student | 0.895 | < .001 |
| | Teacher | 0.923 | 0.024 |
| Des. exp.-YPT | Student | 0.846 | < .001 |
| | Teacher | 0.787 | < .001 |
| Cookbook exp. - RPC | Student | 0.797 | < .001 |
| | Teacher | 0.941 | 0.080 |
| Cookbook exp. -YPT | Student | 0.824 | < .001 |
| | Teacher | 0.905 | 0.008 |
| Int. exp. data - RPC | Student | 0.868 | < .001 |
| | Teacher | 0.965 | 0.366 |
| Int. exp. data -YPT | Student | 0.860 | < .001 |
| | Teacher | 0.727 | < .001 |
| Dev. own. th. mod. - RPC | Student | 0.902 | < .001 |
| | Teacher | 0.935 | 0.056 |
| Dev. own. th. mod. -YPT | Student | 0.882 | < .001 |
| | Teacher | 0.888 | 0.003 |
| Num. sim. - RPC | Student | 0.897 | < .001 |
| | Teacher | 0.872 | 0.001 |
| Num. sim. -YPT | Student | 0.881 | < .001 |
| | Teacher | 0.894 | 0.004 |
| Research in sci. lit. - RPC | Student | 0.871 | < .001 |

The European Commission's support for the production of this publication does not constitute an endorsement of the contents, which reflect the views only of the authors, and the Commission cannot be held responsible for any use which may be made of the information contained therein.

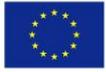

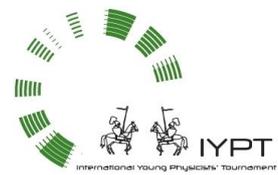

*DEVELOPMENT OF INQUIRY-BASED*
*LEARNING VIA IYPT*

### Test of Normality (Shapiro-Wilk)

|  |  | W | p |
|---|---|---|---|
|  | Teacher | 0.944 | 0.095 |
| Research in sci. lit. -YPT | Student | 0.845 | < .001 |
|  | Teacher | 0.823 | < .001 |
| Crit. ass. - RPC | Student | 0.876 | < .001 |
|  | Teacher | 0.951 | 0.151 |
| Crit. ass. -YPT | Student | 0.869 | < .001 |
|  | Teacher | 0.776 | < .001 |

*Note.* Significant results suggest a deviation from normality.

### Comparison: Hard Skills in RPC and YPT of Students and Teachers

|  | W | p |
|---|---|---|
| **High sch. math. - RPC (Students)** | **734.500** | **0.005** |
| High sch. math. -YPT | 437.000 | 0.159 |
| High sch. phy.- RPC | 1169.000 | 0.660 |
| High sch. phy.-YPT | 998.000 | 0.108 |
| Solv. clos-end. prob. - RPC | 1415.500 | 0.200 |
| **Solv. clos-end. prob. -YPT (Students)** | **1894.500** | **< .001** |
| **Des. exp.- RPC (Students)** | **1833.500** | **< .001** |
| Des. exp.-YPT | 973.000 | 0.073 |
| **Cookbook exp. - RPC (Students)** | **1925.500** | **< .001** |
| **Cookbook exp. -YPT (Students)** | **1697.500** | **0.001** |
| **Int. exp. data - RPC (Students)** | **1581.500** | **0.017** |
| **Int. exp. data -YPT (Teachers)** | **801.500** | **0.004** |
| **Dev. own. th. mod. - RPC (Students)** | **2078.500** | **< .001** |
| Dev. own. th. mod. -YPT | 1155.500 | 0.679 |
| **Num. sim. - RPC (Students)** | **2042.000** | **< .001** |
| Num. sim. -YPT | 1061.000 | 0.241 |
| **Research in sci. lit. - RPC (Students)** | **2006.000** | **< .001** |
| Research in sci. lit. -YPT | 1128.500 | 0.477 |
| **Crit. ass. - RPC (Students)** | **2007.500** | **< .001** |
| **Crit. ass. -YPT (Teachers)** | **816.000** | **0.004** |

*Note.* Mann-Whitney U test. highlighted bold if $p \leq .05$ In parentheses the direction of positive bias group..

Differences in hard skills between RPC and YPT (positive value means better for YPT):

### Group Descriptives: Differences between YPT and RPC (positive value means better in YPT)

|  | Group | N | Mean | SD | SE |
|---|---|---|---|---|---|
| Diff. Math. | Student | 34 | -1.412 | 2.388 | 0.410 |
|  | Teacher | 32 | 0.938 | 2.711 | 0.479 |
| Diff. Phys. | Student | 77 | -0.649 | 1.790 | 0.204 |
|  | Teacher | 32 | -0.156 | 2.112 | 0.373 |
| Diff. Solv. Cl. Pr. | Student | 77 | -0.987 | 1.990 | 0.227 |
|  | Teacher | 32 | -3.094 | 3.383 | 0.598 |
| Diff. Des. Exp. | Student | 77 | 0.779 | 2.516 | 0.287 |
|  | Teacher | 32 | 3.750 | 2.627 | 0.464 |
| Diff. Cookbook | Student | 77 | -0.312 | 1.948 | 0.222 |
|  | Teacher | 32 | -0.344 | 3.525 | 0.623 |
| Diff. Int. Exp. | Student | 77 | 0.130 | 2.582 | 0.294 |
|  | Teacher | 32 | 2.438 | 1.999 | 0.353 |

The European Commission's support for the production of this publication does not constitute an endorsement of the contents, which reflect the views only of the authors, and the Commission cannot be held responsible for any use which may be made of the information contained therein.

**Group Descriptives: Differences between YPT and RPC (positive value means better in YPT)**

|  | Group | N | Mean | SD | SE |
|---|---|---|---|---|---|
| Diff. Dev own theory | Student | 77 | 0.494 | 2.537 | 0.289 |
|  | Teacher | 32 | 4.000 | 2.502 | 0.442 |
| Diff. Num Sim. | Student | 77 | 0.623 | 2.254 | 0.257 |
|  | Teacher | 32 | 4.281 | 2.932 | 0.518 |
| Diff. Research | Student | 77 | 0.987 | 2.526 | 0.288 |
|  | Teacher | 32 | 4.219 | 2.485 | 0.439 |
| Diff. Crit. Ass. | Student | 77 | 0.494 | 2.718 | 0.310 |
|  | Teacher | 32 | 4.375 | 2.537 | 0.448 |

**Test of Normality (Shapiro-Wilk): Differences of hard skills in RPC and YPT**

|  |  | W | p |
|---|---|---|---|
| Diff. Math. | Student | 0.879 | 0.001 |
|  | Teacher | 0.919 | 0.020 |
| Diff. Phys. | Student | 0.887 | < .001 |
|  | Teacher | 0.895 | 0.005 |
| Diff. Solv. Cl. Pr. | Student | 0.839 | < .001 |
|  | Teacher | 0.967 | 0.422 |
| Diff. Des. Exp. | Student | 0.915 | < .001 |
|  | Teacher | 0.973 | 0.594 |
| Diff. Cookbook | Student | 0.865 | < .001 |
|  | Teacher | 0.974 | 0.625 |
| Diff. Int. Exp. | Student | 0.924 | < .001 |
|  | Teacher | 0.963 | 0.334 |
| Diff. Dev own theory | Student | 0.918 | < .001 |
|  | Teacher | 0.961 | 0.302 |
| Diff. Num Sim. | Student | 0.850 | < .001 |
|  | Teacher | 0.971 | 0.521 |
| Diff. Research | Student | 0.859 | < .001 |
|  | Teacher | 0.953 | 0.179 |
| Diff. Crit. Ass. | Student | 0.880 | < .001 |
|  | Teacher | 0.972 | 0.549 |

*Note.* Significant results suggest a deviation from normality.

**Independent Samples T-Test/ Man-Whitney U-tests of the Differences in Hard Skills in YPT and RPC between Students and Teachers**

|  | W | p |
|---|---|---|
| **Diff. Math.** | **278.000** | **< .001** |
| Diff. Phys. | 983.500 | 0.081 |
| **Diff. Solv. Cl. Pr.** | **1741.500** | **< .001** |
| **Diff. Des. Exp.** | **515.500** | **< .001** |
| Diff. Cookbook | 1210.500 | 0.885 |
| **Diff. Int. Exp.** | **551.000** | **< .001** |
| **Diff. Dev own theory.** | **401.500** | **< .001** |
| **Diff. Num Sim.** | **374.500** | **< .001** |
| **Diff. Research** | **407.000** | **< .001** |
| **Diff. Crit. Ass.** | **355.500** | **< .001** |

*Note.* Mann-Whitney U test.

The European Commission's support for the production of this publication does not constitute an endorsement of the contents, which reflect the views only of the authors, and the Commission cannot be held responsible for any use which may be made of the information contained therein.

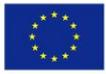

DEVELOPMENT OF INQUIRY-BASED
LEARNING VIA IYPT

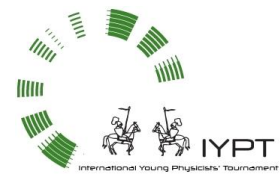

# 4. Supplement: Teachers' Assessment of Hard-Skill Development

## 4.1 Research question

According to our own experience, YPT type competitions can have serious effects on many skills and motivation of most high school students. In order to quantify this experience, we have formulated the following main research question:

*What impact of different teaching platforms do students and teachers attribute to students' hard skill development" (RPC, YPT and Non-YPT competitions)?*

In addition, we were confident that other connections and relationships would become known in the course of the research, but we see these as some welcome side effects.

## 4.2 Hypotheses

Based on our many years of experience in preparing high school students for the IYPT or any YPT competitions, we formulate the following hypotheses:

*1. We do not find significant differences between traditional competitions (Non-YPT) and regular physics classes (RPC) in terms of their impact on most of the hard skills examined.*

*2. YPT competitions have a serious positive effect on certain hard skills compared to the other two learning modes (RCP and Non-YPT competitions).*

The results of the examination of our hypotheses in themselves-, provide only a part of the actual impact test, as it only makes statements from the perspective of teachers. The message of the results of the research should be interpreted in its entirety together with the answers given by the participating students.

## 4.3 Methods

In total, 308 students from nine countries participated in the survey. The largest share of students was from Slovakia (54%), followed by Hungary (23%), the Czech Republic (7%), and Bulgaria (7%).

11 Slovak and 6 Hungarian physics teachers provided the data in January 2021 and by 9 Bulgarian, 3 Slovenian and 4 Czech physics teachers in November 2021. Given the COVID situation, teachers carried out their preparatory work in 2020/2021 mainly online. This is why it is important to mention that most colleagues have been involved in preparing for YPT-type competitions for several years. Teachers had to fill in a questionnaire and answer 16x3 quantitative and 15 qualitative questions about the impact and characteristics of RCP, YPT and Non-YPT competitions.

Data on teachers' answers form the questionnaire were provided in Excel format. For carrying out the empirical analysis, the software JASP[10] was used. First, descriptive analyses on skills as well as a correlation matrix using all variables. Secondly, for testing the hypotheses paired and independent t-tests (and Wilcoxon or Mann-Whitney-tests where needed) were computed. If Wilcoxon or Mann-Whitney-test was used, it is always the relevant result, t-tests are in these cases not relevant.

---

[10] https://jasp-stats.org/

The comparison between the values given for the assessment of the hard skills listed in the Guidelines and 1., 2. and 3. Supplement sections.

## 4.4 Conclusions

Regarding the *first hypothesis* of the study, we can state that we found few significant differences between the effects of traditional classes and competition in the area of the hard skills. RPC and Non-YPT are by teachers quite similar, but by students seems to be RPC more useful:

- by teachers only "Conducting (cookbook) experiments", and "Critical assessments of other results" are better in RPC, and "Independent research in scientific literature" is better in Non-YPT.
- by students seems to be Non-YPT in "Designing experiments", "Interpreting experimental data, data analysis", "Developing own theoretical model", "Numerical simulations", "Independent research in scientific literature", "Critical assessment of others' results" significantly better than RPC activities.

It also seems, that traditional competitions are strongly mathematics-centered and therefore require preparation similar to traditional lessons. Of course, this is both an advantage and a disadvantage, as it does not require special work, knowledge, extra time and energy investment from teachers – as they are complained about it in the case of YPT. It is difficult to increase the number of students who are successful in physics, as mathematics knowledge severely limits the number of students available.

For our *second hypothesis*, the overall effect seems to be more positive: there are several positive significant differences in the effect of YPT-type learning in comparison to RPC, and Non-YPT-type activities.

In the examination of the hard skills in the full sample of students, the platforms RPC and YPT show no significant difference in "High school mathematics", "High school physics" and "Cookbook experiments". The case of "Solving close-ended problems in physics" YPT shows less developmental effect than the other two. It is important to mention that the effect of "High school mathematics" in YPT shows no difference from Non-YPT competitions. However, there are significant positive differences in "Designing experiment", "Interpreting experimental data, data analysis", "Developing own theoretical model", "Numerical simulations", "Independent research in scientific literature", and "Critical assessment of others' results". Teachers only find "High school mathematics" and "Solving close-ended problems in physics" in RPC better, "High school physics" seems to have the same effect in RPC and YPT, all other skills are better in YPT.

The comparison between the 77 students' and 33 teachers' scores shows, that students tend to give extremely higher scores for RPC as the teachers. Although the differences between the scores for RPC and YPT from both students and teacher show, that students and teachers see the effects of RPC and YPT quite similar – with the exception of *High school mathematics*, where students gave more scores for RPC.

The European Commission's support for the production of this publication does not constitute an endorsement of the contents, which reflect the views only of the authors, and the Commission cannot be held responsible for any use which may be made of the information contained therein.

## 4.5 Limitation and future research

The main development potential of our measurement lies in the fact that while the questions examine the perceived effects of teachers and students, it is still lacking what they are actually doing. The initial results are very encouraging, and based on the experience so far, it seems worthwhile to involve more countries in the future and to examine larger samples. The comprehensive interpretation of the results obtained also requires responses from students, this is also particularly important because teacher evaluation alone is often biased, although since we have performed comparative studies with each other, we can hope that this general bias does not have a significant effect on comparative studies.

The European Commission's support for the production of this publication does not constitute an endorsement of the contents, which reflect the views only of the authors, and the Commission cannot be held responsible for any use which may be made of the information contained therein.

The European Commission's support for the production of this publication does not constitute an endorsement of the contents, which reflect the views only of the authors, and the Commission cannot be held responsible for any use which may be made of the information contained therein.

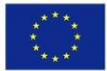
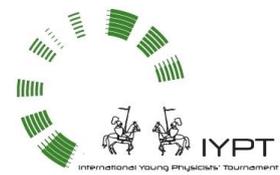

*DEVELOPMENT OF INQUIRY-BASED LEARNING VIA IYPT*

## 4.6. APPENDIX

## Hard Skills

*Assumption tests for "Impact of years to final exam on usefulness of RPC, YPT and other activities"*

| Hard Skills - RPC | Shapiro-Wilk test (p-value) | NCV test (p-value) | Durbin Watson test (p-value) |
|---|---|---|---|
| High school mathematics | 0,000 | 0,428 | 0,094 |
| High school physics | 0,000 | 0,287 | 0,116 |
| Solve close-ended problems | 0,000 | 0,344 | 0,116 |
| Designing experiments | 0,000 | 0,837 | 0,008 |
| Conducting experiment | 0,000 | 0,009 | 0,000 |
| Interpreting experimental data, data analysis | 0,000 | 0,003 | 0,036 |
| Developing own theoretical model | 0,000 | 0,395 | 0,136 |
| Numerical simulations | 0,000 | 0,103 | 0,032 |
| Independent research in scientific literature | 0,000 | 0,158 | 0,230 |
| Critical assessment of others' results | 0,000 | 0,033 | 0,022 |

| Hard Skills - YPT | Shapiro-Wilk test (p-value) | NCV test (p-value) | Durbin Watson test (p-value) |
|---|---|---|---|
| High school mathematics | 0,000 | 0,237 | 0,728 |
| High school physics | 0,000 | 0,982 | 0,762 |
| Solve close-ended problems | 0,000 | 0,402 | 0,978 |
| Designing experiments | 0,000 | 0,469 | 0,212 |
| Conducting experiment | 0,000 | 0,690 | 0,422 |
| Interpreting experimental data, data analysis | 0,000 | 0,175 | 0,808 |
| Developing own theoretical model | 0,000 | 0,056 | 0,056 |
| Numerical simulations | 0,000 | 0,313 | 0,556 |
| Independent research in scientific literature | 0,000 | 0,610 | 0,638 |
| Critical assessment of others' results | 0,000 | 0,856 | 0,976 |

| Hard Skills - Other | Shapiro-Wilk test (p-value) | NCV test (p-value) | Durbin Watson test (p-value) |
|---|---|---|---|
| High school mathematics | 0,000 | 0,016 | 0,998 |
| High school physics | 0,000 | 0,264 | 0,128 |
| Solve close-ended problems | 0,000 | 0,371 | 0,682 |
| Designing experiments | 0,000 | 0,106 | 0,426 |
| Conducting experiment | 0,000 | 0,048 | 0,640 |
| Interpreting experimental data, data analysis | 0,000 | 0,409 | 0,084 |
| Developing own theoretical model | 0,000 | 0,139 | 0,902 |
| Numerical simulations | 0,000 | 0,138 | 0,036 |
| Independent research in scientific literature | 0,000 | 0,344 | 0,104 |
| Critical assessment of others' results | 0,000 | 0,051 | 0,198 |

*Assumption tests for "Impact of physics classes on usefulness of RPC, YPT and other activities"*

The European Commission's support for the production of this publication does not constitute an endorsement of the contents, which reflect the views only of the authors, and the Commission cannot be held responsible for any use which may be made of the information contained therein.

| Hard Skills - RPC | Shapiro-Wilk test (p-value) | NCV test (p-value) | Durbin Watson test (p-value) |
|---|---|---|---|
| High school mathematics | 0,000 | 0,352 | 0,236 |
| High school physics | 0,000 | 0,070 | 0,140 |
| Solve close-ended problems | 0,000 | 0,326 | 0,140 |
| Designing experiments | 0,000 | 0,204 | 0,000 |
| Conducting experiment | 0,000 | 0,751 | 0,000 |
| Interpreting experimental data, data analysis | 0,000 | 0,304 | 0,002 |
| Developing own theoretical model | 0,000 | 0,537 | 0,046 |
| Numerical simulations | 0,000 | 0,412 | 0,004 |
| Independent research in scientific literature | 0,000 | 0,364 | 0,086 |
| Critical assessment of others' results | 0,000 | 0,886 | 0,000 |

| Hard Skills - YPT | Shapiro-Wilk test (p-value) | NCV test (p-value) | Durbin Watson test (p-value) |
|---|---|---|---|
| High school mathematics | 0,000 | 0,289 | 0,642 |
| High school physics | 0,000 | 0,153 | 0,552 |
| Solve close-ended problems | 0,000 | 0,593 | 0,964 |
| Designing experiments | 0,002 | 0,107 | 0,830 |
| Conducting experiment | 0,000 | 0,204 | 0,322 |
| Interpreting experimental data, data analysis | 0,000 | 0,047 | 0,966 |
| Developing own theoretical model | 0,000 | 0,014 | 0,054 |
| Numerical simulations | 0,000 | 0,075 | 0,764 |
| Independent research in scientific literature | 0,000 | 0,331 | 0,632 |
| Critical assessment of others' results | 0,000 | 0,241 | 0,920 |

| Hard Skills - Other | Shapiro-Wilk test (p-value) | NCV test (p-value) | Durbin Watson test (p-value) |
|---|---|---|---|
| High school mathematics | 0,000 | 0,000 | 0,736 |
| High school physics | 0,000 | 0,012 | 0,192 |
| Solve close-ended problems | 0,000 | 0,093 | 0,712 |
| Designing experiments | 0,000 | 0,633 | 0,328 |
| Conducting experiment | 0,000 | 0,185 | 0,976 |
| Interpreting experimental data, data analysis | 0,000 | 0,340 | 0,160 |
| Developing own theoretical model | 0,000 | 0,475 | 0,530 |
| Numerical simulations | 0,000 | 0,830 | 0,088 |
| Independent research in scientific literature | 0,000 | 0,367 | 0,114 |
| Critical assessment of others' results | 0,000 | 0,219 | 0,256 |

*Assumption tests for "Impact of participation in YPT activities on usefulness of RPC, YPT and other activities"*

| Hard Skills - RPC | Shapiro-Wilk test (p-value) | NCV test (p-value) | Durbin Watson test (p-value) |
|---|---|---|---|
| High school mathematics | 0,000 | 0,231 | 0,292 |
| High school physics | 0,000 | 0,195 | 0,530 |

| | Shapiro-Wilk test (p-value) | NCV test (p-value) | Durbin Watson test (p-value) |
|---|---|---|---|
| Solve close-ended problems | 0,000 | 0,000 | 0,052 |
| Designing experiments | 0,000 | 0,425 | 0,092 |
| Conducting experiment | 0,000 | 0,004 | 0,002 |
| Interpreting experimental data, data analysis | 0,000 | 0,326 | 0,400 |
| Developing own theoretical model | 0,000 | 0,523 | 0,502 |
| Numerical simulations | 0,000 | 0,083 | 0,000 |
| Independent research in scientific literature | 0,000 | 0,516 | 0,566 |
| Critical assessment of others' results | 0,000 | 0,878 | 0,006 |

| Hard Skills - YPT | Shapiro-Wilk test (p-value) | NCV test (p-value) | Durbin Watson test (p-value) |
|---|---|---|---|
| High school mathematics | 0,000 | 0,467 | 0,130 |
| High school physics | 0,000 | 0,168 | 0,306 |
| Solve close-ended problems | 0,000 | 0,266 | 0,790 |
| Designing experiments | 0,000 | 0,317 | 0,152 |
| Conducting experiment | 0,000 | 0,244 | 0,088 |
| Interpreting experimental data, data analysis | 0,000 | 0,177 | 0,036 |
| Developing own theoretical model | 0,000 | 0,235 | 0,250 |
| Numerical simulations | 0,000 | 0,297 | 0,420 |
| Independent research in scientific literature | 0,000 | 0,774 | 0,744 |
| Critical assessment of others' results | 0,000 | 0,584 | 0,842 |

| Hard Skills - Other | Shapiro-Wilk test (p-value) | NCV test (p-value) | Durbin Watson test (p-value) |
|---|---|---|---|
| High school mathematics | 0,000 | 0,336 | 0,690 |
| High school physics | 0,000 | 0,710 | 0,010 |
| Solve close-ended problems | 0,000 | 0,539 | 0,488 |
| Designing experiments | 0,000 | 0,950 | 0,078 |
| Conducting experiment | 0,000 | 0,500 | 0,510 |
| Interpreting experimental data, data analysis | 0,000 | 0,445 | 0,876 |
| Developing own theoretical model | 0,000 | 0,054 | 0,558 |
| Numerical simulations | 0,000 | 0,966 | 0,132 |
| Independent research in scientific literature | 0,000 | 0,240 | 0,268 |
| Critical assessment of others' results | 0,000 | 0,365 | 0,742 |

*Assumption tests for "Impact of participation in non-YPT competitions on usefulness of RPC, YPT and other activities"*

| Hard Skills - RPC | Shapiro-Wilk test (p-value) | NCV test (p-value) | Durbin Watson test (p-value) |
|---|---|---|---|
| High school mathematics | 0,000 | 0,628 | 0,106 |
| High school physics | 0,000 | 0,947 | 0,522 |
| Solve close-ended problems | 0,000 | 0,844 | 0,140 |
| Designing experiments | 0,001 | 0,837 | 0,008 |
| Conducting experiment | 0,000 | 0,599 | 0,000 |

| | Shapiro-Wilk test (p-value) | NCV test (p-value) | Durbin Watson test (p-value) |
|---|---|---|---|
| Interpreting experimental data, data analysis | 0,000 | 0,720 | 0,066 |
| Developing own theoretical model | 0,005 | 0,883 | 0,208 |
| Numerical simulations | 0,000 | 0,817 | 0,000 |
| Independent research in scientific literature | 0,000 | 0,912 | 0,290 |
| Critical assessment of others' results | 0,000 | 0,448 | 0,246 |

| Hard Skills - YPT | Shapiro-Wilk test (p-value) | NCV test (p-value) | Durbin Watson test (p-value) |
|---|---|---|---|
| High school mathematics | 0,000 | 0,499 | 0,512 |
| High school physics | 0,000 | 0,341 | 0,410 |
| Solve close-ended problems | 0,000 | 0,748 | 0,566 |
| Designing experiments | 0,000 | 0,731 | 0,038 |
| Conducting experiment | 0,000 | 0,441 | 0,542 |
| Interpreting experimental data, data analysis | 0,000 | 0,420 | 0,272 |
| Developing own theoretical model | 0,000 | 0,721 | 0,212 |
| Numerical simulations | 0,000 | 0,850 | 0,188 |
| Independent research in scientific literature | 0,001 | 0,176 | 0,850 |
| Critical assessment of others' results | 0,000 | 0,440 | 0,490 |

| Hard Skills - Other | Shapiro-Wilk test (p-value) | NCV test (p-value) | Durbin Watson test (p-value) |
|---|---|---|---|
| High school mathematics | 0,000 | 0,961 | 0,768 |
| High school physics | 0,000 | 0,555 | 0,820 |
| Solve close-ended problems | 0,000 | 0,884 | 0,446 |
| Designing experiments | 0,000 | 0,442 | 0,006 |
| Conducting experiment | 0,000 | 0,400 | 0,546 |
| Interpreting experimental data, data analysis | 0,000 | 0,991 | 0,138 |
| Developing own theoretical model | 0,000 | 0,795 | 0,702 |
| Numerical simulations | 0,001 | 0,342 | 0,004 |
| Independent research in scientific literature | 0,000 | 0,005 | 0,142 |
| Critical assessment of others' results | 0,000 | 0,311 | 0,818 |

*Assumption tests for "Impact of RPC, YPT and other activities on self-evaluation"*

| Hard Skills - Self-evaluation | Shapiro-Wilk test (p-value) | NCV test (p-value) | Durbin Watson test (p-value) |
|---|---|---|---|
| High school mathematics | 0,001 | 0,418 | 0,018 |
| High school physics | 0,004 | 0,006 | 0,596 |
| Solve close-ended problems | 0,000 | 0,001 | 0,196 |
| Designing experiments | 0,001 | 0,013 | 0,296 |
| Conducting experiment | 0,000 | 0,002 | 0,888 |
| Interpreting experimental data, data analysis | 0,005 | 0,000 | 0,000 |
| Developing own theoretical model | 0,000 | 0,000 | 0,026 |
| Numerical simulations | 0,000 | 0,296 | 0,092 |
| Independent research in scientific literature | 0,000 | 0,704 | 0,986 |

| Critical assessment of others' results | 0,001 | 0,007 | 0,430 |

# Country differences – Hard Skills

## *Assumption tests for "Across-country differences"*

| Hard Skills – self-evaluation | Shapiro-Wilk test (p-value) | Levene test (p-value) | Durbin Watson test (p-value) |
|---|---|---|---|
| High school mathematics | 0,000 | 0,805 | 0,348 |
| High school physics | 0,000 | 0,663 | 0,890 |
| Solve close-ended problems | 0,000 | 0,987 | 0,964 |
| Designing experiments | 0,000 | 0,866 | 0,050 |
| Conducting experiment | 0,000 | 0,692 | 0,920 |
| Interpreting experimental data, data analysis | 0,000 | 0,363 | 0,394 |
| Developing own theoretical model | 0,000 | 0,113 | 0,362 |
| Numerical simulations | 0,000 | 0,974 | 0,960 |
| Independent research in scientific literature | 0,000 | 0,451 | 0,538 |
| Critical assessment of others' results | 0,000 | 0,113 | 0,820 |

| Hard Skills – RPC | Shapiro-Wilk test (p-value) | Levene test (p-value) | Durbin Watson test (p-value) |
|---|---|---|---|
| High school mathematics | 0,000 | 0,235 | 0,488 |
| High school physics | 0,000 | 0,023 | 0,170 |
| Solve close-ended problems | 0,000 | 0,392 | 0,162 |
| Designing experiments | 0,000 | 0,010 | 0,012 |
| Conducting experiment | 0,000 | 0,000 | 0,002 |
| Interpreting experimental data, data analysis | 0,000 | 0,000 | 0,034 |
| Developing own theoretical model | 0,000 | 0,090 | 0,076 |
| Numerical simulations | 0,000 | 0,381 | 0,254 |
| Independent research in scientific literature | 0,000 | 0,039 | 0,474 |
| Critical assessment of others' results | 0,000 | 0,000 | 0,186 |

| Hard Skills – YPT | Shapiro-Wilk test (p-value) | Levene test (p-value) | Durbin Watson test (p-value) |
|---|---|---|---|
| High school mathematics | 0,000 | 0,510 | 0,744 |
| High school physics | 0,000 | 0,568 | 0,970 |
| Solve close-ended problems | 0,000 | 0,265 | 0,878 |
| Designing experiments | 0,000 | 0,588 | 0,896 |
| Conducting experiment | 0,000 | 0,033 | 0,730 |
| Interpreting experimental data, data analysis | 0,000 | 0,001 | 0,638 |
| Developing own theoretical model | 0,000 | 0,190 | 0,110 |
| Numerical simulations | 0,000 | 0,197 | 0,170 |
| Independent research in scientific literature | 0,000 | 0,114 | 0,440 |
| Critical assessment of others' results | 0,000 | 0,198 | 0,832 |

The European Commission's support for the production of this publication does not constitute an endorsement of the contents, which reflect the views only of the authors, and the Commission cannot be held responsible for any use which may be made of the information contained therein.

| Hard Skills – Other | Shapiro-Wilk test (p-value) | Levene test (p-value) | Durbin Watson test (p-value) |
|---|---|---|---|
| High school mathematics | 0,000 | 0,254 | 0,574 |
| High school physics | 0,000 | 0,256 | 0,214 |
| Solve close-ended problems | 0,000 | 0,995 | 0,808 |
| Designing experiments | 0,000 | 0,015 | 0,268 |
| Conducting experiment | 0,000 | 0,640 | 0,718 |
| Interpreting experimental data, data analysis | 0,000 | 0,725 | 0,196 |
| Developing own theoretical model | 0,000 | 0,440 | 0,702 |
| Numerical simulations | 0,000 | 0,377 | 0,090 |
| Independent research in scientific literature | 0,000 | 0,732 | 0,132 |
| Critical assessment of others' results | 0,000 | 0,010 | 0,258 |